\documentclass[a4paper,12pt]{article}

\usepackage[utf8x]{inputenc}
\usepackage{amsmath,amssymb}
\usepackage{graphicx}
\usepackage{caption}
\usepackage{cite}
\numberwithin{equation}{section}

\usepackage{lipsum}
\usepackage{lmodern}
\usepackage{tcolorbox}

\usepackage{empheq}
\usepackage{mathtools}

\graphicspath{ {images/} }
 \usepackage[toc,page]{appendix}

\usepackage{subcaption}

\usepackage{wrapfig}

\usepackage{xcolor}
\usepackage{amsmath}
\usepackage{amsfonts}
\usepackage{hyperref}
\hypersetup{
    colorlinks=true,
    citecolor=blue,
    linkcolor=blue,
    filecolor=magenta,
    urlcolor=blue}

 \newcommand{\bea}{\begin{eqnarray}}
\newcommand{\eea}{\end{eqnarray}}

\newsavebox{\uuunit}
\sbox{\uuunit}
    {\setlength{\unitlength}{0.825em}
     \begin{picture}(0.6,0.7)
        \thinlines
        \put(0,0){\line(1,0){0.5}}
        \put(0.15,0){\line(0,1){0.7}}
        \put(0.35,0){\line(0,1){0.8}}
       \multiput(0.3,0.8)(-0.04,-0.02){10}{\rule{0.5pt}{0.5pt}}
     \end {picture}}

\begin{document}
 
 

\title{\bf  Phases of  Holographic Interfaces}

\author{Constantin Bachas and Vassilis Papadopoulos}


\maketitle
\begin{center}
\textit{$\,^1$ Laboratoire de Physique  de l'\'Ecole Normale Sup\'eri{e}ure,\\
CNRS, PSL  Research University  and Sorbonne Universit\'es \\
24 rue Lhomond, 75005 Paris, France}
\end{center}

\vskip 1cm

\begin{abstract}
We compute  the  phase diagram of the simplest   holographic bottom-up model
of conformal interfaces. The model consists of a thin domain wall between  three-dimensional
Anti-de Sitter (AdS) vacua,   anchored on a boundary circle. 
 We distinguish five phases depending on  the existence of a black hole, the intersection of its  horizon
with the wall,  and the fate of inertial observers. 
We show  that, like   the Hawking-Page phase transition,   the capture of the wall by the horizon
is also a  first order transition  and comment on its  field-theory  interpretation. 
 The  static  solutions  of the domain-wall equations include  
gravitational avatars  of the  Faraday cage,
 black holes with negative specific heat,  
 and an intriguing  phenomenon of  suspended vacuum bubbles corresponding  to an exotic  
interface/anti-interface fusion. 
  Part of our analysis   overlaps   with  recent work by Simidzija and Van Raamsdonk 
 but the  interpretation is different.

\end{abstract}

\newpage

 {\small\tableofcontents}

 \section{Introduction}
 
Begining  with   the classic  paper  of Coleman and De Lucia \cite{Coleman:1980aw} there have been 
  many  studies  of thin  gravitating domain walls  between   vacua with different 
values of the cosmological constant.
Such  walls   figure    in  models of localized gravity  \cite{Randall:1999vf, Dvali:2000hr, Karch:2000ct}, in holographic duals
of conformal interfaces \cite{Karch:2000gx,DeWolfe:2001pq,Bachas:2001vj}, 
 in efforts to embed inflation in string theory by studying   dynamical  bubbles  
\cite{Freivogel:2005qh,Freivogel:2007fx,Barbon:2010gn,Banerjee:2018qey}, and more recently, 
following  the  ideas in   
refs. \cite{Penington:2019npb,Almheiri:2019psf,Almheiri:2019hni},  
  in toy models of black hole evaporation 
 \cite{Rozali:2019day,Balasubramanian:2020hfs,Chen:2020uac,
 Emparan:2020znc,Bak:2020enw,Chen:2020hmv,Akal:2020twv,Deng:2020ent,Geng:2020fxl}. 
 Besides  being a  simple    form   of matter coupled  to gravity,   domain walls are  
also a  key ingredient \cite{Brown:1988kg} in  effective  descriptions  of the 
  string-theory   landscape    -- see
  \cite{Ooguri:2020sua, Lanza:2020qmt,Bedroya:2020rac}
for  some  recent discussions of domain walls   in this context.\footnote{The above list of  
references is nowhere nearly complete. 
It is only meant as 
 an entry to the vast and growing literature in these subjects.}

    In this paper we  study   a thin static   domain wall  between  Anti-de Sitter (AdS) vacua,   anchored
at the conformal boundary of spacetime.
If a   dual holographic  setup  were to  exist,  it would have  two conformal field theories, 
CFT$_1$ and  CFT$_2$, 
separated by a conformal interface \cite{Karch:2000gx,DeWolfe:2001pq,Bachas:2001vj}. 
We will calculate  the phase diagram of the  system as  function of the AdS radii, the tension of the wall
and the boundary data. 
Several parts of this  analysis have  appeared
before (see below)  but the complete phase diagram has not,
 to the best of our knowledge,  been worked out.
 We will be  interested in phenomena that are hard to see at weak CFT coupling. 
A   broader  motivation, as in much of the  AdS/CFT literature, is   
 understanding    how  the  interior geometry is encoded on the boundary and vice versa,
but we will only briefly allude to this  question  in the present work.

Our analysis is classical in gravity. 
 Different phases   are distinguished  by the presence/absence of a black hole  
and  by the fate of inertial observers,  either those   moving freely  in the bulk or those  bound to the wall.  
 Inertial observers are  a guiding  fixture  of the  analysis, not emphasized in  earlier
works.  In the high-temperature or `hot' phase  all  inertial observers   eventually 
cross  the black-hole  horizon. 
In intermediate or `warm'  phases the wall 
avoids  the horizon,  and may also   shield   bulk observers from falling inside. 
 Such   two-center warm solutions are  gravitational  avatars of the Faraday cage.
 Finally  what differentiates  `cold' horizonless phases is whether all timelike geodesics   
  intersect  inevitably   the wall,   or not.

\begin{figure}[h!]
\centering
\begin{subfigure}{.5\textwidth}
  \centering
  \includegraphics[width=1.2\linewidth]{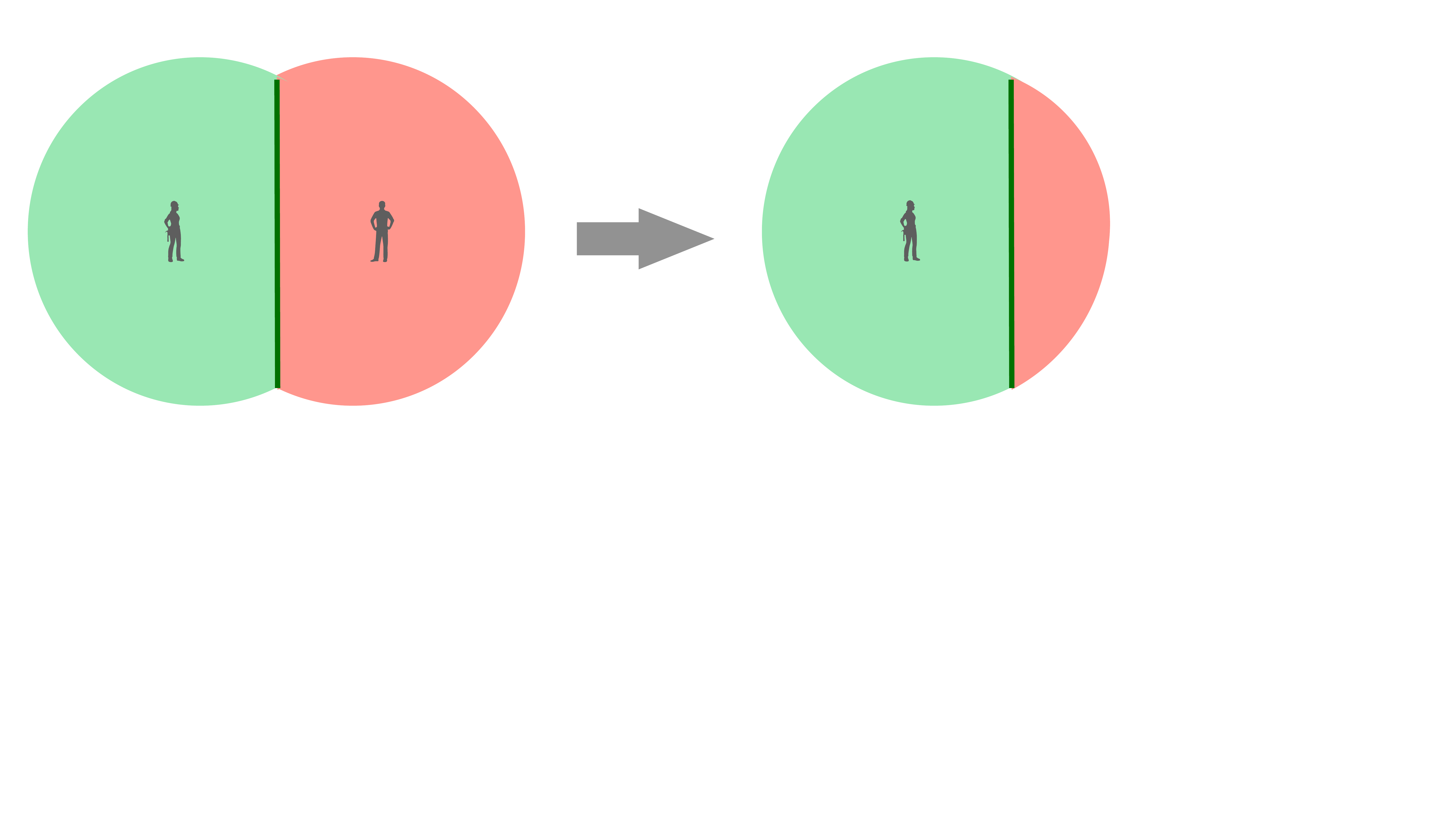}
  \label{fig:sub1}
\end{subfigure}%
\begin{subfigure}{.5\textwidth}
  \centering
  \includegraphics[width=1.2\linewidth]{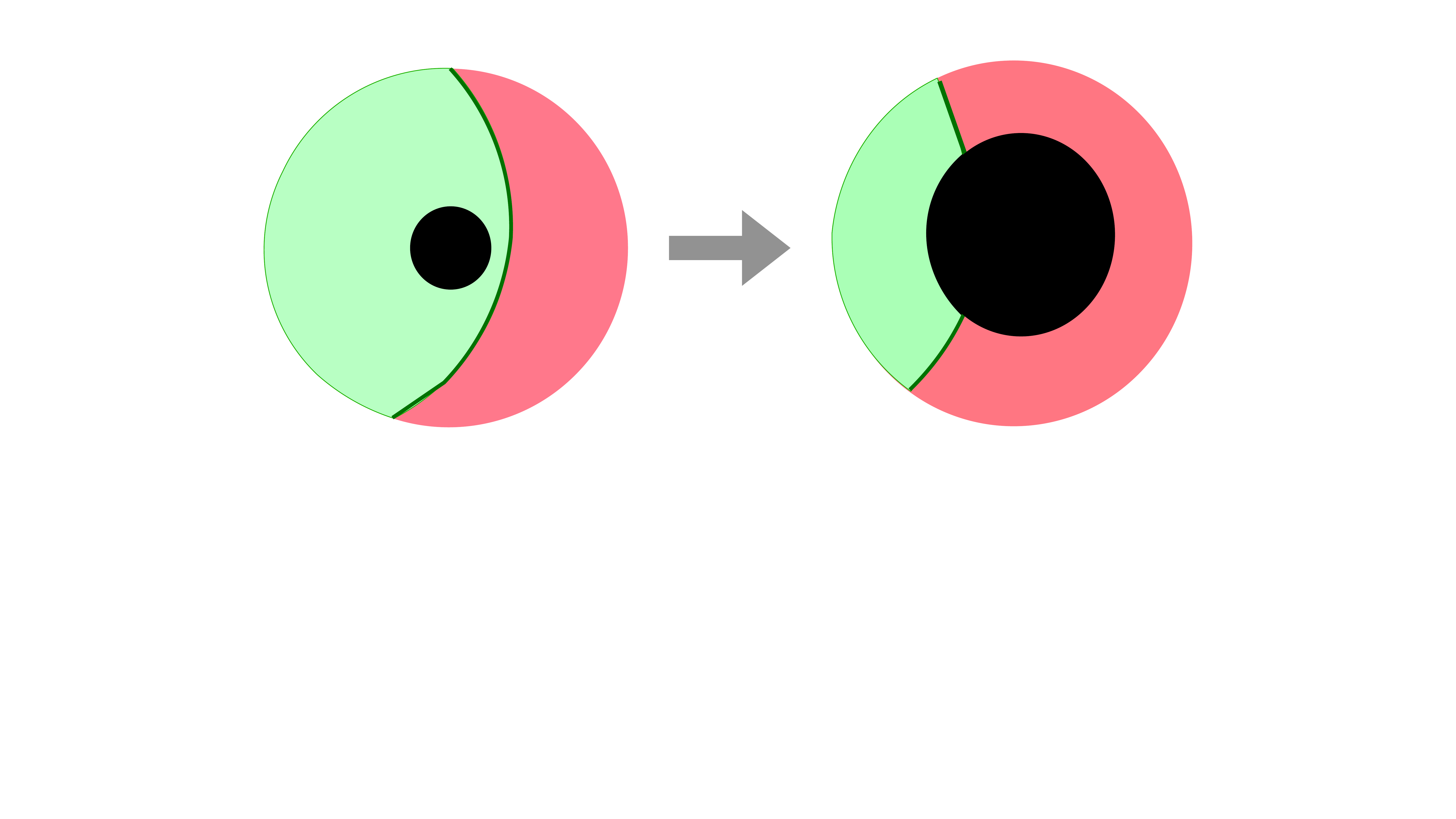}
  \label{fig:sub2}
\end{subfigure}
\vskip -24mm
 \caption{\footnotesize   A domain wall sweeping the center  of the `false AdS vacuum' where an
inertial observer could rest (left), or 
  entering the horizon of a black hole (right).
 }
\label{fig:test}
\end{figure}

Besides  the domain wall and the black hole, the third  actor  in the  problem   is    the
center   of global  AdS   where an inertial observer may   rest. 
The rich phase diagram    is the result of several competing forces: The attraction of  the  AdS trap, with or without
a  black hole in its center,
the tension  of   the wall, and the repulsion between the   domain wall and
 massive particles.
In addition to the first-order Hawking-Page transition  \cite{Hawking:1982dh} 
 that  signals    the formation of a black hole,   
 new phase transitions  occur when   the wall sweeps an AdS rest point   or    when part of  it  enters the 
 horizon, see
 figure \ref{fig:test}. One of our conclusions  is   that  the   latter transition is always first-order.
  \vskip 1mm

  We  work in 2+1 dimensions  because  calculations can be performed  in closed form. 
We  expect,  however,   qualitative features of the phase diagram to
 carry over to higher dimensions. For simplicity we consider a single type of non-intersecting wall,
and only  comment briefly  on extended models  that allow   junctions of different types of wall. 


 The capture of the wall by the black hole   is  related to a transition  analyzed 
in a very  interesting recent paper by Simidzija and 
Van Raamsdonk \cite{Simidzija:2020ukv}, see also \cite{Fu:2019oyc,May:2020tch}. 
  These authors  consider   time-dependent   spherically-symmetric   walls whose intersections 
with the conformal boundary describe 
Hamiltonian quenches in the dual field theory. 
 In this   setting the  boundary is  the   infinite cylinder,   with a stripe  
describing the evolution of the dual
CFT   between  the   quench and    `unquench' times.
 By contrast,   we  are   interested   in   equilibrium configurations. This means that  the  domain-wall  
geometry is static, and the 
 stripes on the conformal  boundary point   in  the time direction. Furthermore, the boundary is not the  cylinder
but  an 
orthogonal torus, adding an extra  parameter to the problem.
  
   Although the  interpretation  is different, 
many of our   formulae are nevertheless   related to those
 of  refs.\,\cite{Fu:2019oyc,Simidzija:2020ukv,May:2020tch}
by swapping  the roles of  boundary space and time  (thereby also swapping the BTZ geometry with thermal
AdS$_3$, see section \ref{sec:2}). 
This    is    fortuitous  
to   2+1 dimensions and   does not carry over to   higher dimensions. 
  
         The  gravitational action of the thin-wall model reads 
\bea\label{modelI}
\, \hskip -15mm  I_{\rm gr}  =    -\frac{1}{2}  \int_{{\mathbb{S}}_1}d^3x 
\sqrt{g_1}\,(R_1+\frac{2}{\ell_1^2}) \hskip -2mm  &-&  \hskip -2mm  \frac{1}{2}
 \int_{{\mathbb{S}}_2}d^3x\sqrt{g_2}\,(R_2+\frac{2}{\ell_2^2})  \nonumber \\
&\,& \hskip -18mm    +\, \lambda \int_{\mathbb{W}} d^2s
     \sqrt{\hat g_w}  \,+\, {\rm GHY\ terms}\,+\, {\rm ct.} \ , \
\eea
where $R_j(g_j)$ are the Ricci scalars of  the  spacetime slices  $\mathbb{S}_j$ on either side of the wall, 
 and  $\hat g_w$ is the induced metric on  the wall's worldvolume. The 
 Gibbons-Hawking-York terms and  counterterms  are given in appendix \ref{app:1}. 
 The action
$ I_{\rm gr}$  depends on  three parameters:  the two AdS radii $\ell_1,\ell_2$ and the wall tension $\lambda$. 
The radii  are related  to the central charges of the dual CFTs \cite{BrH},  and the tension  to the 
entropy   \cite{Azeyanagi:2007qj,Simidzija:2020ukv}  and  to the energy-transport  coefficient  \cite{Bachas:2020yxv}
of the dual interface. Static solutions exist for
\bea
\lambda_{\rm min} < \lambda < \lambda_{\rm max}\,, \qquad {\rm with}\quad
\lambda_{\rm max} = {1\over\ell_1} + {1\over\ell_2}\,;  \ \ 
\lambda_{\rm min} =  \bigl\vert {1\over\ell_1} - {1\over\ell_2}\bigr\vert\ . 
\eea
   The classical phase diagram depends on two dimensionless ratios of 
the above   (e.g. $\ell_2/\ell_1:=b\,$ and 
$\lambda\ell_2:=\kappa$)
and on  the two  parameters that determine the  conformal class 
of the striped boundary torus,  e.g.  $\tau_1:= TL_1$ and $\tau_2:=TL_2$, see  figure \ref{fig:1}. 
Without loss of generality we assume henceforth that $\ell_2\geq \ell_1$, i.e. that
$\mathbb{S}_1$ is the true-vacuum slice.


  An important question is how much of this  analysis has a chance to  carry over to   top-down holographic models, 
where   back-reacting  domain walls are  not   thin.\footnote{Many examples of supergravity domain walls 
 have been worked out in the literature, a representative sample is  
\cite{Bak:2003jk,Gomis:2006cu,Lunin:2007ab,DHoker:2007zhm,DHoker:2007hhe,DHoker:2008lup,
DHoker:2009lky,
Chiodaroli:2010mv,Chiodaroli:2011nr,Bachas:2011xa,Aharony:2011yc,Assel:2011xz,Bak:2011ga,
Bobev:2013yra,Bachas:2013vza,DHoker:2017mds,Lozano:2019ywa,Lozano:2020bxo,Arav:2020asu}. 
None  of these solutions depends, however,  on non-trivial (non-Lagrangian) boundary data, indeed all but one 
 are scale-invariant  AdS$_n$ fibrations.} 
The size  of the  horizon and the number   of  stable rest points are   order parameters that  can be also defined for
thick walls, but  a sharp criterion,  that  decides  whether  a thick domain wall 
enters or avoids the horizon is hard to imagine.  
Nevertheless, the field-theory interpretation of the transition  suggests  that such 
an order parameter
may exist, as we  will 
explain  in section \ref{puzzles}.

  It is worth  stressing  that  the thin-wall model is a minimal gravity dual of   I(nterface)\,CFT    
in the same way that pure Einstein
theory  is a minimal    dual  for homogeneous CFT.  The model  
captures  the two universal boundary operators --  the energy-momentum tensors on either  side of the interface,
as well as their combination refered to as the  displacement operator \cite{Billo:2016cpy}.
Top-down  models  
 have many more operators, some of 
which  correspond to   internal excitations of the domain wall.

 Note also that boundaries,   or end-of-the-world branes (EWBs),   can be considered  as a   limit  of domain walls  
when   one side  becomes  the zero-radius AdS spacetime \cite{Brown:2011gt}.  In this sense   holographic 
B(oundary)\,CFT  \cite{Takayanagi:2011zk,Fujita:2011fp}
can be recovered from  holographic ICFT, though  the limit is subtle  
 and should  be handled with care. 

  A  last remark concerns the 
 Ryu-Takayanagi  surfaces  \cite{Ryu:2006bv,Ryu:2006ef}  that 
delimit  the  entanglement  wedges   of boundary subregions
 \cite{Czech:2012bh,Headrick:2014cta,Wall:2012uf}.\footnote{See e.g.
\cite{VanRaamsdonk:2016exw,Rangamani:2016dms} for reviews.}
 It is clearly of   interest  to study if  these  surfaces intersect   the domain wall, as is done for BCFT
in  ref.\,\cite{Akal:2020twv,Deng:2020ent}.  We hope to return to this  question elsewhere.
\smallskip

  The  plan  of the paper and a summary of our    results  follows. 
In section \ref{sec:2} we review some standard facts about AdS$_3$/CFT$_2$ at finite temperature.
The   wall separates space   in two slices that we color green (true-vacuum side) and 
pink (false-vacuum side). Each  of these comes in one of four topological types described in section \ref{sec:3}. 
In section \ref{sec:4} we solve the matching equations obeyed by a  thin static domain wall,  which we
parametrize conveniently by the blueshift metric factor $g_{tt}$. This section  overlaps substantially 
with ref. \cite{Simidzija:2020ukv} via double Wick rotation  -- a trick specific  to  2+1  dimensions
as earlier noted.

    In  section \ref{sec:5}  we start  analyzing  the  solutions. By studying   the turning point
of the wall we classify the  possible phases, i.e.  the  
topologically distinct solutions.
We rule out in particular centerless geometries, in which no  inertial observer can avoid  the wall,
and  solutions with two black holes  whose merging is prevented by  the wall. 
  
   In  section \ref{eqstate} we write down the equations of state that characterize these phases. 
They  relate the canonical variables $\tau_1, \tau_2$ to  microcanonical variables  
that are  natural  for describing  the interior geometry.
 We also point out the relevance of a critical tension $\lambda_0
= \sqrt{\lambda_{\rm max}\lambda_{\rm min}}\,$ below which the hot solution disappears from a region
of parameter space.
 
In  section \ref{sec:6} we   compute the critical lines for sweeping transitions
in both the cold and the warm phases, and we show that the warm-to-hot transition  is always first-order --
the  domain  wall cannot be lowered continuously to the black horizon. The proof  requires a detailed analysis
of the region  $\mu\approx 1$ with $\lambda \leq \lambda_0$, where  the  hot  and a warm solution  come
arbitrarily close. We also point out some   
 puzzles regarding the   ICFT  interpretation of these phase  transitions.

Section \ref{sec:Faraday}  presents a striking   phenomenon:   bubbles of the true  vacuum suspended
from a point on the conformal boundary of the false vacuum.  This is surprising  from the perspective of
 ICFT, since it implies that the fusion of an interface and  anti-interface does not produce   the trivial (identity)
defect,   as expected from free-field  calculations \cite{Bachas:2007td}, 
but an exotic defect that generates spontaneously a new scale. 

   In section \ref{sec:8} we present numerical plots of the complete phase diagram  in the
canonical ensemble,  for different values of the Lagrangian parameters
$\lambda, \ell_j$. These  plots   confirm our earlier conclusions.  We  point out a   critical threshold 
$b= \ell_2/\ell_1 = 3$, probably an artifact of the thin-wall approximation, 
above which black-hole solutions on the false-vacuum side of the wall cease to exist.  We also
exhibit  coexisting   black-hole solutions, including black holes with negative specific heat. 
 This parallels the discussion of 
black holes in   deformed JT gravity
in ref.  \cite{Witten:2020ert}. 

    Section \ref{outlook} contains  concluding remarks.  In order not to  interrupt
the  flow  of the arguments we relegate  some   detailed calculations
to   four appendices.



 \section{Finite-temperature  AdS/CFT } 
 \label{sec:2}
 
    For completeness   we recall here  some  standard   facts  
    about  AdS/CFT  at finite temperature in three spacetime dimensions.  
      While  doing this    we will be also  setting  notation and conventions.  
 
    \subsection{Coordinates for the  AdS$_3$  black string}
          
           The metric of  the static AdS$_3$    black string    in 2+1 dimensions is
  \bea\label{BB}
  ds^2_{\rm BS}\,  =  \, {\ell^2 dr^2\over r^2 -M\ell^2}  - ({r^2} -M \ell^2 )\, dt^2   +   r^2 dx^2
  \, , 
    \eea  
     where $M>0$,  $\ell$\,  is the radius of   AdS$_3$ and the horizon at  $r^{\rm H} = \ell  \sqrt{M}$
   has   temperature   $T= \sqrt{M}/2\pi$. 
    Length units on the gravity side are such that  $8\pi G=1$. 
         The  dual CFT  lives on the    AdS$_3$  boundary,  at $r= {1/ \epsilon} \to \infty$,  with  conformal 
       coordinates\, $x^\pm \equiv    x\pm t\,  \in \mathbb{R}^2$.
         Its central charge 
       is $c=  12 \pi \ell$ \cite{BrH}.  
    \vskip 1mm

       The holographic  dictionary  becomes
          transparent  in   Fefferman-Graham coordinates, in which any
  asymptotically-Poincar\'e AdS$_3$
          solution  takes the following form
          \cite{Banados:1998gg, Skenderis:1999nb}
     \begin{equation}
     \label{FGmetric}
  ds^2 =  
   {\ell^2 dz^2  \over z^2}  +   {1 \over z^2} \Bigl(dx^+ +  {\ell z^2  } h_-dx^-  \Bigr) 
    \Bigl(dx^-  +  {\ell z^2  } h_+ dx^+ \Bigr) \ . 
   \end{equation}
    Here   $ h_{\pm }  =
  \langle T_{\pm\pm}\rangle $  are  the  expectation values of  the canonically-normalized  
  energy-momentum tensor of the CFT. 
  Note that it  is a special feature of  2+1 dimensions
  that  the Fefferman-Graham expansion  stops at   order $z^2$.
  For  the  static  black string,    $h_+ = h_- =   M\ell/4\,$ 
  giving   $ \langle T_{tt}\rangle  = M\ell/2 = 
   (c/6) \pi T^2$. This is indeed 
    the  energy density of   finite-temperature CFT in two dimensions.
  The relation between $z$ and $r$ is  
      \bea
   {r } = {1\over z}  + {M\ell^2 z\over 4}\ \ \Longleftrightarrow\ \ 
   z = {2\over M\ell^2} \left( r -  \sqrt{{r^2 }  - M\ell^2}\,\right)\ 
    ,  
    \eea
and     the black-string  metric  in  the $(z,t,x)$  coordinates reads
 \bea\label{FGexp} 
 ds^2_{\rm BS}\,  =\,  \ell^2 {dz^2\over z^2} -  \Bigl(   {1\over z}  -  {M\ell^2 z\over 4}   \Bigr)^2 dt^2
 +   \Bigl(   {1\over z}  +  {M\ell^2 z\over 4}   \Bigr)^2 dx^2\ . 
    \eea   
   Note that   $z$  covers  only  the region outside the   
  horizon  ($r > r^{\rm H} $)  and that   near the conformal boundary  $z \approx  r^{-1} $. 

      A  last  change  of coordinates  worth recording, even though we will not use it in
      this paper,     is the one that maps 
       $(z,t,x)$   to  the standard Poincar\'e parametrization of   AdS$_3$. 
      Such a map  is guaranteed to exist because all 
      constant-negative-curvature Einstein manifolds in three dimensions 
       can be obtained from AdS$_3$ by   identifications and excisions. For the case at
       hand\,\footnote{The general transformation, 
for  an arbitrary (conformally-flat) boundary metric and vacuum expectation value $\langle T_{ab}\rangle$, 
 is given in  
       refs.\,\cite{Rooman:2000ei, Krasnov:2001cu, Compere:2015knw}.}   
       the transformation   reads
      \bea
       w^\pm =  \zeta^\pm \left({4- M\ell^2 z^2 \over 4+ M\ell^2 z^2}\right) \, , \quad
        y = {4z (M\zeta^+ \zeta^-)^{1/2}\over 4+ M\ell^2 z^2}\quad
        {\rm with}\quad \zeta^\pm = e^{\sqrt{M} ( x \pm t) }\ . \ \ 
      \eea    
       The reader can check that in these coordinates the metric \eqref{FGexp} becomes
       \bea
       ds^2_{\rm BS}\,  =\,   {\ell^2 dy^2  +  dw^+ dw^-\over y^2} \ ,
       \eea
   i.e. the standard Poincar\'e  form of  AdS$_3$  as advertized.   
  Outside the black  horizon ($ M\ell^2z^2 <4 $) 
    the coordinates $x^\pm \equiv    x\pm t$
  cover only a Rindler wedge of  the  $w^\pm$ plane.

               Since we will be refering to this later, let us  verify   the well-known fact that no inertial
observer can avoid crossing the horizon. In  the proper-time parametrization of the trajectory
a simple calculation gives
\bea\label{rddot}
\ell^2\,  {{\ddot r}\over r} = -1 - M\ell^2 {\dot x}^2
\eea
where dots denote  derivatives with respect to proper time. Since  $M$ is positive 
 there is no centrifugal  acceleration   {\small QED}.  Note that this is a property of the asymptotically AdS
 black hole, not shared by  asymptotically flat black holes  in higher dimension.

  
 \subsection{Hawking-Page transition}
 \label{HP}
 
       From the perspective of the CFT, the temperature $T$ is the only  dimensionful 
       parameter of  the infinite-black-string solution. By a scale transformation we can always set it to one. 
       Things get  more interesting if the black string
       is compactified,   $x \sim x+ L$, thereby  
       converting   the  solution \eqref{BB}
         to the   non-spinning BTZ black hole 
          \cite{Banados:1992wn,Banados:1992gq}.  In addition to the central charge $c$, 
      there is   now a new dimensionless parameter $LT$.  In  the Euclidean geometry   $\tau = i\,  L T$
     is the  complex-structure modulus of the  boundary torus.

  
         At the critical temperature   $T_{\rm HP} = 1/L$ the  theory   undergoes  
  a Hawking-Page phase transition  \cite{Hawking:1982dh, Maldacena:1998bw}. 
  This is seen by comparing the  action of  the  two  competing   
       saddle points  for the   interior geometry:\,\footnote{Thermal AdS$_3$ 
        and    Euclidean 
       BTZ are  part of an infinite SL(2,$\mathbb{Z}$) orbit of gravitational
        instantons,\,\cite{Maldacena:1998bw, Dijkgraaf:2000fq}
       but they are the only dominant ones for an orthogonal torus. 
      Their  regularized Euclidean actions  
       are $I_{\rm TAdS} =   -2\pi^2\ell / \vert\tau\vert$ and $I_{\rm BTZ} =   -2\pi^2\ell   \vert\tau\vert$, see below. 
       } 
         (i) the Euclidean BTZ black hole,  and  (ii) thermal AdS$_3$,  
       whose metric  is the same as  \eqref{BB} but   with  $M$ replaced by 
        $\tilde M=- (2\pi/ L)^2 $.  The difference of 
       free energies  of these two saddle points reads
       \bea\label{FrE}
      F_{\rm BTZ} -   F_{\rm TAdS} 
         =   - 2\pi^2  \ell  \, \bigl ( L T^2   - {1\over L} \bigr)\ . 
       \eea
       Thus  thermal AdS$_3$  is the 
        dominant solution  when   $L T< 1$,  while the  BTZ black hole dominates  when   $L T >  1$. 
                    
       
          Thermal AdS$_3$   and the Euclidean BTZ black hole 
       differ  in  the  choice of  boundary  cycle  that becomes contractible in the   
interior geometry.  The periodicity conditions, respectively  $x \sim x+ 2\pi/ \vert \tilde M\vert^{1/2}$ and 
$t_E \sim t_E +  2\pi/  M^{1/2} $,    ensure regularity when this contractible  cycle  degenerates. 
Below we will    encounter situations in which 
      either the center of  AdS or the BTZ horizon are  excised.  
       In such cases the  regularity  conditions  can be   relaxed. 
 
   \smallskip 
           One other  comment  in order here concerns   the 
            difference of free  energies, eq.\,\eqref{FrE}.  The renormalized
  gravitational action  $I_{\rm gr}$ (where  $I_{\rm gr} = F/T$) 
  is calculated       for the  general  interface model  in appendix\,\ref{app:1}.
  In the case of   a  homogeneous CFT one  can,  however,   obtain the answer  faster. Indeed, from 
      the Fefferman-Graham form 
of the metric,  eq.\,\eqref{FGmetric},  one reads  the   energy  of the  CFT state, 
 \bea\label{U}
E   =\,      L \langle T_{tt}\rangle \, 
  = \, {1 \over 2}\ell M L  \ . 
\eea
For  $M = (2\pi T)^2$ this is   the internal energy of the  high-temperature  state,  as previously noted,  
and for   $M = - (2\pi /L)^2$ it is 
 the Casimir energy of the   vacuum.  The corresponding free energies  obey the thermodynamic identity
 \bea\label{id}
 E   = - T^2 {\partial \over \partial T} \Bigr( {F\over T}\Bigl)\ . 
 \eea
Eqs.\,(\ref{U}) and (\ref{id})  determine  $F$ up to a  term linear  in $T$. This can be argued to
vanish
 both at low $T$,  since the ground state  has no entropy, and at  large $L$ since  $F$
must  be extensive. The final result is  eq.\,\eqref{FrE}. 
        
     Let us finally note that since in empty  AdS  the mass   $M$ is negative, there is   a centrifugal
contribution  in eq.\eqref{rddot}.  An inertial observer may thus either rest at,  or orbit around 
the center  $r=0$. But in the centerless slices that we are about to discuss, all inertial observers
 hit the wall.

  
  \section{Topology of slices} 
 \label{sec:3}

     Consider now   two conformal  field theories, CFT$_1$ and CFT$_2$, coexisting 
     at  thermal equlibrium on a  circle.   
       This  is illustrated in  figure 1. The  horizontal and vertical  axes 
   parametrize   space and Euclidean time. In addition to the central charges
    $c_1, c_2$,  and to the properties of the  
      interfaces  between    the   two CFTs, there are  three  
    more   parameters in this system:  
    the sizes  $L_1, L_2$ of  the regions in which each
    CFT lives,  and the equilibrium  temperature $T$. This gives two  dimensionless parameters,  
    which we can choose
    for instance to be $\tau_1:=TL_1$ and $\tau_2:=TL_2$. 
   
        \begin{figure}[tbh!]
\centering
\vskip  -0.4 cm
\includegraphics[width=.96\textwidth,scale=0.84,clip=true]{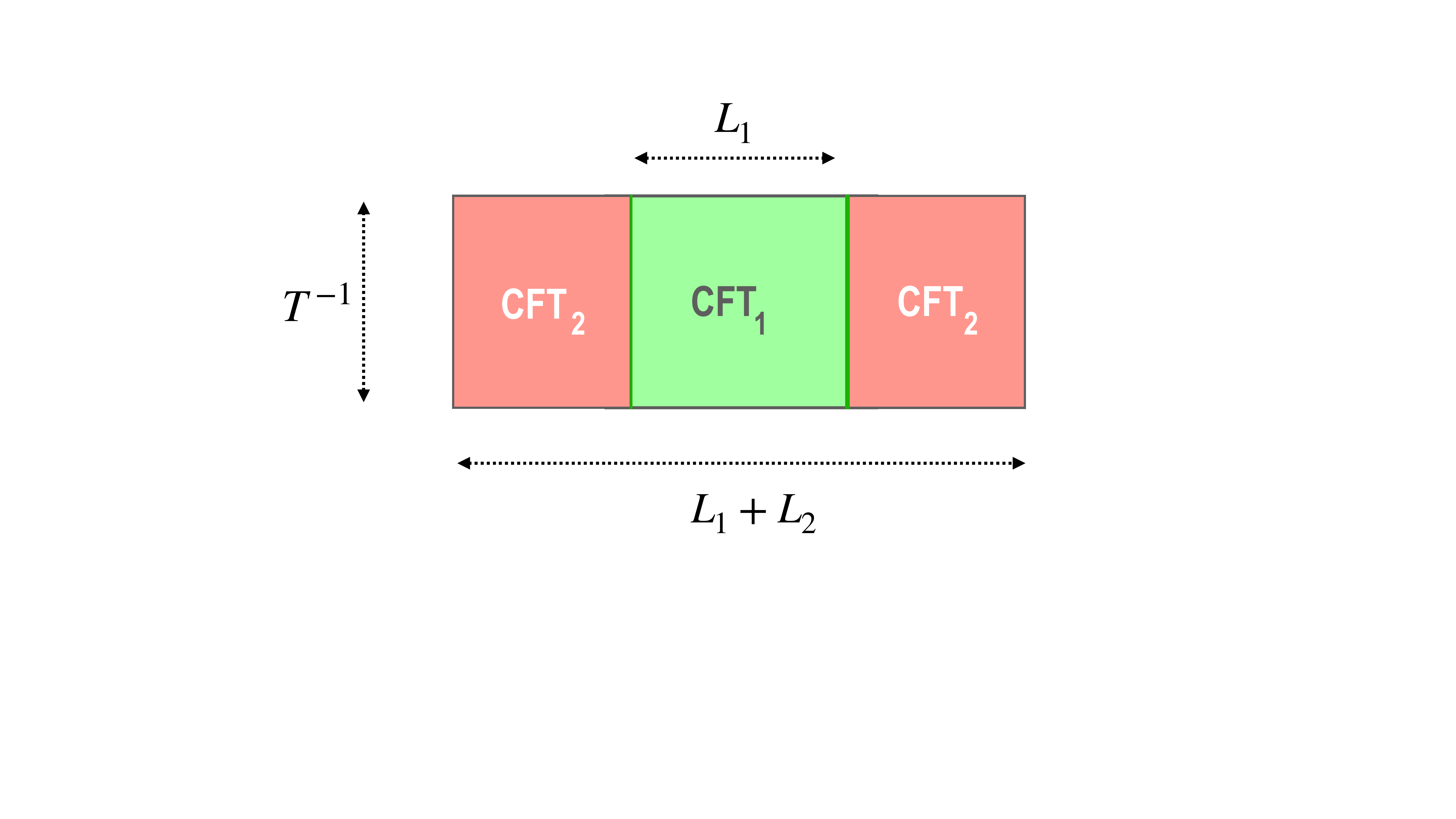}
   \vskip -24mm
   \caption{\footnotesize  The finite-temperature interface CFT at the AdS boundary.
Both space and Euclidean time are compact, so the depicted surface is an orthogonal torus.  }
   \label{fig:1}
 \end{figure}
 
                              The  gravity dual of this  ICFT 
                              features   domain walls, i.e. strings in 2+1 dimensions,\,\footnote{We reserve the word
                               ``interface'' for   the CFT,  and ``domain wall'' or 
       ``string''  for  gravity.  Interfaces  are   anchor points of  domain walls on the AdS
       boundary. The   string  of  our bottom-up  model   should not be confused with 
   the black string 
 responsible for  the interior horizon. In top-down supergravity embeddings 
 the  two types of string  may  be however   interchangeable.  
  }  anchored at the  interfaces  on  the 
       conformal  boundary. We will make the simplifying
     assumption that  the two domain walls differ only in orientation,  and can join smoothly
in the interior of spacetime. Extended models  allowing  junctions of different domain walls are very  interesting
but they are beyond our  present scope. We  will  comment briefly on them in a later section.

\begin{figure}[thb!]
\centering
\includegraphics[width=.99\textwidth,scale=0.84,clip=true]{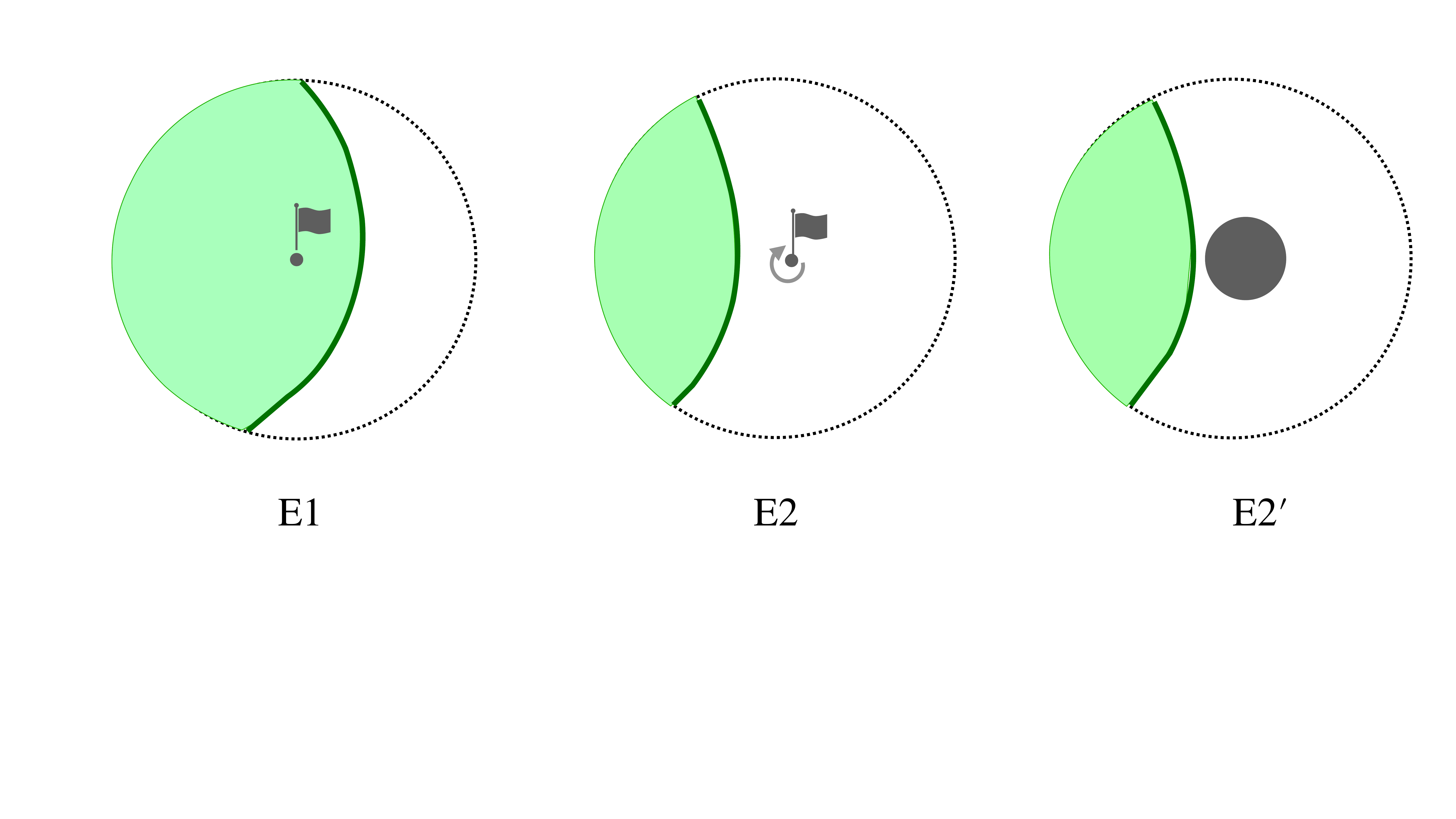}
   \vskip -23mm
 \centering
 \includegraphics[width=.99\textwidth,scale=0.84,clip=true]{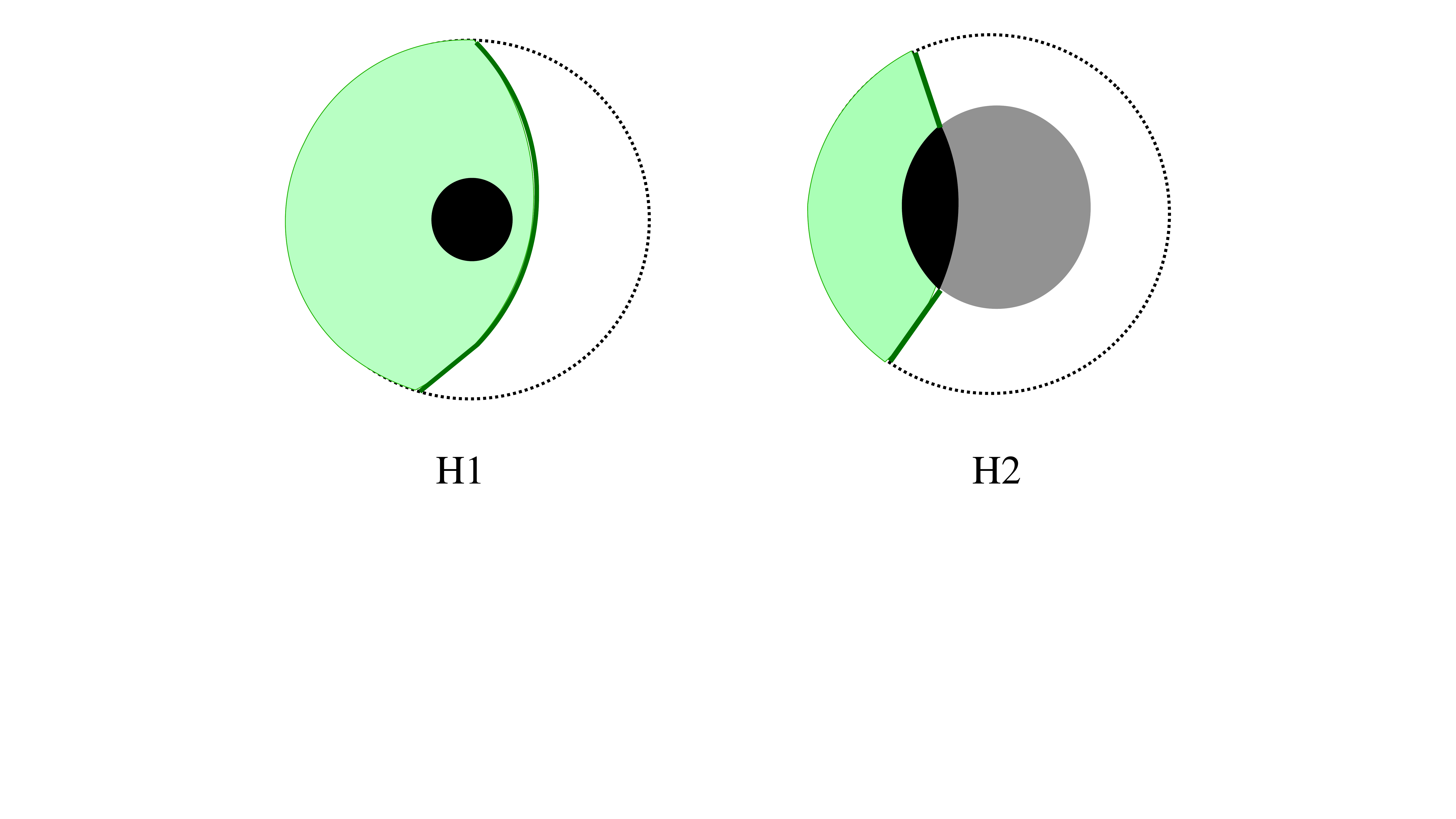}
   \vskip -27mm
   \caption{\footnotesize  The different types of   space-time slice   described in the main
  text.  The actual slice is colored in green, the  complementary region  is excised. 
     The letters `E' and `H' stand for `empty' and `horizon', and the grey flag  denotes the  rest point 
     of an inertial observer.  Note that since this is excised in E2, a conical singularity in its place is permitted.
The centerful slice  E1  can act as a gravitational Faraday cage.
   }
\label{fig:2}
 \end{figure}

      The green and pink   boundary  regions of fig.\,\ref{fig:1},  
  in which  CFT$_1$ and CFT$_2$ live,    extend  in   the interior  to slices  of gravitational 
  solutions   that belong    to one of  
  several   topological  types. 
    These are   illustrated  for the green slice   in  figure \ref{fig:2}.  
    Each slice is either part of thermal  AdS$_3$ with the center, marked by  a grey  flag, 
   included ({\small  E1}) or   excised ({\small  E2}), or part of  the BTZ geometry   with the   horizon 
     excised ({\small  E2$^\prime$}),   included ({\small  H1})
   or   intersecting    the  domain wall ({\small  H2}). 
   The same  options are   available   for the pink   spacetime slice.\footnote{ The Euclidean manifold is
   a (thermal) circle fibration over the fixed-time slice drawn in our figures. The fiber  degenerates at the
   horizon, when one exists.}

 As was explained in section \ref{sec:2}, we may   adopt  the unified parametrization \eqref{BB}  for  
  all   types of   slice,  
   with  $M$  negative for the slices  of type {\small  E1} and {\small  E2}\,  of global  AdS$_3$,    
   and positive for the slices  of type {\small E2$^\prime$, H1}  and  {\small H2}  
    of  the BTZ   spacetime.   We are interested in static configurations which are dual to 
    equilibrium CFT states, so  time   is globally defined and has  fixed imaginary period 
    $t_E \sim t_E +  1/T$. 
The   coordinates $(x,r)$ on the other hand  
   need not be  continuous across the wall. We therefore write the spacetime
  metric   in terms of   two coordinate charts, 
                 \bea\label{charts}
 ds^2 \,  =  \, {\ell_j^2 dr_j^2\over r_j^2 -M_j \ell_j^2}  - ({r_j^2} -M_j \ell_j^2 )\, dt^2   +   r_j^2 dx_j^2
  \qquad {\rm with }\quad  (x_j, r_j) \in \Omega_j \, ,    \, \, 
  \eea
where  $\Omega_1$ is  the range of coordinates
for  the green  slice  and $\Omega_2$ the range of coordinates  for  
  the pink slice. These  ranges are delimited as follows: 
\begin{itemize}
 
\item by the embeddings  of the
static wall    in the two   coordinate systems, 
   $\{x_j(\sigma), r_j(\sigma) \}$,       where $\sigma$ parametrizes 
 the wall\,; 

\item    by  the horizon   
  whenever   the slice contains one,  i.e.\,\,in   cases  {\small  H1}
   and {\small  H2}\,; 

\item by   the  cutoff   surface
     $r_j  \approx  1/\epsilon\to    \infty$ \ .

\end{itemize}

The mass parameters of the slices,  $M_1$ and $M_2$,   are in general   different. 
     Regularity  requires  however  that
       \bea
     M_j  = (2\pi T)^2 \qquad  {\text {for\   slices\  \ {\small{H1, H2}}}} 
    \eea
      that  include a horizon, whereas  $M_j$ is unconstrained for the other slice types. 
  Furthermore,  
  for a slice of type   {\small  E1} 
  in which  the spatial circle is contractible, interior regularity  fixes  the periodicity of $x$, 
       \bea
     x_j \sim x_j + 2\pi/\sqrt{-M_j}\qquad  {\text { in\ case\  {\small  E1}}} \ . 
    \eea
For    {\small  E2,  E2$^\prime$}  and {\small  H2}   the coordinate $x_j$ is not periodic, while 
for   {\small  H1}   its
    period,   proportional  to the horizon  size, is unconstrained.

     Since  the horizon is a closed surface,   a   green slice of type {\small  H2}  can only be paired  with a  
     pink slice
 of the same type. This  is the  topology   that   dominates 
  at very high   temperature when  the black hole
 eats up most of  the bulk spacetime.  
  As the temperature is lowered different   pairs  of the remaining
  slice  types  dominate.   
  The  pairs that correspond to actual solutions of  the domain-wall  equations 
  will be determined  in   section \ref{sec:5}.

       For the time being  let us comment on the  differences  between the horizonless slices
         in the top row  of  fig.\,\ref{fig:2}. 
         What distinguishes  {\small  E1} from the other two is the existence of  the AdS center  (or `refuge')
       where an inertial observer  may  sit at rest. By  contrast, in  
   the slices of type {\small  E2} and {\small  E2$^\prime$} all inertial observers  
  will  inevitably  hit  the domain wall as explained in the previous subsection. 
  This discontinuous behavior 
 differentiates the phases on either side of a  sweeping transition.

    Note   that there is no topological difference between the  slices of type {\small  E2} and {\small  E2$^\prime$}, 
which 
     is  why  we distinguish them only by   a prime. These slices   differ only   in   the sign of $M_j$, or equivalently
      the energy density per degree of freedom   in  the boundary theory.  Together  
{\small  E2} and {\small  E2$^\prime$}
 describe  a continuum 
        ($ -\infty <M_j< \infty$)   of 
    horizonless   slices  with no  rest point.


  \section{Solving the wall equations } 
 \label{sec:4} 
 
  In this section we find  the general solution of the  domain wall equations
 in terms  of   the mass parameters $M_1, M_2$, the 
  AdS radii $\ell_1, \ell_2$,  and the   tension of the wall $\lambda$. 
That a solution always exists for any bulk geometries is a special  feature 
of  2+1 dimensions, as is the double Wick rotation
that relates this part of our analysis to  ref. \cite{Simidzija:2020ukv}. 


\subsection{Matching  conditions} 
\label{sec:41} 

 The matching conditions  at  a thin domain wall  have appeared in numerous studies
of   cosmology and AdS/CFT. They are especially simple in the case
 at hand, where the wall/string  is static and is characterized only by its tension. 
 Matching the induced worldsheet 
metric of  the two charts   \eqref{charts}
 gives one  algebraic and one  first-order differential    equation  for the  
 embedding functions  $x_1(\sigma), r_1(\sigma), x_2(\sigma)$ and $ r_2(\sigma)$: 
\vskip -3mm
\bea\label{00}
  {r_1^2} -M_1 \ell_1^2\,   = \,{r_2^2} -M_2 \ell_2^2\, \, \equiv\, f(\sigma) 
\eea
\bea\label{11} 
{\rm and}\qquad\ \ 
f^{-1} {\ell_1^2\,    r_1^{\prime\, 2}   } +  r_1^2 \, x_1^{\prime\, 2} =
f^{-1} {\ell_2^2\,   r_2^{\prime\, 2} } +  r_2^2\,  x_2^{\prime\, 2} \, \equiv\, g(\sigma)  , 
\eea
where the prime  denotes  a derivative  with respect to $\sigma$. 
We have  defined the  auxiliary  functions $f$ and $g$   in terms of which the induced worldsheet metric reads
$d{\hat s}^2\vert_{\mathbb{W}} =  -f (\sigma)dt^2 + g (\sigma) d\sigma^2$. 
A third  matching equation\footnote{The Israel-Lanczos  
matching  conditions 
 are matrix equations, 
 $[K_{\alpha\beta}] - [\textrm{tr} K] \hat g_{\alpha\beta} =    \lambda  \, \hat g_{\alpha\beta}$,
  where $K_{\alpha\beta}$ is the extrinsic curvature, $\hat g_{\alpha\beta}$ the induced metric,  and 
  brackets  denote   the discontinuity  across the wall. Only the trace part of this equation is non-trivial. 
  The traceless part of $K$  is automatically continuous by virtue of 
   the momentum constraints
$
D^\alpha K_{\alpha \beta} - D_\beta K = 0\, ,
$
where $D_\alpha$ is the covariant derivative with respect to the induced metric.
 Equation \eqref{Israel}  is   the $tt$ component of the matrix equation. 
} 
expresses
  the  discontinuity  of the extrinsic curvature   in terms of the tension, 
 $\lambda$,    of the wall  \cite{Israel,Lanczos}.
 It can be written as follows\,:  
\bea\label{Israel}
 {r_1^2 x_1^\prime \over \ell_1} + 
   {r_2^2 x_2^\prime \over \ell_2} =   \lambda\, \sqrt{fg}\ . 
\eea
 Our  convention is  that 
  $\sigma$ increases as 
 one circles  $\Omega_j$       in the $(x_j, r_j)$ plane
 clockwise. Other   conventions    introduce  
signs  in front of the two terms on the left-hand side of this equation.



  Eqs.\,\eqref{00}--\eqref{Israel} are  three equations for four unknown functions,  but one of these 
  functions   can be
specified  at will using  the string-reparametrization  freedom. 
Furthermore the  equations only  involve
first  derivatives of $x_j$, so the  integration constants    are  irrelevant choices of
the origin of the $x_j$ axes. For given    $\ell_1, \ell_2$ and   $\lambda$,  
the  wall  embedding  functions   $x_j(r_j)$,  are  thus uniquely determined 
 by the  parameteres $M_1$ and $M_2$. 
Different choices  of   ($M_1, M_2$)  may correspond, however,  to the same boundary data  ($L_1, L_2, T$). 
These  are the competing phases of the system.


\subsection{Solution near the  boundary}
\label{sec:42} 

\begin{figure}[tbh!]
\centering
\includegraphics[width=.96\textwidth,scale=0.84,clip=true]{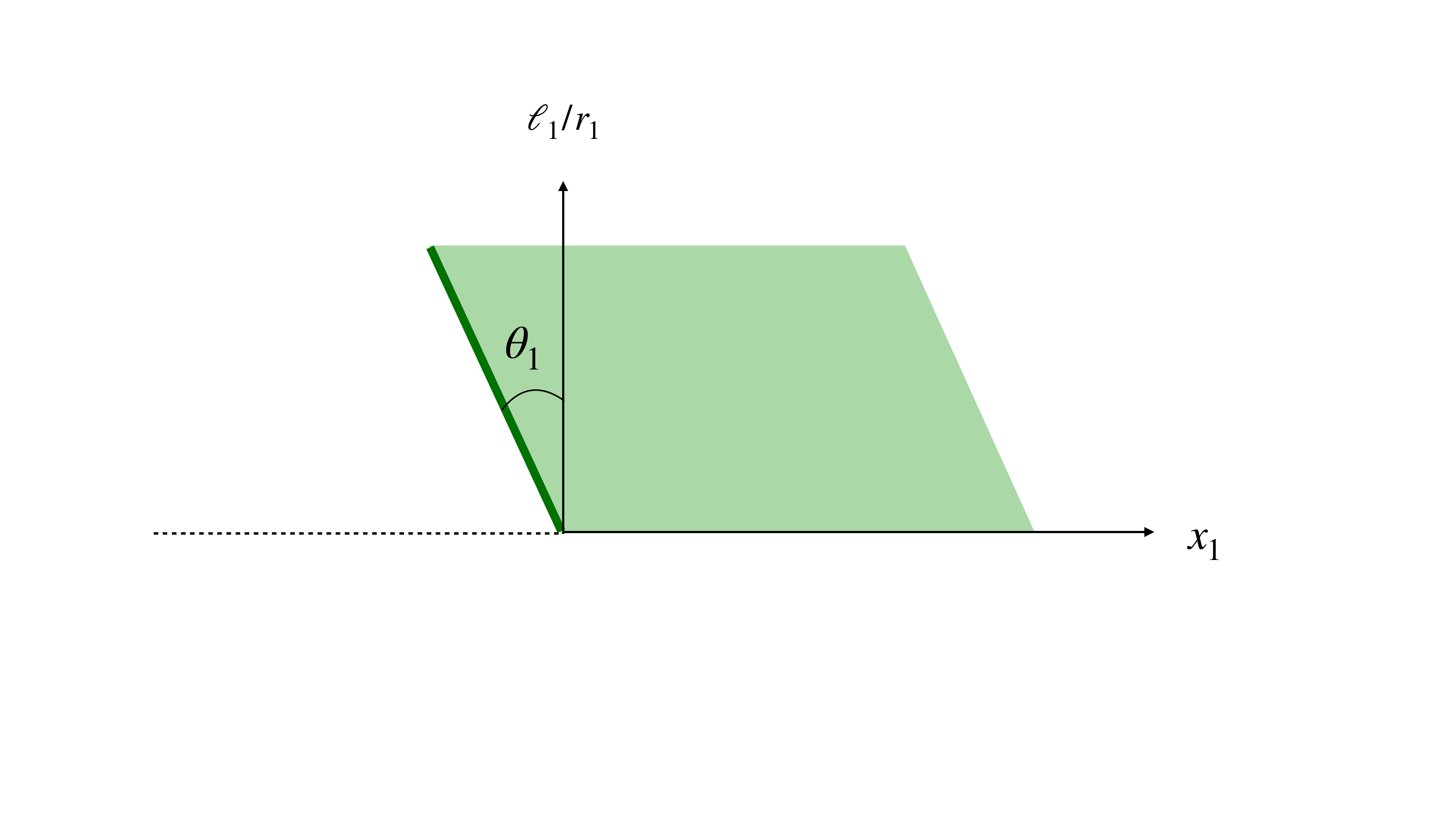}
   \vskip -21mm
   \caption{\footnotesize   Near the AdS  boundary      in the $(x_j, \ell_j/r_j)$ plane 
    the string  is a  straight line 
     subtending  an angle $\theta_j $
    with the normal.}
   \label{fig:3}
 \end{figure}

    Near the conformal boundary, $r_j\to\infty$,   the parameters $M_j$ can be neglected 
    and   the   worldsheet metric asymptotes to 
   AdS$_2$ by virtue of scale invariance. Explicitly the  solution reads   \cite{Bachas:2002nz}
 \bea\label{44}
  r_1 \approx  r_2\,, \qquad  x_j   \,  \approx \,    -    \ell_j\,  (\tan\theta_j) \,  r_j^{-1}\ , 
 \eea
where    $\theta_j $  is  the angle   in the 
 $(x_j \,,  \ell_j/r_j )$ plane  between the  normal to the   boundary and the   interface, see  figure
 \ref{fig:3}.  The  matching eqs.\,\eqref{11} and \eqref{Israel} relate these angles 
   to the  bulk  radii $\ell_j$  and to   the string tension  $\lambda$:
     \bea\label{AdSmatch}
  { \ell_1\over \cos\theta_1} = {\ell_2\over \cos\theta_2 } \, \equiv \,  \ell_{w}   \qquad 
   {\rm and } \qquad  \tan \theta_1  +   \tan\theta_2 = \, {  \lambda }\, \ell_{w} \ , 
   \eea                          
where $\ell_{w}$ is the radius of the AdS$_2$   worldsheet,  and $-{\pi/2} < \theta_j <{\pi/2}$\,.

     \smallskip
  Without loss of generality  we assume that $\ell_1 \leq \ell_2$, so  that 
       CFT$_1$ has the smaller of the two central charges. Its   gravity dual  has the
         lower vacuum energy, i.e. the  green slice is  the `true vacuum'   side of the 
         domain wall and   
   the pink  slice  is  the `false-vacuum'  side.   The  first  eq.\,\eqref{AdSmatch} then  implies that 
    $\vert\tan\theta_1 \vert\geq \vert\tan\theta_2\vert$\, and,  provided   that  the tension is positive,
    the second   eq.\,\eqref{AdSmatch} implies that $\theta_1 >0$. 
        The sign of $\theta_2$,  on the other hand, 
    depends on the  precise value of $\lambda$.  
  Expressing  the tangents in terms of cosines brings indeed  this   equation to  the form
    \bea
 ({1\over \ell_1^{\,2}} -  {1\over \ell_{  w}^{\,2}}  )^{1/2}  \,+ \,  \varepsilon \, 
  \sqrt{{1\over \ell_2^{\,2}} -  {1\over \ell_{w}^{\,2}} }\, =\,    \lambda\ 
  \qquad {\rm with} \quad \varepsilon = {\rm sign}(\theta_2)\ .
 \eea
  Since $\lambda$ is  real we must have   $\, \ell_2 < \ell_{w}< \infty$. Furthermore to 
   each value of  the worldsheet radius $\ell_{w}$ there  correspond
    two values of  the tension $\lambda$,   depending on
  sign($\theta_2$). Explicitly,\,\footnote{It  
 was  argued   in   ref. \cite{Czech:2016nxc} that the walls in the $\lambda <  \lambda_0$ range 
are unstable. But   the  radius instability in this reference   reduces  the action by an amount proportional to 
the infinite volume of AdS$_2$ and 
does not correspond to a normalizable mode. 
 The only normalizable mode of the wall in the thin-brane  model
corresponds  to the displacement operator which is  an   irrelevant 
(dimension = 2) operator  \cite{Bachas:2020yxv}. }
\bea\label{allowedrange}
 \lambda_{\rm min} < \lambda <  \lambda_0    \quad 
   \  {\rm for}\    \varepsilon=-  \qquad  {\rm and} \qquad 
  \lambda_0  <  \lambda  <  \lambda_{\rm max}  \quad 
  {\rm for}\   \varepsilon=+\ , 
\eea 
 where  the three critical tensions read
 \bea\label{3cr}
  \lambda_{\rm min}  =   {1\over \ell_1} - {1\over \ell_2} \ , \qquad
 \lambda_{\rm max} = {1\over \ell_1}+  {1\over \ell_2}\ , \qquad
 \lambda_0  = \sqrt{\lambda_{\rm max}\lambda_{\rm min}}\ . 
\eea
Let us pause here to discuss the  significance  of these critical tensions.
 

 \subsection{Critical tensions}
\label{sec:43}
 
The  meaning of the  critical tensions $\lambda_{\rm min}$ and $\lambda_{\rm max}$   has been    understood
  in  the work of Coleman-De Lucia  \cite{Coleman:1980aw} and Randall-Sundrum 
  \cite{Randall:1999vf}.  
 Below  $\lambda_{\rm min}$   the false vacuum
        is unstable to  nucleation of true-vacuum bubbles, so  the two phases cannot coexist
         in equilibrium.\footnote{Ref.\,\cite{Coleman:1980aw}  actually computes
        the critical tension for  a  domain wall separating  Minkowski from  AdS spacetime. Their result 
        can  be  compared to
         $\lambda_{\rm min}$   in the limit   $\ell_2\to \infty$. }
          The holographic description of such  nucleating  bubbles  raises   fascinating
 questions  in its own right, see e.g. 
  refs.\cite{Freivogel:2005qh,Freivogel:2007fx,Barbon:2010gn}. It has been also advocated  
   that expanding true-vacuum bubbles  could   
  realize accelerating 
  cosmologies  in string theory \cite{Banerjee:2018qey}. Since  our  focus here is on equilibrium  configurations,
   we will not  discuss  these interesting issues any further. 
          
          The maximal tension $\lambda_{\rm max}$ is a stability bound of a different   
          kind.\footnote{Both the  $\lambda = \lambda_{\rm max}$
 and the  $\lambda = \lambda_{\rm min}$ walls  can arise as flat BPS   walls in supergravity theories
coupled to scalars    \cite{Cvetic:1992bf, Cvetic:1996vr}.
  These 
   two  extreme types of flat wall,  called  type II and type III in
  \cite{Cvetic:1996vr}, differ  by  the fact that the   superpotential   avoids, respectively passes through
   zero as  fields
  extrapolate  between the AdS vacua 
      \cite{Ceresole:2006iq}.}  
   For $\lambda >   \lambda_{\rm max}$ the two phases can  coexist, but the   large  tension of the wall forces
  this latter   to inflate \cite{Karch:2000ct}. The phenomenon  is familiar  for gravitating domain
   walls in asymptotically-flat spacetime \cite{Vilenkin:2000jqa},  
     i.e. in the limit  $\ell_1, \ell_2\to \infty$.
    \smallskip
          
       The  meaning   of    $\lambda_0$  is  less clear,  its role will emerge  later.
   For now  note  that it is the turning point
          at which  the worldsheet radius $\ell_{w}(\lambda)$  reaches its minimal value $\ell_2$. 
          Note also that the   range $\lambda_{\rm min}< \lambda<\lambda_0$ only exists 
           for     non-degenerate AdS vacua,  that is  when  $\ell_1$  is 
           strictly smaller  than $\ell_2$. 
           
   Since the wall in this minimal model is described by a single parameter, its tension   $\lambda$, 
all properties of the dual interface depend on it. These include 
 the interface
      entropy,    and   the energy-transport  coefficients.  
      The entropy or $g$-factor,    computed in  
      \cite{Azeyanagi:2007qj,Simidzija:2020ukv},  reads\,\footnote{There  are many 
     calculations of the
      boundary, defect,  and interface entropy in a variety of holographic models --   
      a   partial list is \cite{Chiodaroli:2010ur,Takayanagi:2011zk,
       Fujita:2011fp, Jensen:2013lxa,Erdmenger:2015spo,Gutperle:2016gfe}.  
      The  formula for arbitrary left and right central charges, which we rederive below, 
 was found 
       in  ref.\cite{Simidzija:2020ukv}. 
      }
    \bea\label{vanR}
    {\rm log} \,g_{\rm I}\  = \  2\pi  \ell_1\ell_2   \left[ 
    \lambda_{\rm max}  \,{\rm  tanh}^{-1} \Bigl(  { \lambda \over \lambda_{\rm max}}  \Bigr ) 
   -   \lambda_{\rm min} \,{\rm  tanh}^{-1}  \Bigl(  { \lambda_{\rm min} \over \lambda}  \Bigr ) \right]\ . 
      \eea  
 It varies monotonically  between $-\infty$ and $\infty$ as  $\lambda$ varies  inside  its allowed range
\eqref{allowedrange}. 
    
      The  fraction of   transmitted energy   for waves
     incident on the interface from  the CFT$_1$ side, respectively 
     CFT$_2$ side,  was computed in      \cite{Bachas:2020yxv} with the result
     (reexpressed here in terms of critical tensions)
        \bea\label{Transm} 
       {\cal T}_{1\to 2} =  {\lambda_{\rm max} + \lambda_{\rm min}\over  \lambda_{\rm max} + \lambda}   \,, 
       \qquad
        {\cal T}_{2\to 1} =    {\lambda_{\rm max} -\lambda_{\rm min}\over  \lambda_{\rm max} + \lambda}  \ . 
          \eea
  Note that using   $\lambda_{\rm max} + \lambda_{\rm min} = 2/\ell_1$ and  $\lambda_{\rm max} -
   \lambda_{\rm min} = 2/\ell_2$, one can check that these 
      coefficients obey the detailed-balance condition 
   $c_1 {\cal T}_{1\to 2} = c_2 {\cal T}_{2\to 1}$. The larger of the two transmission coefficients  
      reaches the unitarity bound  when   $\lambda = \lambda_{\rm min}$, 
    and both coefficients  attain their   minimum  when  $\lambda = \lambda_{\rm max}$.  
  Total reflection (from the false-vacuum  to the true-vacuum side)
   is only possible  if $\ell_1/\ell_2 \to 0$,
   i.e. when   the ``true-vacuum'' CFT$_1$ is almost  entirely depleted of  
   degrees of freedom  relative   to CFT$_2$. 
    
      Using   eqs.\,\eqref{3cr} and the Brown-Henneaux formula  one
       can express  the  central charges
      $c_{1,2}$ in terms of the critical tensions $\lambda_{\rm min}$ and $\lambda_{\rm max}$. 
     As we just saw,  $\lambda$ parametrizes two key  properties of the interface.  
The triplet   ($\lambda_{\rm min} ,\,\lambda_{\rm max},\,\lambda)$   of  parameters
   in  the gravitational action defines
therefore the 
  basic data  of the putative dual  ICFT.
 

\subsection{Turning point and horizon}

We will now derive  the general solution of the   equations \eqref{00}\,-\,\eqref{Israel},   
and  then relate the geometric parameters  $M_j$ to the    data
$(T, L_j)$  of the boundary torus shown in fig.\,\ref{fig:1}.   


       A convenient  
    parametrization of  the string  outside any black   horizons 
     is in terms of  the  blueshift factor  of the worldsheet  metric, 
    eq.\eqref{00}, 
        \bea\label{gauge}
         f(\sigma) =   \sigma   \ \ \Longrightarrow\ \  r_j = \sqrt{  \sigma 
           + M_j\ell_j^{\,2}}\ .  
        \eea
        In this parametrization  $d{\hat s}^2\vert_{\mathbb{W}} =  -   \sigma\,  dt^2 + g (\sigma) d\sigma^2$. 
             Let $\sigma_+$ correspond to the minimal value of the blueshift, this   is either zero or positive.
        If  
        $\sigma_+ =0$   the string enters  the   horizon. If on the other hand   $ \sigma_+ >0$ 
         then, as we will confirm  in a minute,  this  is  the  turning point  of $r_j(\sigma)$  where   both  $x_1^\prime$ and 
        $x_2^\prime$     diverge.   
        
\begin{figure}[tbh!]
   \vskip  -3mm
\centering
\includegraphics[width=.96\textwidth,scale=0.84,clip=true]{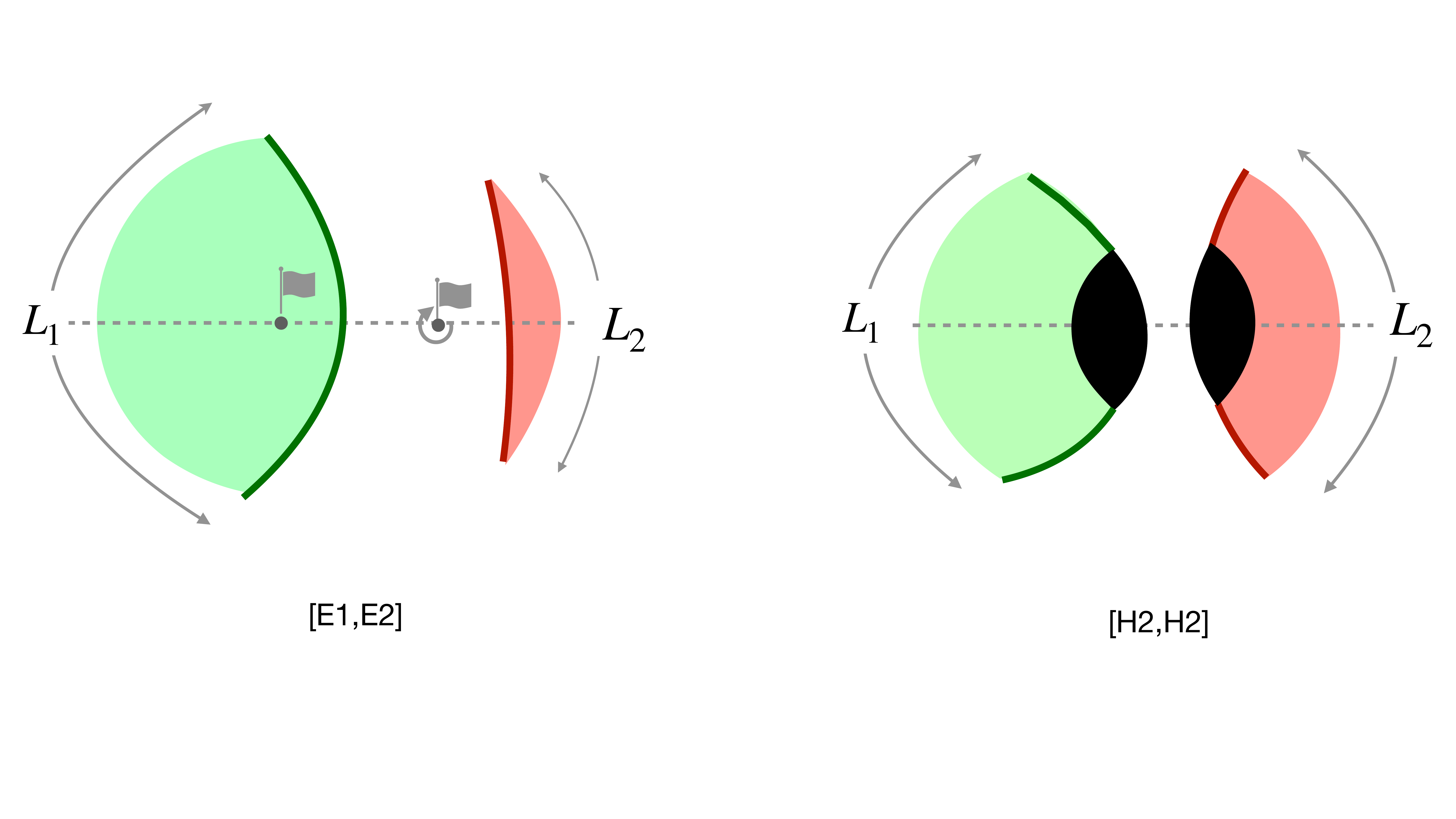}
   \vskip -10mm
   \caption{\footnotesize   Schematic drawing  of  a low-temperature  and a high-temperature  solution, 
    corresponding to pairs of type [E1,E2] and [H2,H2].
    The broken line is the axis of reflection symmetry. 
     The   blueshift parameter  $\vert \sigma\vert $  
     decreases  monotonically   until the string  reaches either   the turning point or the black-hole horizon.}
   \label{fig:4}
 \end{figure}

        A  static string  has (at most)  one  turning point, and is symmetric under
        reflection in   the 
       axis that passes    through the centers  of the  boundary  arcs,\,\footnote{In ref.\cite{Simidzija:2020ukv} 
        this corresponds to
        the time-reflection symmetry of the  instanton solutions.}
          as  illustrated  in  figure  \ref{fig:4}.  
        It follows that the    
         parametrization  is   one-to-two. 
           Henceforth  we focus on the
          half string with  positive $x_j$ (at least near the conformal boundary). The other half string is obtained by   $x_j \to -x_j$.

  Eqs.\,\eqref{gauge}   imply  that $2 r_j  r_j^\prime =1$. Inserting   in eq.\,\eqref{11} gives
 \bea\label{xprime}
  (x_j^\prime)^2 =  r_j^{-2} \Bigl( g(\sigma) - {\ell_j^2 \over 4 \sigma r_j^2} \Bigr) \ . 
 \eea
     Squaring now  twice eq.\,\eqref{Israel} 
     and replacing $(x_j^\prime)^2$ from the above expressions 
   leads to   a quadratic   equation for $g(\sigma)$, 
   the  ${\tiny \sigma\sigma}$ component of the worldsheet metric.
   This equation has a singular solution $g=0$,
   and a  non-trivial one
 \begin{equation}\label{g}
    g(\sigma) =   \lambda^2 \Biggl[  \Bigl(\frac{2r_1  r_2 }{\ell_1 \ell_2 }\Bigr)^2 \hskip -0.4mm   - 
     \Bigl(\frac{r_1^2}{\ell_1^{\, 2}}  + \frac{r_2^2}{\ell_2^{\, 2}} -   \lambda^2 \sigma\Bigr)^2\, 
    \Biggr]^{-1} \hskip -1.9mm 
    =  { \lambda^2  \over A \sigma^2 + 2B\sigma + C}\, ,  
\end{equation}
 where in the second equality we used  eqs.\,\eqref{gauge},  and  
 $$
   A =       (\lambda_{\rm max}^2 - \lambda^2 )  (\lambda^2 - \lambda_{\rm min}^2) 
    \ ; 
$$
 \vskip -8mm
 \bea\label{ABC}
 B =    \lambda^2 (M_1+M_2)  -  \lambda_0 ^2(M_1-M_2)\ ; 
\quad 
C = -   (M_1 - M_2)^2\ . 
\eea        

  We  expressed   the    quadratic polynomial appearing   in the denominator of \eqref{g} 
   in terms of  $M_j, \lambda$ and the critical tensions,  eqs.\,\eqref{3cr}, in order to
 render   manifest the fact that for
 $ \lambda$  in the allowed range, $ \lambda_{\rm min}< \lambda  < \lambda_{\rm max}$,
     the coefficient $A$ is positive.  
  This is required for    $g(\sigma) $ to be positive near the boundary where   $\sigma\to\infty$.
     In addition,    $ AC  \leq   0$ which ensures   that the two roots of the 
      denominator  in  eq.\,\eqref{g}  
       \bea\label{spm}
       \sigma_\pm   ={ -B \pm  (B^2 - AC)^{1/2}\over A}  
       \eea
       are real, and that the larger root $\sigma_+$  is non-negative.  
       Inserting  eq.\,\eqref{g} in   eq.\,\eqref{xprime}  and fixing the sign of the square root 
     near the conformal boundary 
     gives after a little  algebra
  \begin{subequations}\label{soln}
 \bea\label{solna}
\frac{x_1'}{\ell_1}  \, = \, -  \frac{\sigma\, (\lambda ^2 + \lambda _{0}^2)+M_1-M_2}
{{2(\sigma + M_1\ell_1^{\, 2})}\,\sqrt{A \sigma ( \sigma- \sigma_+)(\sigma-\sigma_-)}} \ \ , 
\eea \vskip -3mm
\bea\label{solnb}
\frac{x_2'}{\ell_2}  \, =  \,  - \frac{\sigma\,(\lambda ^2 -  \lambda _{0}^2)+M_2-M_1}
{{2(\sigma + M_2\ell_2^{\, 2})}\,\sqrt{A \sigma ( \sigma- \sigma_+)(\sigma-\sigma_-)}} \ \  . 
\eea     
  \end{subequations}
                   We may  now confirm our  earlier claim that  if $\sigma_+ > 0$  then 
           both $x_1^\prime \propto dx_1/dr_1$  and $x_2^\prime \propto dx_2/dr_2$
           diverge  at this point. 
  Furthermore, since   $\sigma + M_j\ell_j^{\, 2}  = r_j^2$ is  positive,\footnote{Except for the measure-zero set of 
  solutions in which the string passes through the center of global AdS$_3$.}  
    the $x_j^\prime$ are   finite at all $\sigma >\sigma_+$. Thus $\sigma_+$
  is the unique turning point of the string,  as  advertized.

          Eqs.\,\eqref{gauge} and \eqref{soln} give    the general solution  of the string equations
   for  arbitrary mass  parameters  $M_1, M_2$ of the green and pink   slices.
 These   must be determined   by  interior
   regularity,  and by the Dirichlet  conditions  
    at the conformal boundary. Explicitly,  the  boundary  conditions 
    for the  different slice types  of    figure
    \ref{fig:2} 
    read: 
 \begin{subequations}\label{Dir}
   \begin{equation}\label{Dira}
    L_j \ =     \, 2  \int_{\,\sigma_+}^\infty d\sigma \,  x_j^\prime\,   \qquad
 \qquad   {\text  {for \  \  {\small E2, E2$^\prime$}}}\ ;
 \end{equation}
  \begin{equation}\label{Dirb}
   L_j \ =   \  n  P_j  +  \,  2  \int_{\,\sigma_+}^\infty d\sigma \,  x_j^\prime\,   \qquad 
    {\text{ for \  \  {\small E1, H1}}}\ ; 
 \end{equation}
  \begin{equation}\label{Dirc}
     L_j\  =    \   \Delta x_j\bigl\vert_{\rm Hor} \, + \,   2  \int_{\,\sigma_+}^\infty d\sigma \,  x_j^\prime\, 
     \qquad\quad   {\text{  for \  \  {\small H2}}}\ .     
   \end{equation}
  \end{subequations}
 The  integrals in these equations are   the   opening arcs,   $\Delta x_j$,    between the two  
 endpoints of a  half string.  
 They can be expressed as 
 complete elliptic integrals of the first, second and third kind,  
 see appendix \ref{app:2}.  For the   slices  {\small E1, H1}
 where   $x_j$  is a periodic coordinate,  
 we have  denoted by   $P_j >0$ its period, and by $n$ the  string winding number. 
   Finally for strings   entering   the horizon we denote by  
   $\Delta x_j\vert_{\rm Hor}$  the 
    opening arc  between the two  horizon-entry points  in the  $j$th coordinate chart.

Possible phases of the ICFT for  given   torus parameters  $T, L_j$ must be solutions 
to one  pair of  conditions\,\eqref{Dir}.
  Apart from    interior regularity,  we will also require  that the 
  string  does not self-intersect.
In principle, two string  bits
 intersecting at an angle $\not= \pi$ could   join into  another  string.   
  Such string junctions would be  the gravitational counterparts 
 of interface fusion  \cite{Bachas:2007td},  and allowing them  would  make  the holographic model much
 richer.\footnote{Generically  the intersection point  in one slice will correspond to two points that must be
identified in the other slice;  this  may  impose further conditions.}  
To keep, however,  our discussion  simple  we will only allow   a single type
 of  domain wall in this  work.  

  The reader can easily  convince herself  that  to avoid string intersections
  we must have $P_j > L_j$ and  
   $n=1$  in  \eqref{Dirb},  and  
  $\Delta x_j\bigl\vert_{\rm Hor} >0$   in \eqref{Dirc}.


\section{Phases: cold, hot $\&$ warm}
   \label{sec:5}        
               
                     Among  the five slice types of figure \ref{fig:2},  
       {\small H2} stands apart  because  it  can only pair  with itself.  This
              is because    a  horizon   is a closed surface, so it
                cannot end on the domain wall.\footnote{Except possibly in the limiting  case
where the wall is the boundary of space.} 
      We will  now show  that
       the  matching equations    actually  rule out  several
       other pairs   among   the  remaining slice  types.

     One   pair that is easy  to  exclude is    [{\small H1,H1}], i.e.\,\,solutions that   describe
     two black holes  sitting   on either side
   of the wall.  Interior regularity would require  in this case  $M_1 = M_2 = (2\pi T)^2$. 
    But   eqs.\,\eqref{ABC} and \eqref{spm} 
   then imply that    $\sigma_+=0$,  so  the  wall cannot avoid the
    horizon leading to a contradiction. 
\newpage

This gives our first no-go lemma:
 \vspace{-2mm}
 \begin{center}  
   \begin{tcolorbox}[width=13.5cm]
     { Two   black holes} 
     on either side of a static  domain wall are ruled out. 
\end{tcolorbox}
\end{center}    
   \vspace{-2.5mm}
    Note by  contrast that   superheavy   domain walls  ($\lambda > \lambda_{\rm max}$)   
        inflate and  could   thus  prevent  the black holes from  coalescing.\footnote{Asymptotically-flat 
        domain walls, which have been studied a lot in the context of Grand Unification  \cite{Vilenkin:2000jqa},
 are automatically 
        in this range.} 

  \smallskip  
 A second class of pairs   one  can exclude are the
     `centerless geometries'   
      [{\small E2,E2}],   [{\small E2,E2$^\prime$}],
       [{\small E2$^\prime$,E2}] and    [{\small E2$^\prime$,E2$^\prime$}].  
       We use the word  `centerless'  for 
        geometries that contain  neither a  center of global AdS, nor
       a black hole in its place (see  fig.\,\ref{fig:2}).
               If such solutions existed, all inertial observers would necessarily hit
       the domain wall since there would be  neither a  center where  to rest,  
  nor  a  horizon  where    to escape.\footnote{In  the double-Wick rotated context  of   Simidzija and Van Raamsdonk
the   [{E2,E2}] geometries  give   traversible wormholes  \cite{Simidzija:2020ukv}. 
}

        The argument   excluding  such  solutions is based on a simple  observation: 
                        What distinguishes the centerless
       slices   {\small E2} and {\small E2$^\prime$} 
       from those with  a AdS center ({\small E1})  or  a black hole  ({\small H1}) 
        is the sign of   $x_j^\prime$  at the turning point,   
        \bea\label{signs5}
           {\rm sign} \bigl(x_j^\prime\bigl\vert_{\sigma\approx \sigma_+} \bigr) \
           = \begin{cases}  +  \quad {\rm for \ {\small E2}, {\small E2}}^\prime\,, \\
           -  \quad  {\rm for \ {\small E1}, {\small H1}}\ . 
           \end{cases}
        \eea
        Now from eqs.\,\eqref{soln}  one has \vskip -4mm
       \bea
        (\sigma + M_1 \ell_1^{\,2}) \,\frac{x_1'}{\ell_1}  + 
         (\sigma + M_2 \ell_2^{\,2}) \,\frac{x_2'}{\ell_2} \, <  \, 0\ , 
       \eea
       so  both  $x_j^\prime$  cannot be  simultaneously positive. This holds for  all $\sigma$, and 
         hence also  near the turning point   {\small{QED}}. This is our second no-go lemma:
     \vspace{-1mm}
   \begin{center}  
   \begin{tcolorbox}[width=13.2cm]
  `{Centerless}'    static spacetimes  in which  all
   inertial observers  would   inevitably hit  the domain wall are ruled out.
  \end{tcolorbox}
\end{center}
    \vspace{-2mm} 
 
 We can    actually   exploit this   argument further. 
  As is clear   from eq.\,\eqref{solna},  if 
   $M_1>M_2$ then    $x^\prime_1$ is manifestly  negative, i.e.\,the 
  green slice  is  of type {\small E1} or {\small H1}. The 
    pairs    [{\small E2$^\prime$,E1}] and [{\small E2$^\prime$,E2}]
     for which the above   inequality   is automatic are thus ruled out. 
     One can also   show that 
     $x^\prime_2\vert_{\sigma\approx \sigma_+}$  is  negative if  $M_2 > 0 > M_1$. 
       This is obvious   from eq.\,\eqref{solnb}   in   the range $\lambda > \lambda_0$,   and less obvious
     but  also true  as  can be checked by explicit calculation for  $\lambda < \lambda_0$.\footnote{ 
     The tedious algebra is  straightforward and not particularly instructive, so we chose not to present it here.
     We did it with mathematica but also tested it numerically.} 
      The pairs  [{\small E1,E2$^\prime$}] and [{\small E2,E2$^\prime$}] for which the 
      above mass inequality is  automatic, are thus also  excluded.
      
          Recall   that    the energy density 
          of   the $j$th CFT  reads  $\langle T_{tt}\rangle = {1\over 2}\ell_j M_j$. 
       Ruling out  all pairs  of  {\small E2}$^\prime$ with    {\small E1} or {\small E2} 
           implies  therefore   that  in the ground state the energy density must be    everywhere 
          negative. When   one $L_j$ is much smaller than the other,  the Casimir energy scales like
        $E_0 \sim \#/L_j$. The fact that the coefficient $\#$ is negative means that 
           the   Casimir force is attractive,    in agreement with general theorems   
            \cite{Kenneth:2006vr, Bachas:2006ti}. This is the third no-go lemma:

        \begin{figure}[t!]
   \vskip  -10mm
\centering
 \includegraphics[width=.96\textwidth,scale=0.84,clip=true]{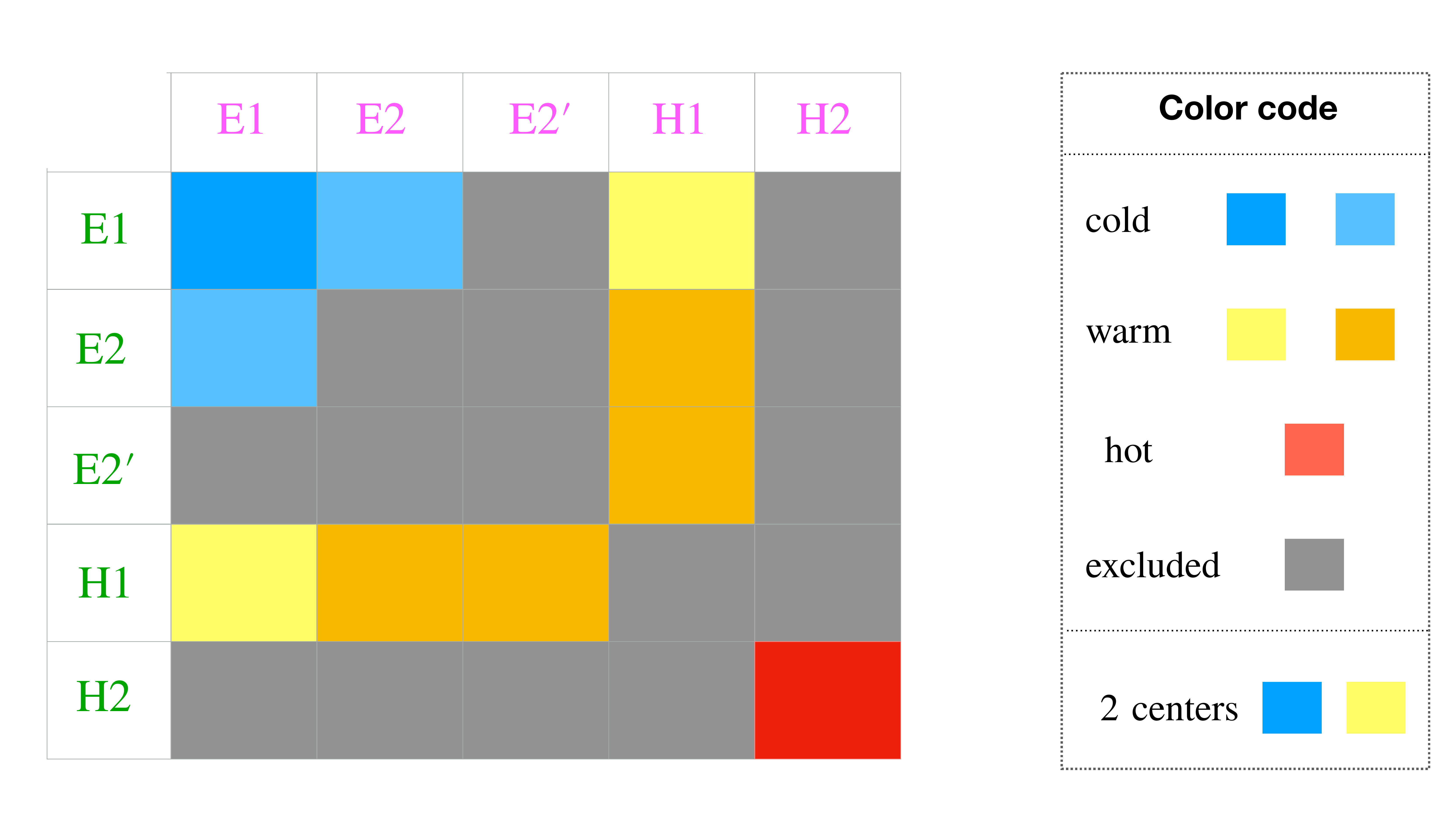}
\vskip -2mm
    \caption{\footnotesize  
  Phases of the domain-wall spacetime.  The   type of the green  slice 
   labels  the rows  of the table, and that  of the pink  slice  the columns. 
   In the hot (red) phase
   the wall   enters  the black-hole  horizon,  while in the warm   (yellow)  phases  it avoids it.
   The cold (blue) phases have  no black hole. Geometries in which an inertial observer   is attracted to
   two  different centers are indicated by a  different shade (light yellow or darker blue).}
    \label{fig:5}
    \end{figure}

   \begin{center}      \vspace{-1.1mm}
   \begin{tcolorbox}[width=13.7cm]
   A slice of   global AdS$_3$  cannot be paired with   a horizonless  BTZ slice.  
This implies that in  the  ground state  of the putative dual ICFT
the  energy density is everywhere negative. 
 \end{tcolorbox}
\end{center}
 \vskip -3mm

  We have collected for  convenience  all these conclusions  in
 fig.\,\ref{fig:5}.  The table  shows the eligible slice pairs, or the allowed topologies of
  static-domain-wall spacetimes.  It also defines a color code for phase diagrams.

The light yellow  phases  that feature  a wall  between the  black hole and 
an AdS restpoint  are  the gravitational avatars of the Faraday cage. 
Such solutions  are easier to construct for  larger $\lambda$. 
Domain walls lighter than $\lambda_0$, in particular,   can never    shield  
 from   a  black hole in the   `true-vacuum'  side.  
Indeed, as follows easily from  eq.\,\eqref{solnb},  for $\lambda <  \lambda_0$ and $M_1>0> M_2$,  
the sign of $x_2^\prime\vert_{\sigma\approx\sigma_+}$  is  positive, so   geometries  of type 
       [{\small H1,E1}] are excluded.


\section{Equations of state}\label{eqstate}

   The   different colors in figure \ref{fig:5}  describe  different  phases of the system, since
 the corresponding  geometries are topologically distinct. 
They differ in how  the wall, the horizon (if one exists) and  
  inertial observers intersect or avoid each other.
       
  Let us now think thermodynamics. For fixed  Lagrangian parameters   $\lambda, \ell_j$, 
the canonical  variables that determine the state of the system
are the temperature $T$ and the volumes $L_1, L_2$. 
Because of scale invariance only  two  dimensionless ratios matter:
\bea\label{names}
\tau_1 := TL_1\ , \quad \tau_2:= TL_2\qquad {\rm or}\quad 
 \quad \gamma  := {L_1\over L_2} = {\tau_1\over \tau_2}\ .
\eea
The microcanonical  variables,  the energy density and  the entropy of each  subsystem, 
  read (see section \ref{sec:2},  and recall that $8\pi G=1$)
\bea\label{othervars}
 {E_j\over L_j} = {\ell_j\over 2}  M_j\qquad {\rm and}\qquad
S_j =  {r^{\rm H}_j \Delta x_j\vert_{\rm Hor} \over 4 G } =  2\pi \ell_j \sqrt{M_j}\, \Delta x_j\vert_{\rm Hor} \ .\ \ 
\eea
These are the natural  parameters of the interior geometry.  The
 entropies are scale invariant. The other  key 
  dimensionless variable   is the mass ratio, viz.    the
ratio of energy densities per degree
of freedom in the two  CFTs \vskip -6mm
\bea
\mu  := {M_2\over M_1} \ . 
\eea
 When several    phases   coexist  the dominant one
is the one with  the lowest  free energy  $F = \sum_j (E_j - TS_j)$.  As a  sanity check,
we   rederive  $F$    from the renormalized on-shell gravitational action   in appendix \ref{app:1}.

The Dirichlet  conditions, eqs.\,\eqref{Dir},  give for each type of geometry two relations 
 among   the above    variables  that  play the role of equations of
state.\footnote{In homogeneous systems there is  a single
  equation of state. Here we have one equation
for each subsystem.}   They relate the natural interior parameters  $S_j$ and  $\mu$ to
the variables $\tau_j$ and $\gamma$ of the boundary torus. 
Note that in each phase of the system  since  for horizonless slices $S_j=0$ and for slices with horizon 
$M_j = (2\pi T)^2$.  In computing the phase diagram
 we  will have to invert  these equations of state.


 \subsection{High-$T$ phase}
 \label{sec:61}

  For fixed $L_j$ and   very high temperature
                    the black hole  grows  so large   that it   eats  away  a piece of the domain wall and the  AdS rest points.
The dominant solution is thus  
                    of type {\small [H2,H2]} and   regularity fixes  the  mass parameters in  both  slices, 
                    $M_1=M_2= (2\pi T)^2$. 
The boundary   
                    conditions \eqref{Dirc}   reduce in this case to simple equations for the opening  horizon arcs 
 $\Delta x_j\vert_{\rm Hor}$. Performing explicitly the integrals 
                    (see appendix \ref{app:2}) gives 
  \begin{subequations}\label{HighTarcs}
    \bea\label{HighTarcsa} 
  L_1 - \Delta x_1\bigl\vert_{\rm Hor}\  =\   - {1\over \pi T}\, {\rm  tanh}^{-1}\left( {\ell_1  (\lambda^2 + \lambda^2_0)
  \over 2\lambda}\right)\ , 
  \eea\vskip -5mm
    \bea\label{HighTarcsb}
  L_2 - \Delta x_2\bigl\vert_{\rm Hor}\  =\   - {1\over \pi T}\, {\rm  tanh}^{-1}\left( {\ell_2 (\lambda^2 -  \lambda^2_0)
  \over 2\lambda}\right)\ . 
  \eea  
\end{subequations}  
For consistency we must have   
   $\Delta x_j\vert_{\rm Hor}>0$, which   
     is automatic
   if  $\lambda > \lambda_0$. 
   If   $\lambda < \lambda_0$, on the other hand,   positivity of $\Delta x_2\vert_{\rm Hor}$  puts   a lower bound 
on $\tau_2$, 
   \bea\label{tau2star}
    \tau_2  \, \geq  \, 
    {1\over  \pi  }\, {\rm tanh}^{-1}\left( {\ell_2 (\lambda_0^2 -  \lambda^2) \over 2\lambda}\right)\ :=\, \tau_2^*\  .   
   \eea
We see here a first interpretation of the critical tension $\lambda_0$ encountered  in section \ref{sec:43}. 
For walls lighter than $\lambda_0$ there is a region of parameter space where the hot solution  ceases
to exist, even as a metastable phase. 
\smallskip

 The total energy and  entropy in the  high-$T$  phase read
\bea\label{hotEF}
  E_{\rm [hot]}\,= \,  {1\over 2}(\ell_1 L_1M_1 + \ell_2 L_2M_2 ) \,=\, 2\pi^2  T^2 \,(  \ell_1 L_1 +  \ell_2 L_2) \ , 
\eea
\bea\label{hotS}
  S_{\rm [hot]} =  4\pi^2  T \bigl(  \ell_1 \Delta x_1\bigl\vert_{\rm Hor} + \,
  \ell_2 \Delta x_2\bigl\vert_{\rm Hor}\bigr)   =   
  4\pi^2  T^2 (  \ell_1 L_1 +  \ell_2 L_2)  + 2 \log g_I \ ,  \ \  \ 
 \eea

\medskip
\noindent where $\log g_{\rm I}\,$ is given by  eq.\,\eqref{vanR} and the rightmost  expression of the entropy
 follows from 
eqs.\,\eqref{HighTarcs} and  a  straightforward reshuffling of the arctangent functions. This is a satisfying result.
Indeed, the first term in the right-hand side of \eqref{hotS} is the thermal   entropy  of the two CFTs (being
 extensive these entropies  cannot depend  on the ratio $L_1/L_2$), 
while the second term is the entropy of the two  interfaces on the circle. 
The Bekenstein-Hawking formula captures  nicely both contributions.

Eqs.\eqref{HighTarcs}  and \eqref{hotS}  show    that  shifting  the $L_j$ at fixed $T$  
 does not change the 
entropy  if  and only if 
  $\ell_1\,\delta L_1 = -\ell_2\, \delta L_2$.  Moving 
 in particular  
a defect (for which $\ell_1=\ell_2$) without changing the volume 
$L_1+L_2$   is an adiabatic process, while moving a
more general  interface 
 generates/absorbs  entropy by  modifying   the density of degrees of freedom.


   \subsection{Low-$T$ phase(s)}
\label{sec:62} 

     Consider next  the ground state   of the system,  at $T=0$.
The  dual  geometry  belongs to one of the three  horizonless types: 
 the double-center geometry  {\small [E1, E1]}, 
       or  the single-center ones  {\small [E1, E2]} and  
       {\small [E2, E1]}  (see fig.\,\ref{fig:5}).  
Here the entropies $S_j=0$, and the only relevant dimensionless variables are the volume and energy-density
ratios,  $\gamma$ and $\mu$.  Note that they are both  positive
since  $L_j>0$ and 
 $M_j<0$ for both $j$. 

  The   Dirichlet  conditions  (\ref{soln}) for  horizonless geometries  read
   \bea\label{61}
 \sqrt{\vert M_1\vert}\,  L_1 = 2\pi  \, \delta_{{\mathbb S}_1, {\rm E1}}- f_1(\mu)\ , \qquad  
 \sqrt{\vert M_1\vert}\,  L_2 = {2\pi \over  \sqrt{\mu}}\,\delta_{{\mathbb S}_2, {\rm E1}} - f_2(\mu)\ ,
        \eea         
where  $\delta_{{\mathbb S}_j, {\rm E1}} = 1$
  if the $j$th slice is of type {\small E1} and $\delta_{{\mathbb S}_j, {\rm E1}}=0$  otherwise, and 
   \begin{subequations}\label{mu}
\bea\label{mua}
  f_1(\mu) \,=\,    
   {\ell_1\over \sqrt{A} } 
    \int_{s_+}^\infty   ds  {
     s(\lambda^2+\lambda_0^2)  - 1 + \mu  \over
   (s  -  \ell_1^{\,2}) \sqrt{s(s-s_+)(s-s_-) } }  \ , \  \  \
\eea  \vskip -1mm
\bea\label{mub}
  f_2(\mu) \,=\,    
   {\ell_2\over \sqrt{A} } 
    \int_{s_+}^\infty    ds  {
     s(\lambda^2 - \lambda_0^2)  + 1  -  \mu  \over
   (s -  \mu \ell_2^{\,2} ) \sqrt{s(s-s_+)(s-s_-) } }  \ , \   \  \ 
\eea
 \end{subequations}
  with  
  \bea\label{s++cold}
  A\, s_\pm = \lambda^2 (1+\mu ) - \lambda_0^2 (1- \mu ) \pm 2\lambda 
  \sqrt{{1-\mu \over \ell_2^2} + {\mu^2 -\mu  \over \ell_1^2}  + \mu \lambda^2 }\ \ . 
  \eea
 The dummy integration variable $s$ is the  appropriately 
  rescaled blueshift factor of the string worldsheet, $s= \sigma/\vert M_1\vert$\,. 
\smallskip 

 Dividing the two sides of  eqs.\,\eqref{61}   gives 
     $\gamma$ as  a function of $\mu$  for each of the three possible
  topologies.\footnote{The functions  $f_{j}(\mu)$ 
 are combinations of   complete elliptic integrals of the first, second and third kind,  
   see appendix \ref{app:2}. The value $\mu=1$ gives $\gamma = 1$,  
  corresponding  to the
  scale-invariant AdS$_2$ string worldsheet. The known  supersymmetric top-down solutions  live at
this special point in phase  space.
}    
If the  ground state of the putative dual  quantum-mechanical system was unique, 
we  should find a single slice-pair type and value of $\mu$
for  each value of $\gamma$. 
Numerical plots  show that this is indeed  the case.
Specifically, we   found that $\gamma(\mu)$  is  a monotonically-increasing
 function of $\mu$ for any given  slice pair, 
and that it changes continuously from one  type of pair    to another. 
We will return  to these 
 branch-changing  `sweeping transitions'    in   section \ref{sec:6}. Let us stress that the  uniqueness
of the cold solution  did not have to be automatic 
in classical gravity,  nor in the dual large-$N$ quantum mechanics.

For  most of the $(\ell_j, \lambda)$ parameter space, as 
 $\gamma$ ranges   in $(0, \infty)$   the mass ratio   $\mu$   covers also  the entire range 
$(0, \infty)$.  
However,   if  $\ell_1 <  \ell_2$ (strict inequality)  and  for
sufficiently light  domain walls,  we  found that $\gamma$   vanishes     at  some  positive   
$\mu= \mu_0(\lambda, \ell_j)$.   Below this critical value   $\gamma$   becomes
negative signaling that the wall self-intersects and the solution must be discarded.
 This leads to a  striking  phenomenon  that we discuss   in section \ref{sec:Faraday}.


\subsection{Warm  phases}

The last set  of solutions  of the model are the yellow- or orange-coloured  ones
in   fig.\,\ref{fig:5}. Here 
 the string avoids the horizon, so 
the slice pair  is of type {\small [H1,X]} or {\small [X,H1]} with {\small X} one of the  
horizonless types: {\small E1, E2} or {\small E2}$^\prime$. 
 
Assume first  that the black hole is on the  green side of the wall, so  that
   $M_1 = (2\pi T)^2$.  In terms of  $\mu$  the  Dirichlet 
   conditions  (\ref{Dira}, \ref{Dirb})  read: 
   \bea\label{tildeDir}
    2\pi T \Delta x_1\bigl\vert_{{\rm Hor}}   - \, 2\pi \tau_1 = \tilde f_1(\mu)\ ,   \qquad
   2\pi \tau_2  =  {2\pi\over \sqrt{-\mu}} \,\delta_{{\mathbb S}_2, {\rm E1}} - \tilde f_2(\mu) \ , 
   \eea
where
   \begin{subequations}\label{tildemu}
\bea\label{tildemua}
  \tilde f_1(\mu) \,=\,    
   {\ell_1\over \sqrt{A} } 
    \int_{\tilde s_+}^\infty   ds  {
     s(\lambda^2+\lambda_0^2) +1  -\mu  \over
   (s  +   \ell_1^{\,2}) \sqrt{s(s- \tilde s_+)(s- \tilde s_-) } }  \ , \  \  \
\eea  \vskip -1mm
\bea\label{tildemub}
  \tilde f_2(\mu) \,=\,    
   {\ell_2\over \sqrt{A} } 
    \int_{\tilde s_+}^\infty    ds  {
     s(\lambda^2 - \lambda_0^2)  - 1  +  \mu  \over
   (s  +  \mu \ell_2^{\,2} ) \sqrt{s(s-\tilde s_+)(s-\tilde s_-) } }  \ ,  \   \  \ 
\eea
 \end{subequations}
 and the  roots  $\tilde s_\pm = \sigma_\pm/M_1$ inside  the square root are  given by   
  \bea\label{s++greensidebh}
  A\, \tilde s_\pm = - \lambda^2 (1+\mu ) + \lambda_0^2 (1- \mu ) \pm 2\lambda 
  \sqrt{{1-\mu \over \ell_2^2} + {\mu^2 -\mu  \over \ell_1^2}  + \mu \lambda^2 }\ \ . 
  \eea
In the first condition \eqref{tildeDir} we have 
used the fact that  the period of the  green  slice that contains 
the horizon  is 
$P_1 =  \Delta x_1\vert_{{\rm Hor}}$.

If the black hole is on the  pink  side of the wall,  the  conditions  take a similar form in terms of the inverse
mass ratio  $\hat\mu = \mu^{-1} = M_1/M_2$, 
   \bea\label{hatDir}
  2\pi T \Delta x_2\bigl\vert_{{\rm Hor}}   - \,  2\pi \tau_2 = {\hat  f}_2(\hat \mu)\ ,   \quad
   2\pi \tau_1  =  {2\pi \over  \sqrt{-\hat \mu}} \,\delta_{{\mathbb S}_1, {\rm E1}} - \hat f_1(\hat \mu) \ ,
   \eea
where   here
  \begin{subequations}\label{hatmu}
\bea\label{hatmua}
  \hat f_1(\hat \mu) \,=\,    
   {\ell_1\over \sqrt{A} } 
    \int_{\hat s_+}^\infty   ds  {
     s(\lambda^2+\lambda_0^2) +  \hat\mu   - 1  \over
   (s  +  \hat\mu  \ell_1^{\,2}) \sqrt{s(s- \hat s_+)(s- \hat s_-) } }  \ , \  \  \
\eea  \vskip -1mm
\bea\label{hatmub}
  \hat f_2(\mu) \,=\,    
   {\ell_2\over \sqrt{A} } 
    \int_{\hat  s_+}^\infty    ds  {
     s(\lambda^2 - \lambda_0^2)  - \hat \mu   +  1  \over
   (s  +    \ell_2^{\,2} ) \sqrt{s(s-\hat s_+)(s-\hat  s_-) } }  \ . \   \  \ 
\eea
 \end{subequations}
 and the  roots  $\hat  s_\pm = \sigma_\pm/M_2$ inside  the square root are  given by   
  \bea\label{s++pinksidebh}
  A\, \hat  s_\pm = - \lambda^2 (\hat\mu + 1) + \lambda_0^2 (\hat\mu - 1 ) \pm 2\lambda 
  \sqrt{{\hat\mu^2- \hat \mu \over \ell_2^2} + {1 -\hat \mu  \over \ell_1^2}  + \hat \mu \lambda^2 }\ \ . 
  \eea

The  functions $\tilde f_j$ and $\hat f_j$,  as well as   the $f_j$ of 
  the cold phase,  derive from the same basic  formulae  (\ref{solna}, \ref{solnb})  and differ only 
 by  a few   signs.  We 
 chose to write them out   separately because these signs are important. 
Note also that while in  cold   solutions  $\mu$ is 
always positive, here $\mu$ and its inverse $\hat\mu$  can have either sign.

  All  the values of \,$\mu$\, and\, $\hat\mu$\,  do not,  however,  correspond to admissible   solutions. 
For a pair   of type {\small [H1,X]} we must demand  (i)  that  the right-hand sides in 
  \eqref{tildeDir}
be positive -- the non-intersection requirement,  
and (ii)  that  $x_1^\prime\vert_{\sigma\approx \sigma_+}$  be negative -- 
 the  turning point condition  \eqref{signs5}.  Likewise for solutions   of type {\small [X, H1]}
we must demand that the right-hand sides in 
  \eqref{hatDir} be positive and that $x_2^\prime\vert_{\sigma\approx \sigma_+}$  be negative. 

 The turning-point requirement  is easy to implement. In the  {\small [H1,X]} case, 
 $x_1^\prime\vert_{\sigma\approx \sigma_+}$ is negative when the numerator
of the integrand 
 in  \eqref{tildemua}, evaluated at at $s=\tilde s_+$,    is positive.  
Likewise for  
the  {\small [X,H1]} pairs,  
$\,x_2^\prime\vert_{\sigma\approx \sigma_+}$ is negative when the numerator 
of the integrand in \eqref{hatmub}, evaluated at at $s=\hat s_+$,
  is positive.  After  a little algebra these conditions take  a simple form
 \bea\label{618}
{\rm for}\ \ {\rm [H1,X]} \quad  \mu\in (-\infty, 1]\ ; 
\qquad 
{\rm for}\ \  {\rm  [X,H1]}   \quad  \hat \mu = \mu^{-1}\in (-\infty, 1]   \ . \ \  \ 
\eea
Recalling that $\mu = \hat\mu^{-1} = M_2/M_1$, we conclude that  
  in all the cases 
the energy density per degree of freedom  in the horizonless  slice
is lower than the  corresponding density  in the black hole slice. 

This agrees with physical intuition:  the energy density per degree of freedom 
  in the cooler  CFT   is less than the thermal density
 $\pi  T^2/6 $  --  the interfaces did not let the theory  thermalize. 
 When $\mu\to 1$ or $\hat\mu \to  1$, the   wall  enters  the horizon
and the energy   is equipartitioned.
 \smallskip

     This completes our  discussion of the equations of state.  
To summarize, these equations relate 
 the parameters of the interior geometry ($\mu, S_j$) to   those
of the conformal boundary  ($\gamma , \tau_j$).  The relation 
  involves  elementary functions  in  the hot phase,
and was reduced   to a single function  $\gamma(\mu)$,  that
can be readily plotted, in  the cold phases.  
 Furthermore  {\it {at  any given point in parameter space  the hot and cold solutions,
when they exist,  are unique.}} 
The excluded regions are   $\tau_2< \tau_2^*(\lambda, \ell_j)$  for the hot solutions,  and   
 $\mu > \mu_0(\lambda, \ell_j)$ for the cold solution with  $\mu_0$   the point where
$\gamma =0$. 
 
     In  warm phases  the story  is richer  since more than one  solutions typically coexist
at any given value of $(\gamma, \tau_j)$.  Some   solutions  have  negative specific heat, as we
will discuss later. 
To find the  parameter regions where different solutions exist requires  inverting 
 the relation  between ($\gamma, \tau_j$)
  and ($\mu, S_j$). We will do this analytically in some limiting cases, and
numerically  to  compute   the full phase diagram in section \ref{sec:8}.

 
 \section{Phase transitions}
 \label{sec:6}

The  transitions between different phases are of  three  kinds:
    
    \begin{itemize}
    
    \item \underline{Hawking-Page}  transitions describing  the formation of a  black hole. 
    These  transitions  from the cold  to the  hot  or warm phases of  
    fig.\,\ref{fig:5} are always first order;

    \item \underline{Warm-to-hot}   transitions  during which  part of   the   wall  is captured by the   horizon. 
  We will show that   these   transitions   
are also  first-order;

     \item  \underline{Sweeping}   transitions where
the   wall
    sweeps away   a  center  of global AdS, i.e.\,\,a rest point for  inertial observers. 
These are continuous transitions 
between  the one-  and two-center phases   of  fig.\,\ref{fig:5}.

    \end{itemize}
     
       It is  instructive    to picture these  transitions  by plotting the metric
       factor $g_{tt}$  while  traversing 
       space  
       along  the   axis of reflection symmetry.  The   curve changes qualitatively 
as shown   in  figure \ref{fig:7}, 
    illustrating   the topological nature  of the
  transitions on the gravity side. 
      
             \begin{figure}[tbh!]
\centering
\vskip -0.6 cm
\includegraphics[width=.96\textwidth,scale=0.84,clip=true]{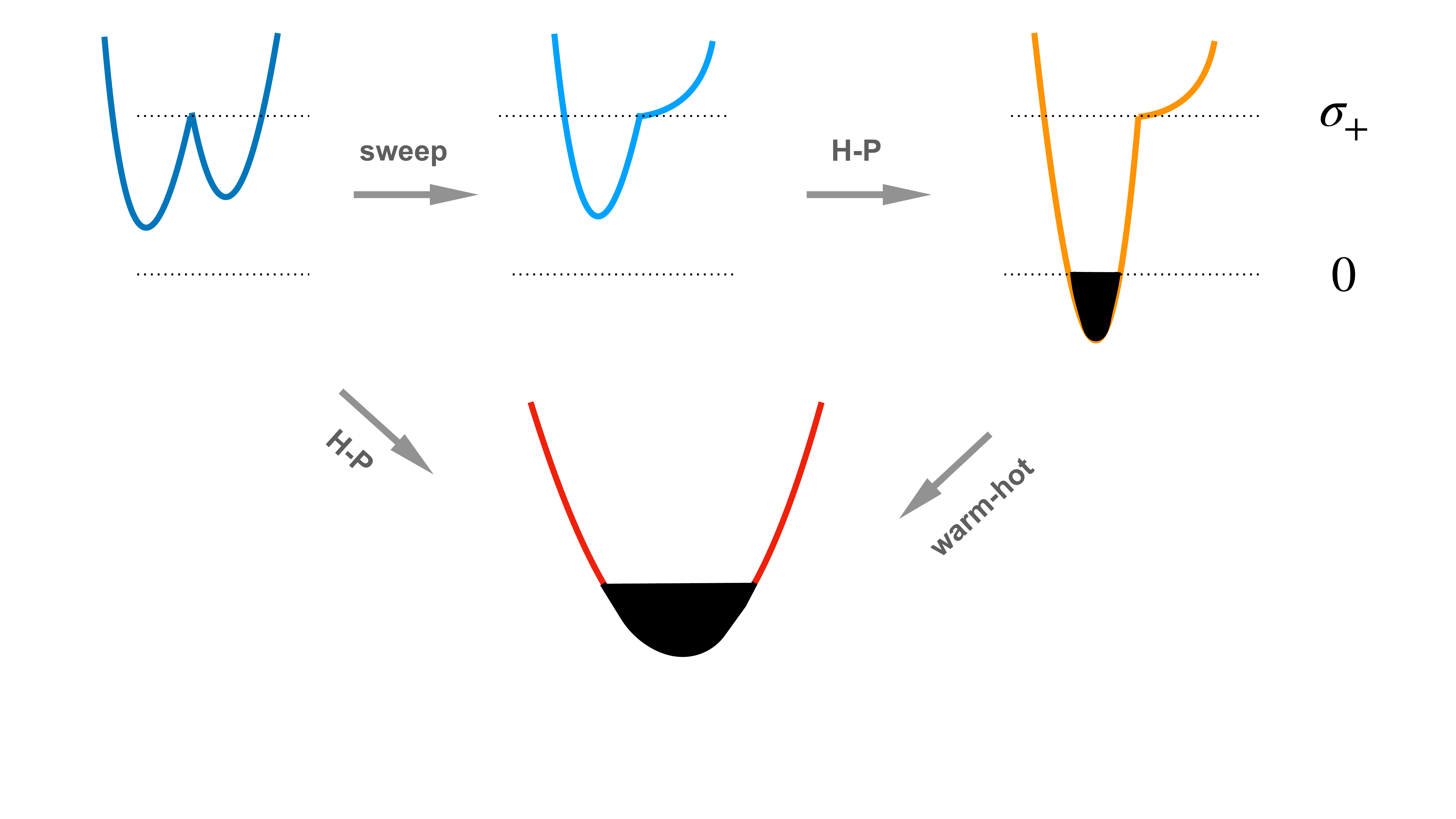}
  \vskip -6mm
      \caption{\footnotesize  Curves of the blueshift  factor $g_{tt}$ as one traverses space  along the
    $Z_2$ symmetry axis. The  color code is the same as in fig.\,\ref{fig:5}. 
 The wall is located at the turning point  $g_{tt}=\sigma_+$ where the curve is discontinuous. The grey arrows 
 indicate possible transitions. The blackened parts of the curves are  regions behind the horizon.
    }
    \label{fig:8}    
  \end{figure}

    Before embarking in numerical plots,   we will first do  the following   things: 
 (i) Comment on the ICFT interpretation of these  transitions; 
  (ii) Compute the  sweeping transitions analytically;  and (iii) Prove 
 that the warm-to-hot transitions are  first order, i.e.  that one cannot lower 
the wall to   the horizon continuously by varying the boundary data.


\subsection{ICFT  interpretation}
\label{puzzles}

       When   a holographic  dual exists,  Witten has argued that the appearance of a  black hole  at the 
Hawking-Page (HP) transition
signals  deconfinement in  the gauge theory \cite{Witten:1998zw}. 
Assuming   this interpretation\,\footnote{There is an extensive literature on the subject 
including  \cite{Sundborg:1999ue,Aharony:2003sx},  studies  specific 
to   two dimensions \cite{Aharony:2005ew,Keller:2011xi},  and 
 recent discussions
 in relation with  the superconformal index  in  $N=4$ super Yang Mills 
  \cite{Cabo-Bizet:2018ehj,Choi:2018hmj,Choi:2018vbz,Copetti:2020dil}. 
For an introductory  review see \cite{Marsano:2006kr}.
}
 leads to the conclusion  that in
 warm phases   a  confined theory  coexists with a deconfined one.  We will see below  
 that such   coexistence is easier when the confined theory is CFT$_2$, i.e. the  
theory  with the  larger central charge.\footnote{Even though for homogeneous 2-dimensional CFTs  the critical
temperature,  $\tau_{\rm HP}= 1$,  does not depend on  the central charge
by virtue of modular invariance.} 
  This is natural from the gravitational perspective. 
  Solutions of type {\small [H1, X]} are more likely than solutions of 
type {\small [X, H1]} because a  black hole forms more readily on the 
`true-vacuum' side of the wall. 
    We will actually provide  some  evidence later that if   $c_2 > 3c_1$ 
there are no equilibrium  phases at all  in  which   CFT$_2$ is deconfined while  
 CFT$_1$   stays  confined. 
 
        The question that jumps to one's mind is what happens for thick walls, where one  expects 
a warm-to-hot crossover rather than a sharp transition. One possibility is that the coexistence
of confined and deconfined phases  is impossible in microscopic holographic models.  Alternatively, 
an  appropriately defined Polyakov loop \cite{Witten:1998zw} 
could provide  a sharp order parameter for this transition.

   For sweeping transitions the puzzle is  the other way around.  Here  a  sharp order parameter 
exists in classical gravity --  it is the number of rest points for inertial observers. This   can be  defined 
 both for  thin-  and  for thick-wall geometriess. The interpretation on the field theory side
is however unclear.  The transitions 
 could be    related to properties of the  low-lying spectrum at infinite $N$, 
or 
  to the entanglement structure of the ground state.  

We   
  leave   these  questions  open for  future work. 


\subsection{Sweeping transitions}
\label{sec:71}

Sweeping transitions  are continuous transitions that happen at 
fixed  values of the  mass ratio $\mu$.  We will prove these statements
here. 

Assume for now continuity, and let the $j$th slice go from type {\small E1} to type  {\small E2}. 
The transition  occurs  when the string turning point
and  the center of  the $j$th AdS   slice coincide, i.e. when  
\bea\label{rj}
r_j(\sigma_+)  = \sqrt{ \sigma_+ + M_j\ell_j^2} \, =\,  0\ . 
\eea
Clearly this has a solution only if $M_j<0$. 
Inserting  in \eqref{rj}    the expressions \eqref{spm}\,-\,\eqref{ABC} for $\sigma_+$  
gives two  equations for the critical values of $\mu$
with the following solutions
  \bea\label{mucr}
 \mu_1^* =    { 1 - \ell_2^{\,2} \lambda^2\over   \ell_2^2/ \ell_1^2    }  \quad\ 
 {\rm and} \quad\ 
   \mu_2^* =     {\ell_1^2/ \ell_2^2 \over 1 - \ell_1^{\,2} \lambda^2}    \ \  .  
  \eea 

In the low-$T$  phases  both $M_j$ are negative and $\mu$ is positive.  
Furthermore, a little  algebra  shows   that  for all   
 $\lambda\in (\lambda_{\rm min},\, \lambda_{\rm max})$\,  the following is true
   \bea
    x_1^\prime\bigl\vert_{\sigma\approx\sigma_+} <0   \quad  {\rm at}\ \  \mu\gg 1 \ 
   \quad  {\rm and} \quad 
   x_2^\prime\bigl\vert_{\sigma\approx\sigma_+} <0  \quad  {\rm at}\ \ \mu\ll 1 \,.
    \ \ 
   \eea
This means that for  $\mu\gg 1$ the green slice is of type {\small E1}, and for 
$\mu\ll 1$ the pink  slice is of type {\small E1}. 
A sweeping transition can occur if the critical mass ratios \eqref{mucr}  are  in the allowed range. 
We   distinguish three regimes of  $\lambda$: 
\begin{itemize}
\item \,  {  Heavy}\,{($\lambda > 1/\ell_1$):} 
None of the $\mu_j^*\,$  is positive, so the  solution  is   of type {\small [E1,E1]} for all $\mu$,
i.e.\,cold  solutions are  always double-center\,;

 \item \, { Intermediate}\,{(${1/ \ell_1}> \lambda > 1/\ell_2$):} 
   Only $\mu_2^*\,$
  is positive. If  this   is  inside the  range  of  non-intersecting walls,
  the solution  goes   from {\small [E1,E2]} 
  at large $\mu\,$,  to    {\small [E1,E1]} 
   at small $\mu$. Otherwise the geometry  is  always of the single-center type  {\small [E1,E2]}\,;

\item \,{  Light}\,{($\lambda < 1/\ell_2$):}   Both  $\mu_1^*\,$ and $\mu_2^*\,$
  are  positive, so there is   the possibility of two sweeping transitions:  
 from   {\small [E2,E1]}  at small $\mu$   to {\small [E1,E2]}  at  large $\mu$
  passing  through the double-center 
  type {\small [E1,E1]}\,. Note that since $\lambda_{\rm min} = 1/\ell_1 - 1/\ell_2$,  this    range 
 of $\lambda$ only exists
  if  $\ell_2< 2\ell_1$,  i.e.\,\,when  CFT$_2$  has no more than   twice the number of   degrees
  of freedom of  the more depleted   CFT$_1$.  
  \end{itemize}           

 \noindent      
We can now confirm that sweeping transitions are continuous, not only in terms of the mass ratio  $\mu$
but also in terms of the ratio of volumes $\gamma$. To this end we expand the relations  \eqref{61}
around the above critical points  and show that the $L_j$ vary  indeed continuously  across the transition.
The calculations can be found   in  appendix \,\ref{app:3}\,.

  For  the warm phases we proceed along  similar lines.  
One of the two $M_j$ is now equal to $(2\pi T)^2 >0$, 
so sweeping transitions may  only occur for negative  $\mu$.   Consider first  warm
solutions of type {\small [H1,X]} with 
the  black hole in  the `true vacuum'  side.  A little calculation shows that  
$\,x_2^\prime\vert_{\sigma\approx\sigma_+}$  is  negative,
i.e.  {\small X=E1},\,
 if and only if 
\bea
 \lambda > {1\over \ell_1}\qquad {\rm and}\qquad  \mu <  \mu_2^* 
<0  \ . 
\eea
Recall that when {\small X=E1}  some inertial observers can be shielded from the black hole by taking
refuge at the  restpoint of the pink slice. We see that this is only possible 
for  heavy walls  
($\lambda >1/\ell_1$)  and for $\mu <\mu_2^*$.  
A sweeping transition  
{\small [H1,E1]} $\to$ {\small [H1,E2]} takes place  at  $\mu= \mu_2^*$.

 Consider  finally a black hole  in the `false vacuum' side, namely
 warm solutions  of  type  {\small [X,H1]}. 
Here   $\,x_1^\prime\vert_{\sigma\approx\sigma_+}$ is negative, i.e. 
  $X$ has a rest point, if and only if the following conditions are satisfied
\bea\label{BHsweep}
 \lambda > {1\over \ell_2}\qquad {\rm and}\qquad \hat  \mu := \mu^{-1}  <  (\mu_1^*)^{-1} <0\ . 
 \eea
Shielding from the black hole looks here easier,  both heavy and intermediate-tension
walls can do   it.  In reality,  however,  we have found  that solutions with the black hole in the `false vacuum' side
are rare, and that the above inequality pushes  $\hat \mu$ outside the  admissible range.  
 
 The  general trend emerging from the    analysis is that  the heavier the wall 
the  more likely are  the two-center geometries.  A suggestive  calculation actually  shows that
\bea
  {\partial \sigma_+ \over \partial\lambda}\Bigl\vert_{M_j\ {\rm fixed}} \  \ {\rm is}\ 
\begin{cases} {\textrm {  positive\ \  for \ two-center\ solutions}}\\ \,
  {\textrm {negative\ \  for \ single-center\ solutions}}
\end{cases}
\eea
where the word `center' here includes  both  an  AdS restpoint and  a black hole. 
At fixed energy densities a  single center is  therefore pulled closer to  a heavier wall, while 
two centers are instead pushed away. It might be interesting to also compute 
${\partial \sigma_+ /\partial\lambda}$ and  
${\partial V/\partial\lambda}$ at $L_j$ fixed,  where $V$ is the regularized  volume of  the interior space.
In the special case of  the vacuum solution with 
an
AdS$_2$ wall,  the volume  (and the associated complexity  \cite{Chapman:2018bqj}) 
can be  seen to grow with  the tension $\lambda$. 
 


\subsection{Warm-to-hot  transitions}
\label{sec:72}
   
   In   warm-to-hot transitions   the thin  domain wall   enters   
the black-hole   horizon.  
One may have expected this to happen continuously,
  i.e.  to be able to
  lower   the wall to the horizon  smoothly,  
 by  slowly varying the  boundary data   $L_j, T$. We will now show that,
if  the tension $\lambda$ is  fixed, 
the transition is actually always  first order.

  Note first that in warm solutions   the  slice  that  contains the black hole
has   $M_j= (2\pi T)^2$. 
 If  the  string  turning point  
 approaches   continuously  the horizon, then    $\sigma_+ \to 0$. 
  From eqs.\,(\ref{ABC}, \ref{spm}) we see that   this  can happen
if and only if $(M_1-M_2)\to 0$,  which implies in passing that  the solution must necessarily  be 
of type {\small [H1,E2$^\prime$]} or  {\small [E2$^\prime$,H1]}. 
Expanding   around  this putative point where the wall touches the horizon 
 we set 
   \bea
        {M_1 - M_2 \over M_1+M_2} := \delta    \quad {\rm with}\quad  \vert \delta\vert \ll 1     \  \   \Longrightarrow\  \ 
        \sigma_+ \approx  \biggl( {2\pi T\over \lambda}\biggr)^2 \delta^2  \, . 
         \eea 
Recalling that the horizonless slice has the smaller $M_j$ we see that 
for  positive $\delta$  the black hole must be  in the green slice and $\mu = 1 - 2\delta + {\cal O}(\delta^2)$, while for 
negative $\delta$  the black hole is  in the pink  slice and  $\hat \mu = 1+  2\delta + {\cal O}(\delta^2)$. 
\smallskip

        The second option  can be  immediately ruled out  since   it is impossible to satisfy the  
boundary conditions \eqref{hatDir}.  Indeed, 
 $\hat f_1(\hat\mu \approx 1)$ is manifetsly positive, as is clear  from eq.\,\eqref{hatmua}, 
 and  we have assumed that  $\mathbb{S}_1$  is  of type  {\small E2$^\prime$}.
   Thus the second condition \eqref{hatDir} cannot be obeyed.
    By the same reasonning we see that for  $\delta$ positive, and since now   
$\mathbb{S}_2$  is  of type  {\small E2$^\prime$}, we need that $\tilde  f_2(\mu \approx 1)$ be negative. 
As is clear  from the expression \eqref{tildemub} this  implies that  $\lambda <\lambda_0$. 

The upshot of the discussion  is that a warm solution  arbitrarily close to the hot solution
may exist only if $\lambda <\lambda_0$ {\it and} if   the black hole is on the true-vacuum side. 

  It is   easy to see that under these conditions the  two branches of solution indeed  meet at  $\mu =1$, 
$\Delta x_2\vert_{{\rm Hor}}=0$ and  hence from \eqref{HighTarcsb}
\bea\label{tau2starr}
\tau_2 \, = \, {1\over \pi  }\, {\rm tanh}^{-1}\left( {\ell_2 (\lambda_0^2 -  \lambda^2 )
  \over 2\lambda}\right) \,:=\,     \tau_2^*\, . 
\eea
Recall from section \ref{sec:61}
 that this  is the limiting value for the existence of the hot solution -- the solution ceases to exist 
at   $\tau_2<  \tau_2^*\, . $ The nearby warm solution could in principle  take over in this forbidden range, 
provided that $\tau_2(\delta)$ decreases as  $\delta$ moves away from zero. It actually turns out
that $\tau_2(\delta)$  initially  increases  for small   $\delta$, so  this last possibility  for a continuous
warm-to-hot transition is also ruled out.

  To see why this is so, expand   \eqref{tildeDir} and \eqref{tildemub} around $\mu=1$, 
$$
 \tilde s_+ = {\delta^2\over \lambda^2} + {\cal O}(\delta^3)\,,\quad
 \tilde s_-  =   - {4\lambda^2\over A}\Bigl(1-\delta (1+ {\lambda_0^2\over \lambda^2})\Bigr) 
   + {\cal O}(\delta^2)\,,\quad
$$
and shift  the  integration variable $s := y+\tilde s_+$ so that \eqref{tildemub} reads
\bea\label{delta}
2\pi\,\tau_2(\delta)  =  {\ell_2\over\sqrt{A}}\int_0^\infty  dy \left[  { y(\lambda_0^2 - \lambda^2) + 2\delta\over
(y+ \mu\ell_2^2) \sqrt{y(y+  \tilde s_+) (y - \tilde s_-) } } + {\cal O}(\delta^2)
\right]\ . 
\eea
We  neglected in the integrand all contributions  of ${\cal O}(\delta^2)$ except for the  
$ \tilde s_+$   in the denominator
that regulates
the logarithmic divergence of the ${\cal O}(\delta \log\delta)$ correction. 
 Now use the   inequalities
$$
 { y(\lambda_0^2 - \lambda^2) + 2\delta\over
 \sqrt{(y+  \tilde s_+) (y - \tilde s_-) } } >   { y(\lambda_0^2 - \lambda^2) + 2\delta\over
 \sqrt{ (y+  \delta^2/\lambda^2) (y +4\lambda^2/A ) } } >
{\sqrt{ y} (\lambda_0^2 - \lambda^2)  \over
 \sqrt{  (y +4\lambda^2/A ) } }\ , 
$$
where the second one  is equivalent to $2\delta > (\lambda_0^2/\lambda^2 - 1)\delta^2$, which 
is true  for small enough $\delta$. Plugging in \eqref{delta} shows that 
$\tau_2(\delta)  > \tau_2(0)$
 at the   leading order  in $\delta$, 
proving   our claim. 

            \begin{figure}[tbh!]
\centering
\includegraphics[width=.82\textwidth,scale=0.84,clip=true]{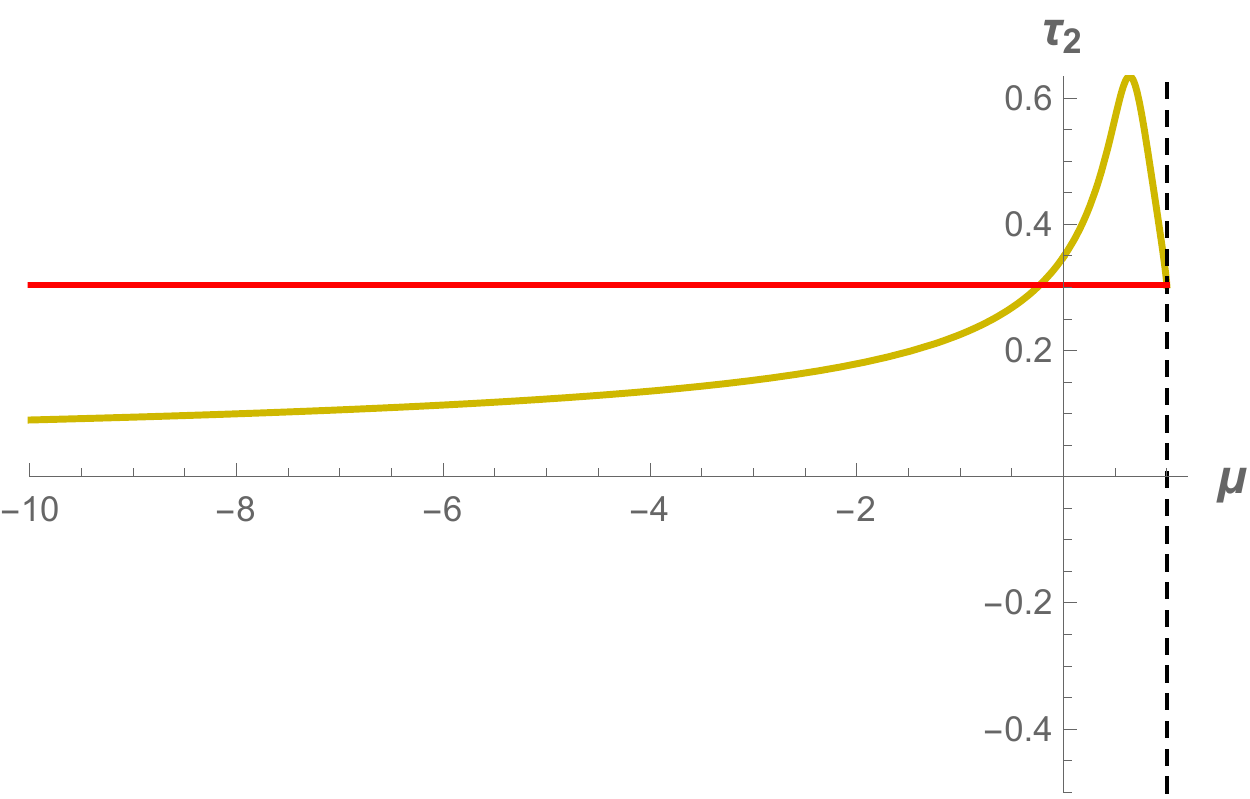}
       \caption{\footnotesize  The function $\tau_2(\mu) $  in the 
{\small[H1,E2]} and  {\small[H1,E2$^\prime$]}
branches of solutions, for $2\ell_2 = 3\ell_1 $ and $\lambda = {3/5\ell_2} < \lambda_0 = {\sqrt{5}/ 2\ell_2}$\,. 
 The red line indicates the bound  $\tau_2^*$ below   which the hot solution ceases to exist. }
    \label{fig:tau2}    
  \end{figure}

\vskip -2mm

A typical   $\tau_2(\mu)$ \,  in the  {\small[H1,E2]} and  {\small[H1,E2$^\prime$]}
branch of solutions, and 
  for $\lambda <\lambda_0$, is plotted   in figure \ref{fig:tau2}.  The function 
 grows initially as  $\mu$ moves away from 1,
reaches a maximum value and then turns around and goes to zero as $  \mu \to -\infty$. 
The red line indicates the limiting value  $\tau_2^*$ below which there is no   hot solution.
For  $\tau_2$ slightly above  $\tau_2^*$ we see that there are three coexisting black holes,
 the hot and two warm ones.
  For $\tau_2 < \tau_2^*$, on the other hand,  only one warm solution survives, but it  describes a  wall  
at a finite distance from the horizon. Whether this is the dominant solution or not,  the transition is
therefore necessarily  first order.

\section{Exotic fusion and bubbles}
\label{sec:Faraday}

   Before proceeding to  the  phase diagram,  we pause   here to discuss  the   peculiar phenomenon
 announced   earlier,   in section \ref{sec:62}. This   arises in the limits
$ \gamma= L_1/L_2\to 0$ or $\gamma \to \infty$,  with $L_1+L_2$ and $T$ kept fixed. In these limits 
 the conformal boundary of one slice   shrinks  to a point.

Consider for definiteness  the limit $L_1 \to 0$. 
In the language of the   dual field theory 
the interface and anti-interface fuse in this limit  into a defect of CFT$_2$.  The naive expectation, based on
free-field calculations  \cite{Bachas:2007td,Bachas:2012bj,Bachas:2013ora}, 
 is that this is
 the trivial (or identity)  defect. 
Accordingly, the green  interior  slice should  recede to   the conformal   boundary, leaving  as the
  only remnant
a  (divergent)  Casimir energy.  

We have found that this    expectation is not always   borne out    as we will now explain.
      
  Suppose first that the surviving   CFT$_2$ is in its ground state, and that the result of the
interface-antiinterface  fusion is the 
expected trivial defect. 
 The geometry   should   in this case  approach  global AdS$_3$ of radius $\ell_2$,   with  
$M_2$  tending to  $ -(2\pi/L_2)^2$,  see section \ref{sec:2}. 
Furthermore,  $\sigma_+$  should   go to infinity   in order for the green slice to  shrink towards
 the ultraviolet region.
As seen  from eqs.\,(\ref{spm},\,\ref{ABC})  this requires 
 $M_1 \to -\infty$,  so that  $\mu\,$ should vanish together with  $\gamma$. 
    This is indeed  what happens in  much of the 
  $(\lambda, \ell_1, \ell_2)$ parameter space. One finds  
  $\mu \sim \gamma^{2}\to 0$,   a scaling compatible   with 
the expected    Casimir energy $\sim \#/L_1$.

\begin{figure}[t!]
\centering
\vskip -0.1 cm
\includegraphics[width=.98\textwidth,scale=0.84,clip=true]{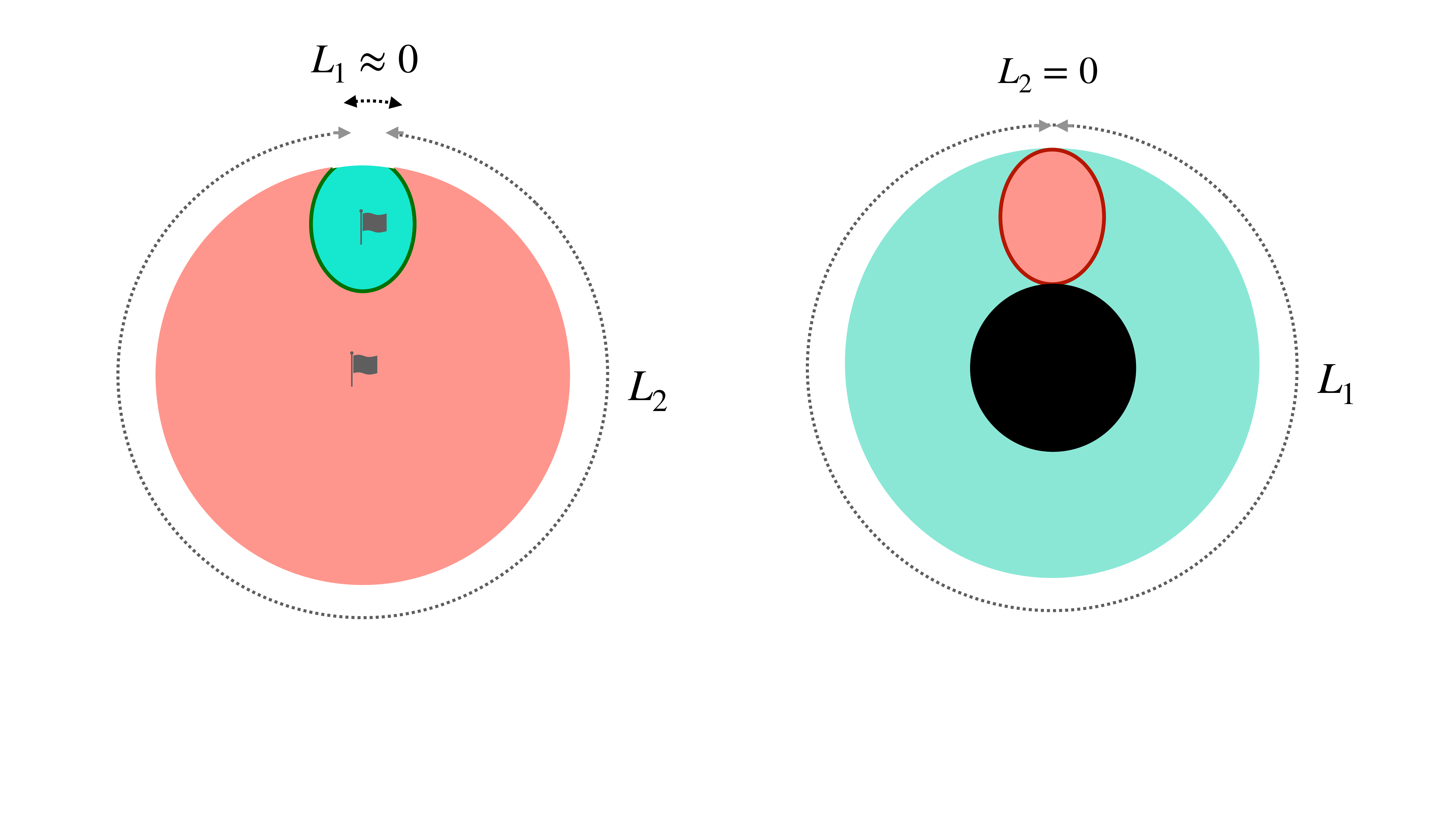}
\vskip -17mm
    \caption{\footnotesize  Left: A bubble of true vacuum  that survives inside the false vacuum despite  the 
fact that its conformal  boundary  shrinks to a point.   Right: A bubble of false vacuum with $\lambda =\lambda_0$
inscribed  between the boundary and the horizon of a black hole.
    }
    \label{fig:7}    
  \end{figure}

      Nevertheless,  sometimes
$\gamma$ vanishes at   finite $\mu_0$.
In such   cases, as $\mu\to\mu_0$  the green slice does not disappear 
even though   its conformal boundary has shrunk to a point.  This is illustrated
  by the left  figure \ref{fig:7}, which shows  a  static bubble of `true vacuum'   suspended from  a point 
on the  boundary of the `false vacuum'.\footnote {
These are static solutions,  not to be confused with `bags of gold'
 which are  cosmologies  glued onto the backside
of 
a Schwarzschild-AdS spacetime, see e.g.\,\cite{Marolf:2008tx,Fu:2019oyc}. 
The phenomenon is reminiscent of  spacetimes  that realize
 `wedge'  or codimension-2 holography, like those  in refs.\,\cite{Bachas:2017rch,Akal:2020wfl,Miao:2020oey}. 
} 
To convince ourselves  that the phenomenon  is real, we give  an analytic proof in appendix \ref{app:4}
of the existence of
such suspended  bubbles in at least one  region of parameters  
  ($\ell_2 > \ell_1$ and $\lambda\approx \lambda_{\rm min}>0$). 
Furthermore, since the vacuum solution for a given $\gamma$ is unique, there is no other competing
solution. In   the example of appendix \ref{app:4}, in particular,  $\gamma$ is 
finite and negative at $\mu=0$.

\noindent

    In the language  of field theory this is a striking phenomenon. It implies 
  that interface and anti-interface do not annihilate, but fuse
into an exotic defect  generating  spontaneously a new  scale in the process.    This is 
the  blueshift at the tip of  the bubble,    $\sigma_+(\mu_0, L_2)$, or better the corresponding frequency 
scale $r_2(\sigma_+)$  in  the D(efect)CFT.

The   phenomenon is  
not   symmetric under the exchange  $1\leftrightarrow 2$. Static bubbles of the  false vacuum 
(pink)  spacetime inside  the  true (green) vacuum   do not seem to
 exist. We  proved    this analytically for $\lambda < \lambda_0$, and 
 numerically for all other  values of the tension.  
We have also found that the
 suspended  green bubble  can   be of type {\small E1}, i.e. have a center.
 The redshift factor $g_{tt}$ 
inside the bubble can  even be lower than   in the surrounding  space,  so that the bubble
hosts the excitations of lowest energy. We did not show  this  analytically, but the numerical
evidence is compeling. 

\begin{figure}[h!] 
 \centering
\begin{subfigure}{.45\textwidth}
  \centering
  \includegraphics[width=1\linewidth]{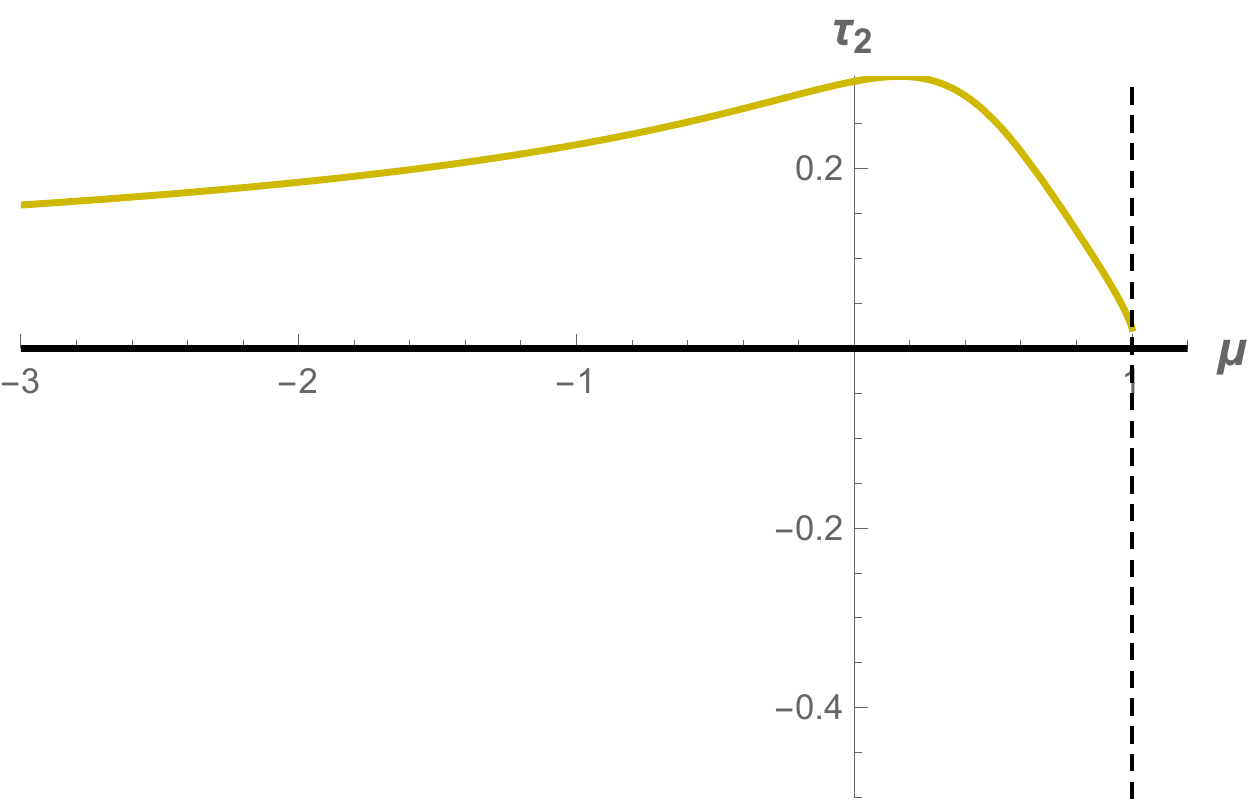}
 \end{subfigure}%
\hspace{10mm}
\begin{subfigure}{.45\textwidth}
  \centering
  \includegraphics[width=1\linewidth]{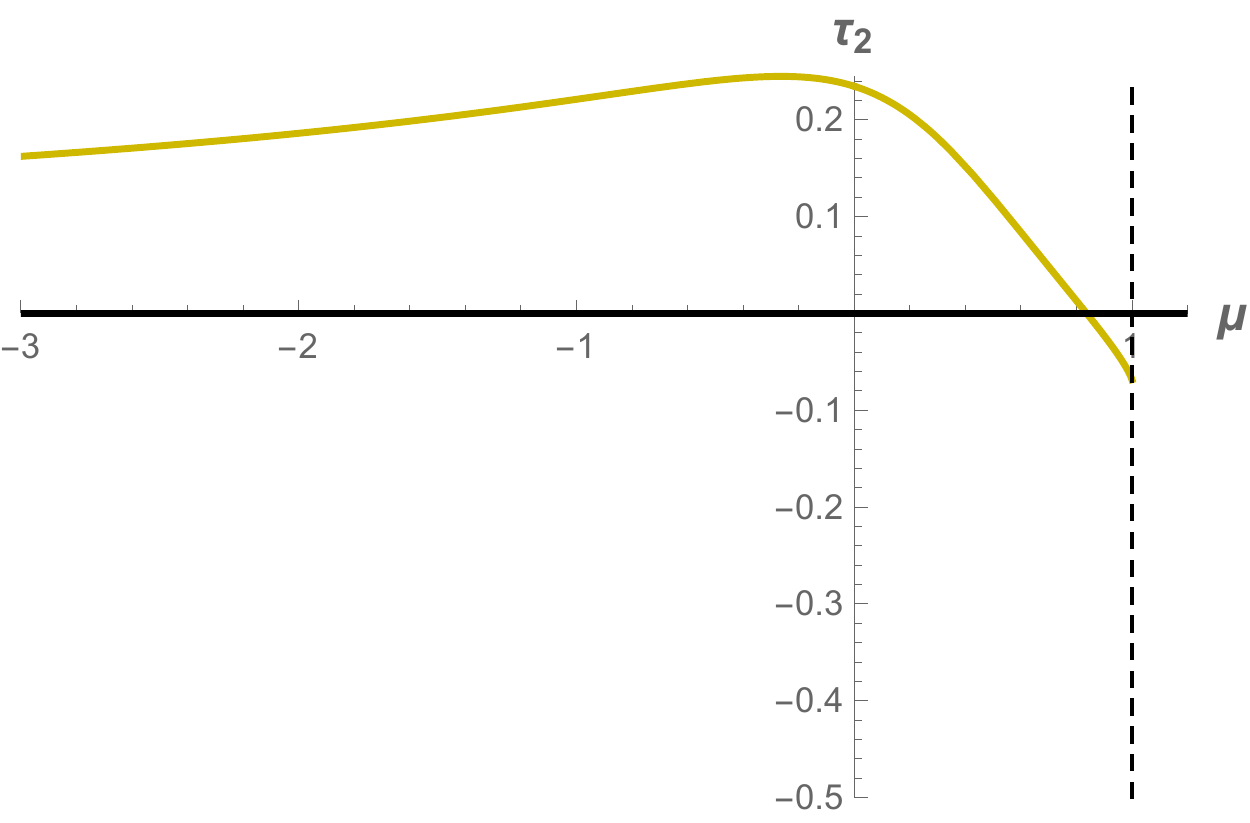}
 \end{subfigure}
 \caption{\footnotesize   Plots of the  function $\tau_2(\mu)$ in the 
 [H1,E2] or  [H1,E2$^\prime$] branch of solutions for $\ell_2/\ell_1= 1.5$. The critical tension  is 
$\lambda_0 \ell_2 \approx 1.12$.  The curve on the left is for $ \lambda_0 \ell_2 = 1.05$, and the
curve on the right for $\lambda_0 \ell_2 = 1.35$. 
 }
\label{fig:app}
\end{figure}

  Do    suspended  bubbles  also exist when 
the   surrounding  spacetime contains  a   black hole\,? 
The answer is affirmative as one can show
  semi-analytically 
by focussing on the region $\lambda \approx \lambda_0$. We have seen in 
  the previous  section that near this critical tension there exist
 warm solutions of type {\small[H1,E2$^\prime$]}
with the wall arbitrarily close the horizon. 
Let us consider the function $\tau_2(\mu, \lambda)$ 
 given in this branch of solutions by  eqs.\,\eqref{tildemub} and \eqref{tildeDir}  
(with $\mathbb{S}_2 \not=${\small E1}). 
This is a continuous function in both arguments, so as 
   $\lambda$ increases past $\lambda_0$, 
$\tau_2(1)$  goes from positive to negative  with
 the overall shape of the
function varying  smoothly.   This is illustrated in
figure \ref{fig:app}, where we plot $\tau_2(\mu)$ for  $\lambda$ slightly below and slightly above
$\lambda_0$.  It  should be  clear from these plots that  for  $\lambda > \lambda_0$ (the plot on the right)
$\tau_2$ vanishes at  a finite  $\mu\approx 1$. This is a  warm bubble solution, as  advertized.

    We have found more generally that warm  bubbles  can  be also  of type {\small E1}, thus acting as
a suspended Faraday cage that protects  inertial observers from falling towards the horizon of the
black hole.     Contrary,  however,  to what happened for  the ground state, warm   bubble solutions are  not unique. 
There is always a  competing solution at $\mu\to -\infty$, and it is the dominant one  by virtue of 
its divergent negative Casimir energy.   A  stability analysis would show 
if warm  bubble solutions can be metastable and long-lived, but this is  beyond our present  scope.
 
   As for  warm   bubbles of type {\small[X, H1]},  that is with the black hole
in the false-vacuum slice, 
 these also 
 exist but only if  $\ell_2<3\ell_1$.   Indeed, as we will see   in a moment, when  $\ell_2> 3\ell_1$
the wall cannot  avoid a  horizon  located on the false-vacuum side.

   Finally   simple inspection of  fig.\,\ref{fig:app} shows that by varying the tension,  the   bubble solutions for 
 $\lambda > \lambda_0$  go over smoothly to the hot solution at  $\lambda=\lambda_0$.  
At  this critical tension the bubble is inscribed between the horizon and the conformal boundary, 
 as    in  figure \ref{fig:7}.  
 This gives another    meaning to   $\lambda_0$:   Only walls with  this  tension  
may  touch  the horizon without falling inside.


\section{Phase diagrams} 
\label{sec:8}

  In this last section of the paper we present numerical plots of the phase diagram of the
model. We work in the canonical ensemble, so the  variables are the temperature and volumes, or
by scale invariance  two of the dimensionless ratios  defined in \eqref{names}.   
We choose these to be  
$\,\tau  = \tau_1+\tau_2 = T(L_1+L_2)$ and
$\,\gamma = L_1/L_2$. The color  code is   as in  fig.\,\ref{fig:5}. 

  We plot  the  phase diagram   for different   values of  the action parameters 
$\ell_1, \ell_2, \lambda$. 
   Since our analysis is classical in gravity, Newton's constant $G$ plays no role. Only two
dimensionless ratios   matter,\footnote{Dimensionless in gravity, not in the dual  ICFT. 
} 
for instance
\bea\label{kappa}
  b := {\ell_2\over \ell_1} = {c_2\over c_1} \geq 1 \qquad {\rm and}\quad  \kappa := \lambda \ell_2 \in (b-1, b+1)\ . 
\eea
The value  $b=1$ corresponds to a defect  CFT, while  $b\gg 1$ 
is the opposite  ``near void" limit in which the  degrees of freedom of  CFT$_2$ ovewhelm  
those of CFT$_1$. The true vacuum approaches in this limit the infinite-radius
AdS,  and/or  the  false  vacuum approaches 
flat spacetime. 
 The critical tension $\lambda_0$ corresponds to $\kappa_0 = \sqrt{b^2-1}$.

   To  plot  the  phase diagrams we   solved numerically for  $\mu$ in terms  of the boundary data
$(\gamma, \tau)$
and for all  types of slice pair,  and  compared their  free energies when solutions of different type coexist.
   As explained in the introduction, although the interpretation is different,  our  diagrams are related to
the ones of Simidzija and Van Raamsdonk
\cite{Simidzija:2020ukv} by double-Wick rotation (special to 2+1 dimensions). Since time in  this reference 
is non-compact, only  the boundaries of our phase diagrams,  
at $\gamma = 0$ or $\gamma= \infty$, can be compared. The roles of thermal AdS and BTZ are 
also exchanged


   \subsection{Defect CFT}    

\begin{figure}[h!]
 \centering
\begin{subfigure}{.40\textwidth}
  \centering
  \includegraphics[width=1\linewidth]{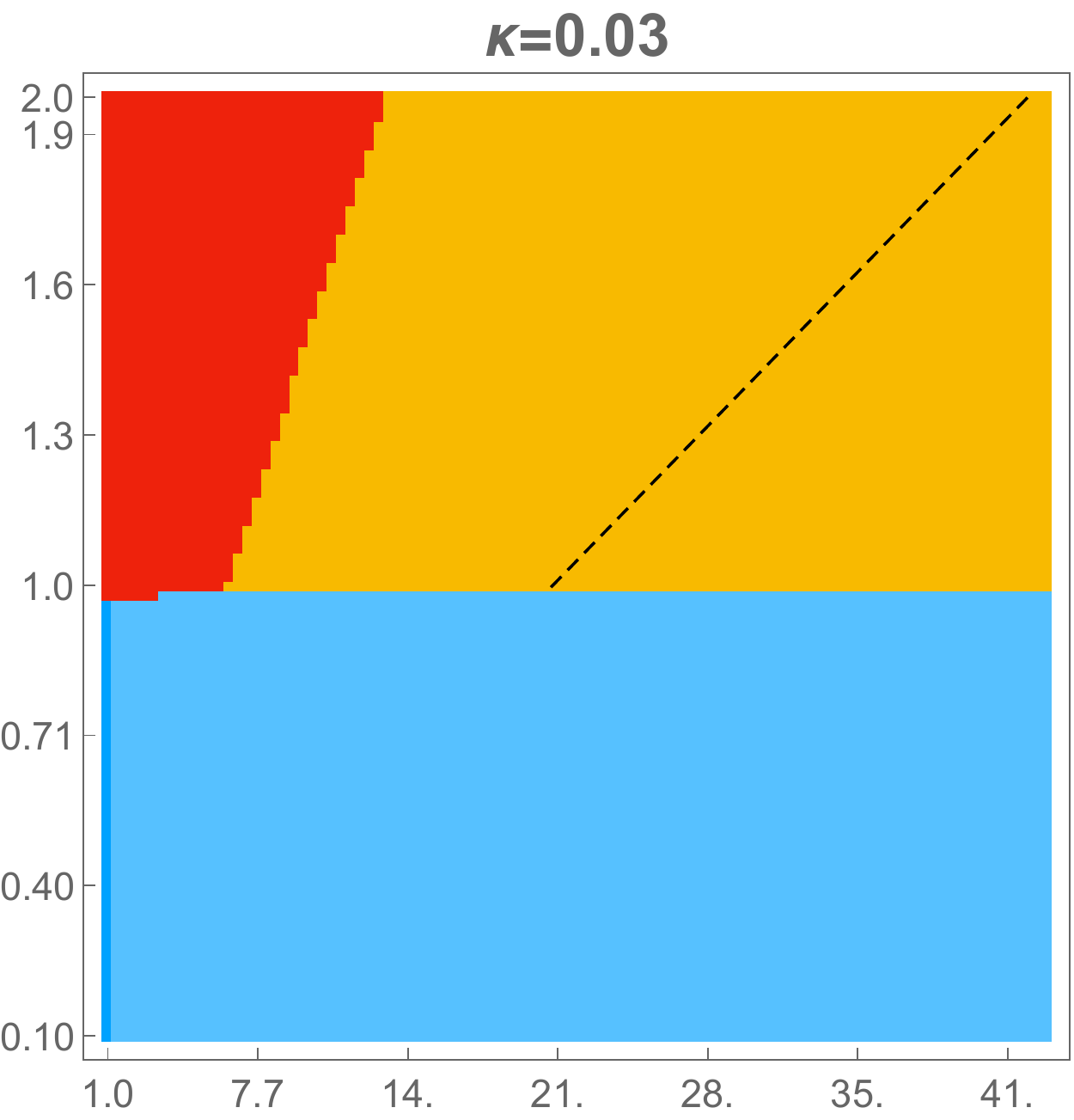}
 \end{subfigure}%
\hspace{19mm}
\begin{subfigure}{.4\textwidth}
  \centering
  \includegraphics[width=1\linewidth]{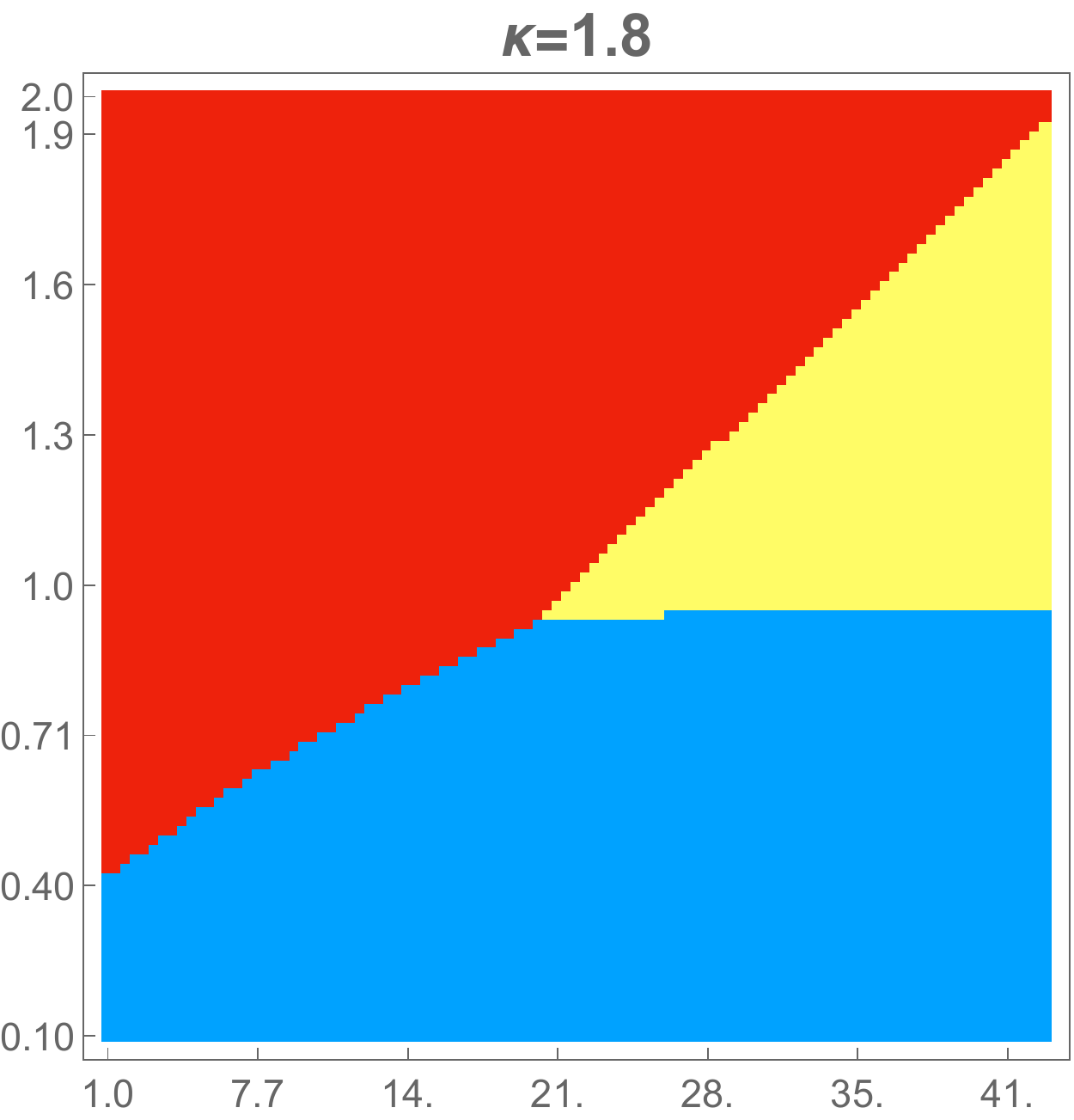}
 \end{subfigure}
 \caption{\footnotesize   Phase diagrams  of  a very light (left)  and a very heavy (right)
domain wall between   degenerate  vacua ($b=1$). The horizontal and vertical axes are $\gamma$ and $\tau$.
The broken  line in the left diagram separates  solutions of type [H1, E2$^\prime$] and  [H1,E2] that   only
differ in the sign of the energy of  the horizonless slice. The color code is as in fig.\,\ref{fig:5}.
 }
 \label{fig:9}
\end{figure}

  Consider first $b=1$. By symmetry,  we may restrict  in this case
to  $\gamma \geq 1$.
 Figure\,\,\ref{fig:9} presents  the phase diagram in the $(\gamma, \tau)$ plane  
 for a   very light   ($\kappa = 0.03$) and a very heavy ($\kappa = 1.8$)
  domain wall. 
For the light, nearly  tensionless,  wall   the phase diagram  
approaches that  of a homogeneous CFT. The low-$T$  solution is   single-center,
and the 
  Hawking-Page (HP)  transition occurs at $\tau \approx  1$. 
 Light   domain walls follow closely geodesic curves, and   
   avoid  the horizon in a large region of parameter 
space.\,\footnote{One can  compute this phase  diagram analytically
  by expanding in  powers of  $\lambda$.}  

Comparing  the left with the right figure \ref{fig:9} shows that heavy walls facilitate 
 the formation of the black hole  
  and have   a harder time staying outside.  Indeed,  in the right figure the 
HP  transition occurs   at  lower $T$,   and the  warm phase   recedes to $L_1\gg L_2$. 
Furthermore, both the cold and the warm solutions  have  now an additional  AdS restpoint. 
This confirms  the   intuition that heavier walls repel  probe masses  more strongly, 
and    can shield them from 
 falling inside  the black hole.

\begin{figure}[h!] 
 \centering
\begin{subfigure}{.40\textwidth}
  \centering
  \includegraphics[width=1\linewidth]{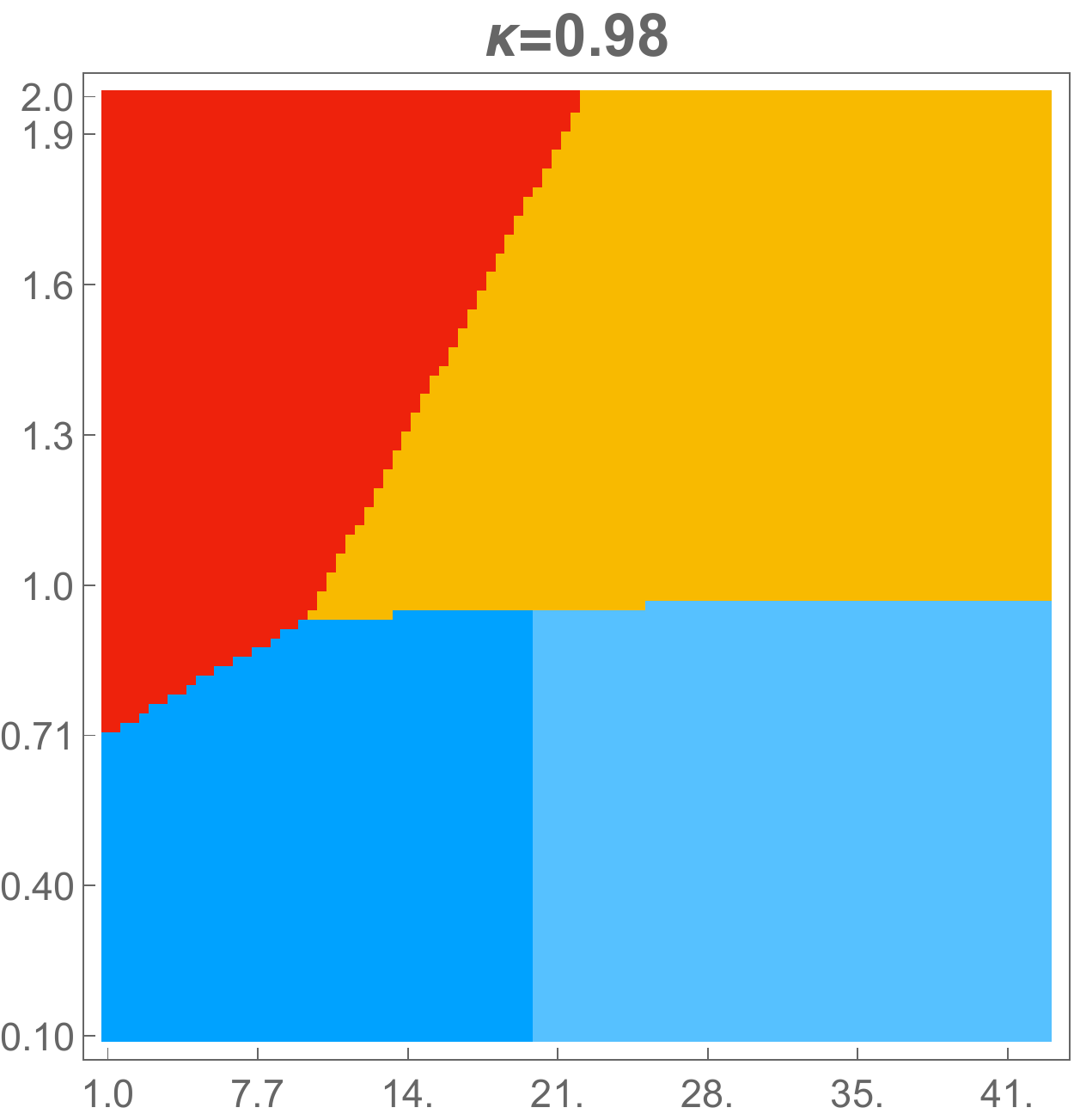}
 \end{subfigure}%
\hspace{19mm}
\begin{subfigure}{.4\textwidth}
  \centering
  \includegraphics[width=1\linewidth]{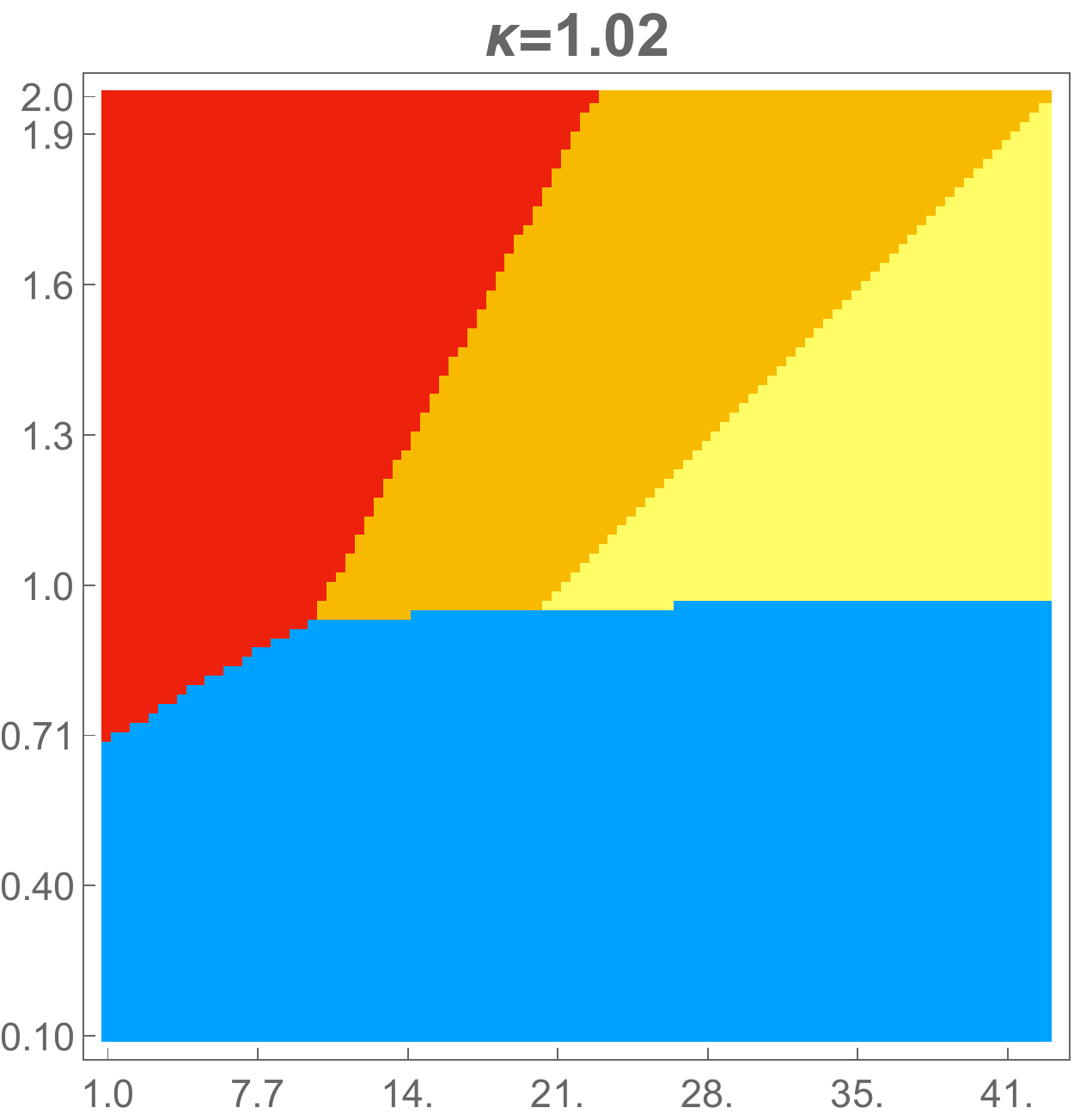}
 \end{subfigure}
 \caption{\footnotesize   Phase diagrams  for  intermediate-tension walls exhibiting sweeping
transitions.  On the left   a restpoint of the vacuum solution  is swept away as  $\gamma$ increases
beyond a critical value.  On the right   the same happens   in the warm solution but 
for decreasing  $\gamma$.  In these diagrams, the one-center warm solution is always [H1,E2].
 }
\label{fig:10}
\end{figure}

  The  transition  that sweeps  away this AdS restpoint   is shown   explicitly  in the 
  phase diagrams  of 
 figure  \ref{fig:10}.  Recall from the analysis of section \ref{sec:71} that in the low-$T$  phase
such  transitions  happen  for   $\lambda < 1/\ell_1 \Longrightarrow \kappa < b = 1$.  Furthermore,
the transitions  take  place at the critical mass ratios  $\mu_j^*\,$,  given by eq.\,\eqref{mucr}.
Since 
in cold  solutions the relation between $\mu$ and $\gamma$ is one-to-one, the  
dark-light blue critical lines
are  lines  of constant $\gamma$. 
These  statements are in perfect agreement with the findings of   fig.\,\ref{fig:10}. 

Warm  solutions of type {\small [H1,E1]}, respectively   {\small [E1,H1]}, 
 exist for  tensions
$\lambda >  1/\ell_1 \Longrightarrow \kappa > b$,   respectively 
$\lambda >  1/\ell_2 \Longrightarrow \kappa > 1$. In the case of  a defect  these two ranges coincide.
The stable black hole forms in the larger of the two slices, i.e.  for $\gamma >1$
in the $j=1$ slice. The sweeping transition
occurs at  the critical mass ratio  $\mu_2^* = (b^2-\kappa^2)^{-1}$,  which through 
 eqs.\,\eqref{tildeDir} and \eqref{tildemua} corresponds to a fixed value of 
$\tau_2$. Since $\tau = \tau_2(1+\gamma)$, the critical  orange-yellow 
line is  a straight line in the $(\gamma, \tau)$ plane,
in accordance again with the findings  of  fig.\,\ref{fig:10}.

 A noteworthy fact is the rapidity of these  transitions as function of $\kappa$.  
For  $\kappa$  a little below or above the critical
value the single-center cold, respectively warm phases almost disappear. 
Note also the cold-to-warm transitions are always near  $\tau \approx 1$. This is the critical value
for Hawking-Page transitions in the homogeneous case,  as expected at large $\gamma$ when  the  $j=1$
slice  covers most of space.

  The critical curves  for the cold-to-hot   and warm-to-hot transitions 
also look linear 
 in the above figures, but this is an  illusion. Since the  transitions are
first order we must compare   free energies. 
 Equating  for example  the hot and cold free energies 
gives   after some   rearrangements (and with   $\ell_1=\ell_2:= \ell$)  
\bea\label{criticalnotline}
 2\pi^2   \tau +  {2\over \ell} \log g_I =   {1\over 2\tau_1}   \vert M_1\vert L_1^2\, (1+ {\mu\over \gamma})\ . 
\eea
Now $\vert M_1\vert L_1^2$ can be   expressed in terms of $\mu$ through eq.\,(\ref{61}, \ref{mua}),
and $\mu$ in the cold phase  is a function of $\gamma$. Furthermore 
$ \log g_I/\ell= 4\pi {\rm tanh}^{-1} (\kappa/2)$  is  constant, see eq.\,\eqref{vanR}, and 
$\tau_1 = \tau/(1+\gamma^{-1})$. 
Thus   \eqref{criticalnotline} can be written as  a relation $\tau= \tau_{\rm hc}(\gamma)$, and we have
 verified  numerically that  $\tau_{\rm hc}$ is {not}  a linear function of $\gamma$. 
 

   \subsection{Non-degenerate vacua}    
 
\begin{figure}[h!] 
 \centering
\begin{subfigure}{.47\textwidth}
  \centering
  \includegraphics[width=1\linewidth]{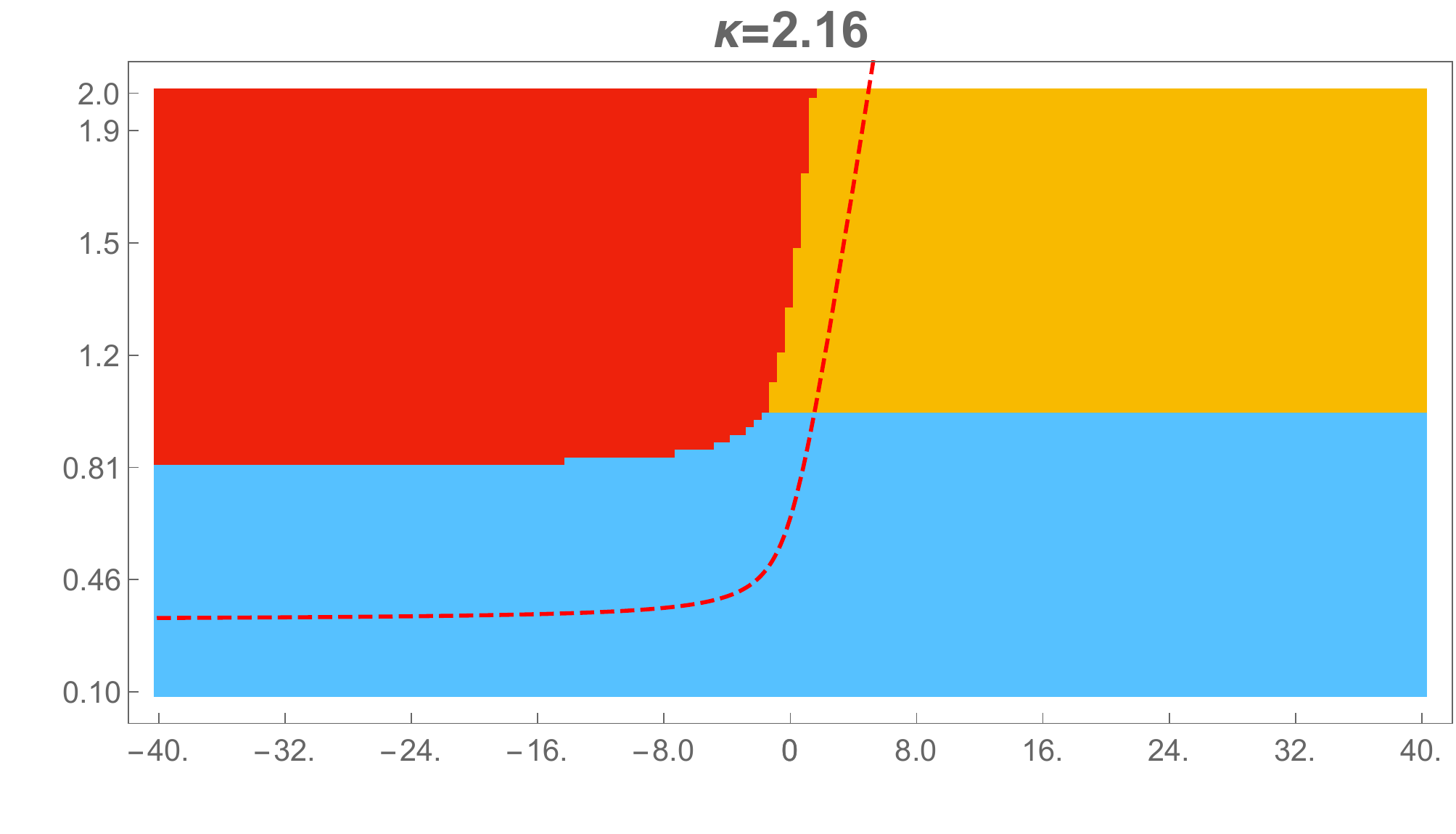}
 \end{subfigure}%
\hspace{4mm}
\begin{subfigure}{.47\textwidth}
  \centering
  \includegraphics[width=1\linewidth]{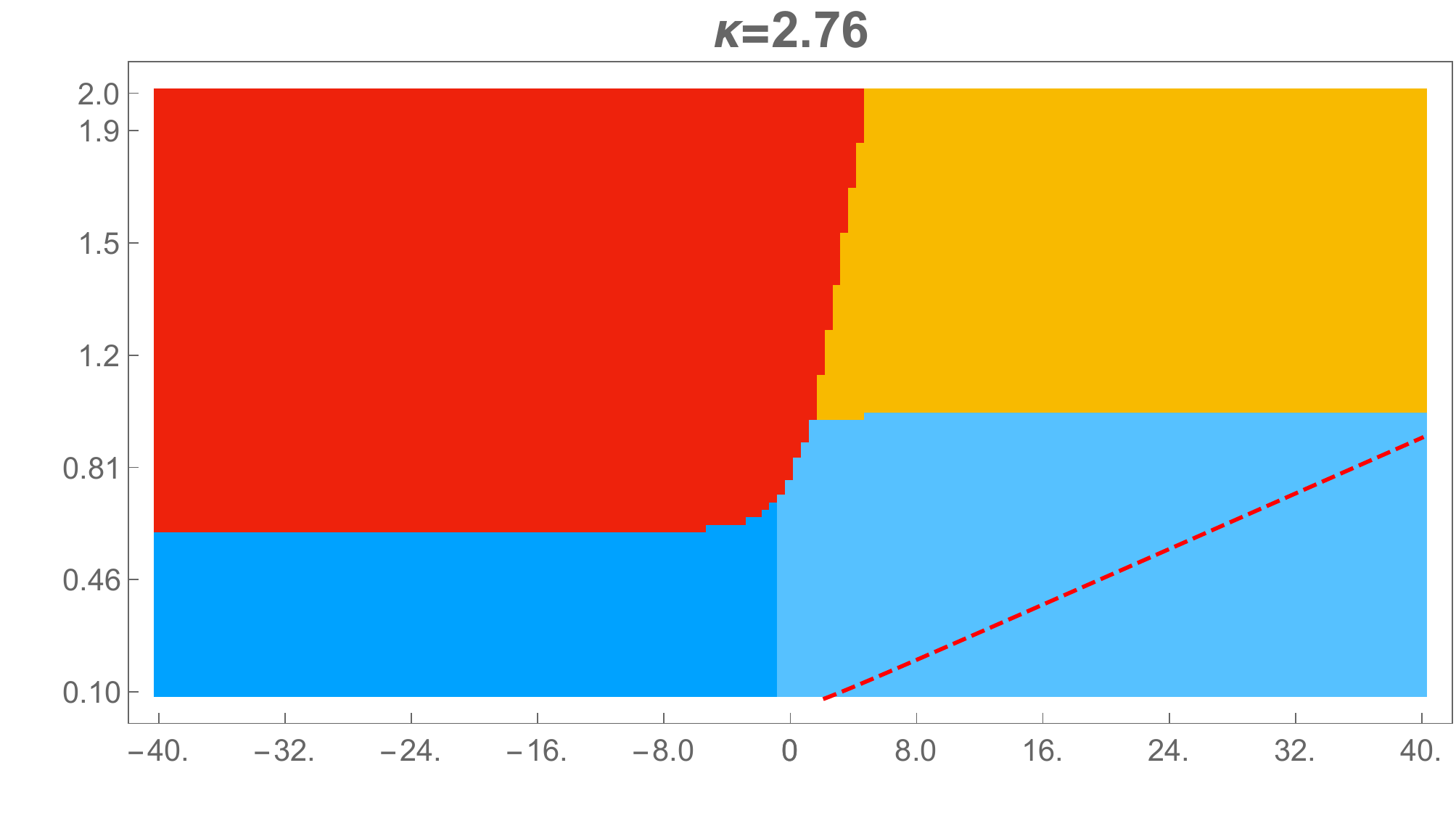}
 \end{subfigure}
\vskip 6mm
 \begin{subfigure}{.47\textwidth}
  \centering
  \includegraphics[width=1\linewidth]{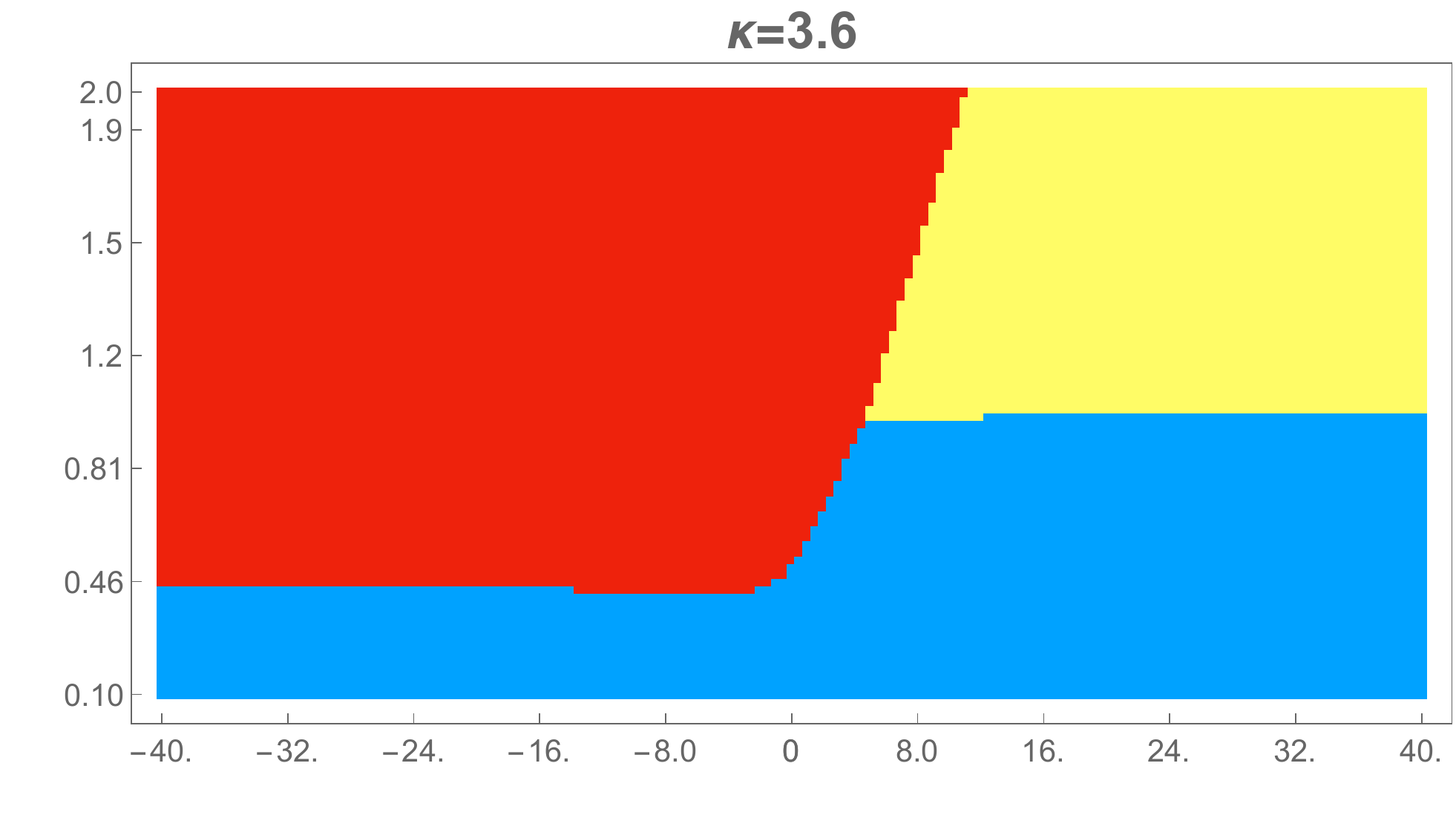}
 \end{subfigure}%
\hspace{4mm}
\begin{subfigure}{.47\textwidth}
  \centering
  \includegraphics[width=1\linewidth]{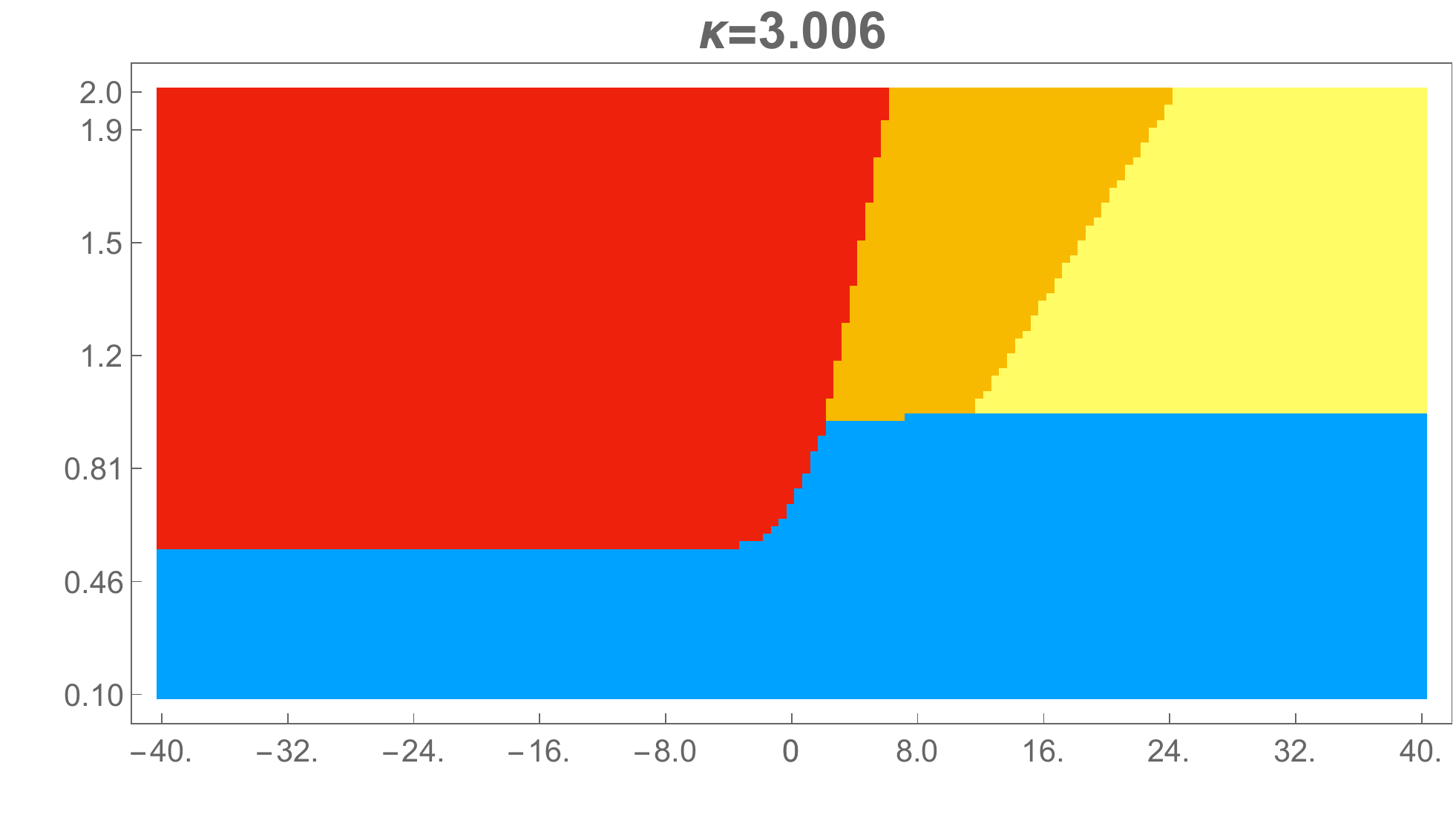}
 \end{subfigure}
 \caption{\footnotesize   Phase diagrams  for  $b=3$, and  values of the  tension that increase
 from the top-left figure clockwise.
The  horizontal  and vertical axes are  $\alpha:= \gamma - \gamma^{-1}$ and $\tau$.
The broken red curve is the bound $\tau = \tau_2^*(1+\gamma)$  below which the 
 hot solution does not exist
 (there is no such bound  in the lower panels in which  the tension $\lambda > \lambda_0$)}.
Note the absence of a warm phase in the left  ($\gamma <1$)  region of the diagrams.
For the heaviest wall all non-hot solutions are double-center.
\label{fig:11}
\end{figure}

     Figure \ref{fig:11}   presents   the phase diagram  in the  case of non-degenerate AdS vacua, 
$b=\ell_2/\ell_1 = c_2/c_1 = 3$,  and  for  different values of the tension in the allowed range,  $\kappa\in (2,4)$.  
Since there is no $\gamma \to  \gamma^{-1}$  symmetry, $\gamma$ here varies  between  0 to $\infty$.
To avoid squeezing the $\gamma \in (0,1)$ region,
we  use for  horizontal axis  $\alpha:= \gamma - \gamma^{-1}$.  This   is almost linear
 in the larger of $\gamma$ or $\gamma^{-1}$, when either  of these  is large,  but  the region 
 $\gamma \approx  1$ 
is  distorted   compared to   figs.\,\ref{fig:9} and \ref{fig:10} of the previous section.
 
 The most notable  new feature in  these phase diagrams 
is the absence of a warm phase
in the region  $\gamma <1$.   This shows that it is impossible to keep the wall outside
the black hole when the  latter forms on the false-vacuum side. 
   From the perspective of  the dual  ICFT, see section \ref{puzzles}, 
the absence of  {\small [X,H1]}-type  solutions  means that 
no interfaces, however heavy,  
can  keep  CFT$_1$ in the confined phase   
if  CFT$_2$ (the theory with larger central charge)  has already   deconfined. 

    We suspect that this is    a feature of  the thin-brane  
 model which does not allow   interfaces  to be perfectly-reflecting \cite{Bachas:2020yxv}.

      Warm solutions  with the horizon in the pink slice 
 appear   to altogether disappear  above the  critical ratio of central charges
$b_c=3$.\footnote{This critical value
was also noticed in  ref.\,\cite{Simidzija:2020ukv}, who also note  that multiple branes can evade the bound
confirming the intuition that it is a feaure specific to  thin branes. 
As a matter of fact, although  {\small [X,H1]} solutions do exist 
for  $b <3$ as we show below, they have  very large  $\gamma$, outside the range of our  
numerical plots, unless $b$ is very close to 1.
} 
 The boundary conditions corresponding to  topologies  of type {\small [X,H1]} are 
given by eqs.\, \eqref{hatDir}. We plotted the right-hand side of the second condition \eqref{hatDir}
for different   values of $\lambda$ and  $\mu$ in their allowed range, and found no solution
with positive $\tau_1$  for $b>3$. 
Analytic evidence for the existence of a strict  $b_c=3$ bound can be found by considering 
the limit of a maximally isolating wall, $\lambda\approx \lambda_{\rm max}$, and  of a shrinking 
green slice   $\hat \mu\to -\infty$.  In this limit, the right-hand side of \eqref{hatDir} can be computed 
in closed form with the result
\bea
\tau_1(\hat \mu)  =  {\pi\over \sqrt{-\hat\mu}} \biggl(2 - \sqrt{1+ {\ell_2\over \ell_1}} \,\,\biggr)  + {\rm subleading}\ .
\eea
We took  {\small X=E1} as dictated by the  analysis of sweeping transitions, see 
section \ref{sec:71} and in particular eq.\,\eqref{BHsweep}. 
This limiting   $\tau_1(\hat \mu)$
 is negative for $b>3$,  and positive for $b<3$ 
 where  warm  {\small [E1,H1]} solutions do exist, as claimed.

   An interesting corollary is that end-of-the-world branes cannot  avoid the horizon of a black hole, 
since  the near-void limit,  $\ell_1 \ll \ell_2$, is in the range that has no {\small [X,H1]}  solutions.


  \subsection{Unstable black holes} 

      The   phase diagrams  in figs.\,\ref{fig:9}, \ref{fig:10}, \ref{fig:11} show the   solution
with the  lowest free energy in various  regions of parameter space. 
Typically, this dominant phase  coexists with  solutions that describe unstable or metastable
 black-holes which   are ubiquitous
 in the  thin-wall model.\footnote{For a similar discussion
of   deformed   JT gravity see   ref.\cite{Witten:2020ert}. Note that 
in  the  absence of a   domain wall,  the only static black hole solution  of pure 
Einstein gravity in  2+1 dimensions  is  the non-spinning
BTZ black hole. 
}

\begin{figure}[h!] 
 \centering
   \includegraphics[width=1.04\linewidth]{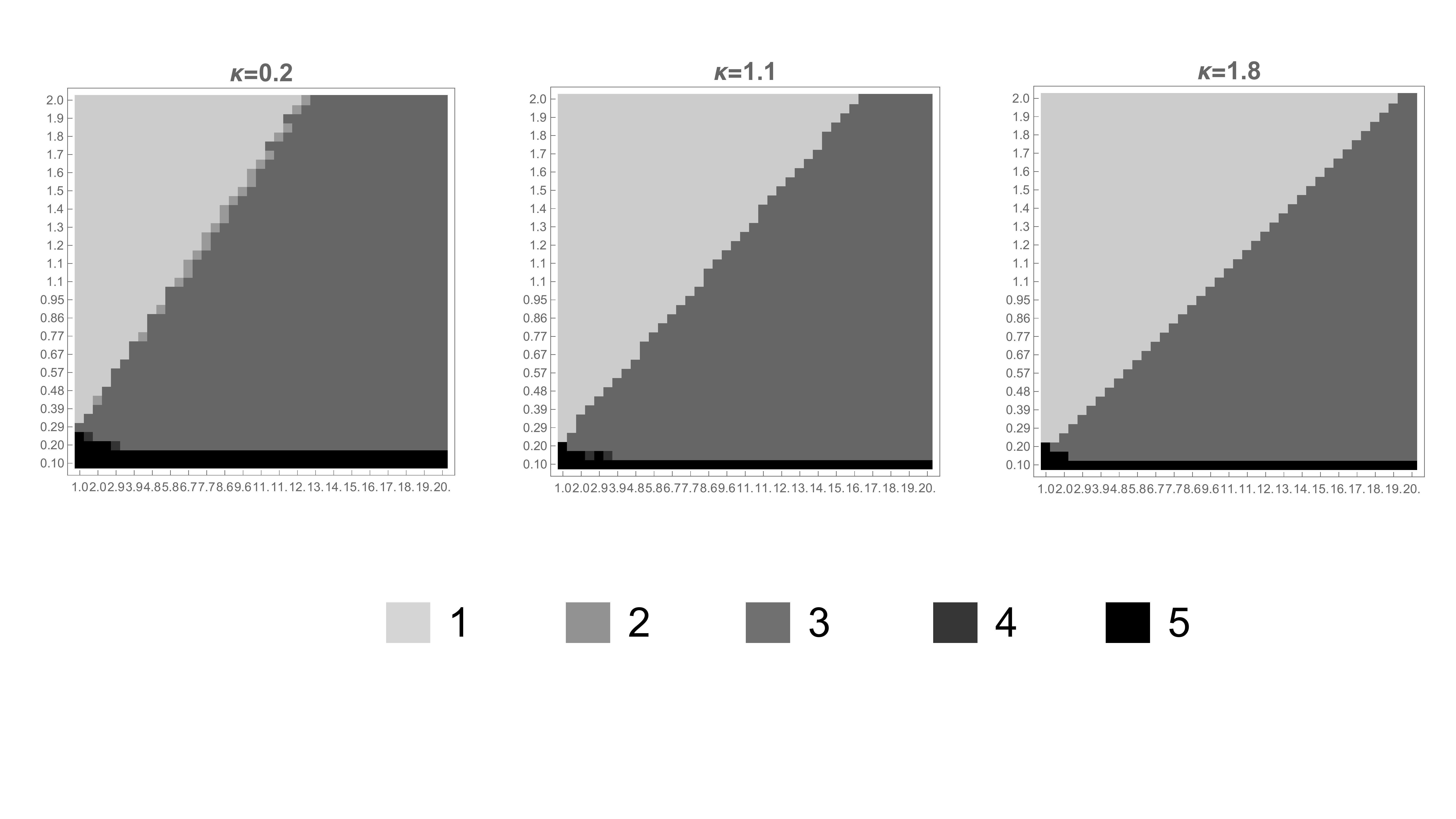}
\vskip -11mm
   \caption{\footnotesize  The number of independent black hole solutions in the $(\gamma, \tau)$ 
parameter space  for $b=1$, and three values of the tension $(\kappa =  0.2;\ 1.1; {\rm and} \    1.8)$.
The darker the shade the larger the number of black holes.
}
 \label{fig:13}
\end{figure}

    Figure  \ref{fig:13} shows   the number
of  black hole solutions in the degenerate case,  $b=1$, for  small, intermediate and large wall tension, 
 and in different regions  of the $(\tau, \gamma)$ parameter space.  
The axes are the same as   in figs.\,\ref{fig:9} and \ref{fig:10}
but  the range of $\gamma$ is halved.  At sufficiently high temperature the growing horizon captures
 the wall, and the only solution is the hot solution.  We see
 
\noindent  however that in 
  a large region of intermediate
temperatures  the hot solution coexists with two warm solutions. 
Finally at  very  low temperature the hot solution coexists with  four  other black-hole solutions, 
 two on either side of the wall.  
The dominant phase  in this region  is  vacuum, so the  black holes  play no role
in the canonical ensemble.

      The hot solution exists almost everywhere, except when $\lambda <\lambda_0$
and $\tau = \tau_2(1+\gamma) < \tau_2^*(1+\gamma)$  with $\tau_2^*$ given by eq.\,\eqref{tau2star}.
It has positive specific heat even when it is not the dominant phase. 
For warm black holes, on the other hand, the  specific heat can have  either sign. 
   One can see  this  semi-analytically by focussing once again  on  our favourite   near-critical region  
 $\lambda \approx \lambda_0$.  Simple inspection of fig.\,\ref{fig:tau2} shows that  in some range 
 $ \tau_2^* < \tau_2 < \tau_2^{\rm max}$ the hot solution coexists with two  nearby warm solutions. 
 At the maximum $\tau_2^{\rm max}$,  where $d\tau_2/d\mu=0$, the warm  solutions merge and
then disappear. Since the black hole is in the $j=1$ slice, $M_1= (2\pi T)^2$ and their 
 energy reads
\bea
 E_{\rm [warm]} \,=\,  {1\over 2}(\ell_1 M_1 L_1 + \ell_2 M_2 L_2)\,=\,
 2\pi^2 T^2 L_2 \, \bigl(  \ell_1  \gamma  + \ell_2 {\mu } \,\bigr)\ . 
\eea
Taking a derivative with respect to $T$ with  $L_1, L_2$  kept fixed 
 we  obtain 
  \bea\label{sheat}
{d\over dT} E_{\rm [warm]} \,=\,  {2\over T}E_{\rm [warm]}
 + 
 2\pi^2 T^2 \,  L_2^{\,2}\, \ell_2\, {d \mu\over d\tau_2} \ . 
\eea
Near $\tau_2^{\rm max}$ the dominant contribution to this expression comes from the 
derivative ${d \mu/  d\tau_2}$   which 
jumps  from $-\infty$ to $+\infty$. It follows that  the warm black hole with the higher mass has negative
specific heat,   and should decay  
to its companion black hole either classically or in the quantum theory.\footnote{We have 
 verified  numerically that   the black holes with negative specific heat are never the ones 
with  lowest free energy, a conclusion similar to the one reached in 
  deformed JT gravity in ref.\,\cite{Witten:2020ert}. 
}
 
 It would be very interesting to calculate this decay process, but we leave this for future work.

  One last comment  concerns  transitions from the double-center vacuum geometries,
 of type {\small [E1,E1]},
to warm solutions where the wall avoids the horizon.  One can ask what side of the wall does the
black hole choose. A natural guess is that   it  forms in the deepest of the two AdS  wells.
The relative depth is the ratio of blueshift factors at the two rest points, 
     \bea
   {\mathfrak  R} := \sqrt{g_{tt}\vert_{r_1=0}\over g_{tt}\vert_{r_2=0}}\,  =   \,{\ell_2\over \ell_1 \sqrt{\mu(\gamma)}} \,\ . 
        \eea
One expects the black hole to form in the $j=1$ (green) slice if  ${\mathfrak  R}<1$ and
in the $j=2$ (red) slice if  ${\mathfrak  R} > 1$. Our numerical plots confirmed in all cases this expectation.



\section{Outlook}\label{outlook}

   One  urgent  question,   already noted in the introduction,
 is  how much of this   analysis   will  survive  in top-down interface 
 models,  where gravitating  domain walls are typically thick. 
The order parameters of the Hawking-Page  and  sweeping transitions -- the area of the horizon and
 the number of rest points for inertial observers,  do not depend on the  assumption of a thin wall
and could go through.  The warm-to-hot transition,  on the other hand,  
may be   replaced by a crossover, since there is no sharp criterion to decide if a thick wall
enters or avoids the horizon.  As discussed, however,  in section \ref{puzzles}  a sharp
order parameter, such as a Polyakov loop,  may be suggested by the field theory side
of the correspondence.  

    One  other  question left open in the present work is the entanglement structure of the equilibrium
states.  Indeed, a   guiding thread of   our  paper  were the intersections  of 
the domain wall with  the black hole horizon and the trajectories of inertial observers. The  
 Ryu-Takayanagi (RT) surfaces  \cite{Ryu:2006bv,Ryu:2006ef} are 
another natural class of curves  whose intersection with the wall should be studied
      along  
lines similar to   ref.\,\cite{Akal:2020twv,Deng:2020ent} for BCFT.

Simple extensions of  the minimal model, such as
the addition of a Chern-Simons field (see e.g.\,\cite{Zhao:2020qmn}) might also be worth exploring.  

    Last  but not least,  the simplicity of the model and its  rich spectrum of black holes  
make it a promising ground where to try to shed some more light on the recent  exciting developments 
related to black hole evaporation, islands and the 
Page curve \cite{Penington:2019npb,Almheiri:2019psf,Almheiri:2019hni}. 
We hope to return to some of these questions in the near future. 

\vskip 0.5cm

\noindent {\bf\large Aknowledgements}
\smallskip

   We are grateful to  Mark Van Raamsdonk  for his  critical reading of  a  preliminary draft
of this  paper and for many useful comments.   Many  thanks also to   Panos Betzios, Shira Chapman, 
   Dongsheng Ge, Elias Kiritsis, Ioannis Lavdas, Bruno Le Floch, 
Emil Martinec, Olga Papadoulaki  and Giuseppe Policastro
for  discussions  during the course of this work.

 

\newcommand{\h}{\hat}
\newcommand{\eps}{\epsilon}
\newcommand{\lam}{\lambda}
\renewcommand{\d}{\delta}
\renewcommand{\th}{\theta}
\newcommand{\al}{\alpha}
\newcommand{\z}{\zeta}
\newcommand{\g}{\gamma}
\newcommand{\be}{\beta}
\newcommand{\co}{\nabla}
\newcommand{\pa}{\partial}
\newcommand{\we}{\wedge}
\newcommand{\si}{\sigma}
\newcommand{\ti}[1]{\Tilde{#1}}
\renewcommand{\cal}[1]{\mathcal{#1}}

 \appendix

\newcommand{\s}{\sigma}

\newcommand{\gam}{\gamma}
 \newcommand{\sig}{\sigma}
  
   
\section{Renormalized on-shell action}
\label{app:1}

The Euclidean action of  the holographic-interface  model, in units $8\pi G=1$, 
is the sum of  bulk, brane, boundary and corner contributions, see e.g.\cite{Takayanagi:2019tvn}
 \bea\label{B1}
I_{\rm gr} =    -\frac{1}{2}&\hskip -2.5mm \int_{{\mathbb  S}_1}d^3x 
\sqrt{g_1}\,(R_1+\frac{2}{\ell_1^2})  -\frac{1}{2}
 \int_{{\mathbb  S}_2}d^3x\sqrt{g_2}\,(R_2+\frac{2}{\ell_2^2}) +   \lam \int_{\mathbb   W} d^2s
     \sqrt{\hat g_w} \ \ \ \  \ \  \nonumber  \\
    & \hskip -6mm  + \int_{\partial {\mathbb  S}_1} d^2s \sqrt{\hat g_1}\, K_1 
     + \int_{\partial {\mathbb  S}_2} d^2s \sqrt{\hat g_2}\, K_2 
    \ + \ \int_{\rm C} (\theta - \pi)  \sqrt{\hat g_c} + {\rm  c.t. }\    
     \eea
 where the counterterms,  abbreviated  above by c.t.,  read \cite{Balasubramanian:1999re}
 \bea 
  {\rm c.t.} =  \,  {1\over \ell_1}\int_{{\rm B}_1} \sqrt{\hat g_1}  
  \, +  {1\over \ell_2}\int_{{\rm B}_2} \sqrt{\hat g_2} \, -\,  
  \int_{{\rm B}_1\cap {\rm B}_2} (\theta_1+\theta_2)  \sqrt{\hat g_c} \   .  
\eea
 Here  $\mathbb  S_j$   are the  spacetime  slices whose boundary 
 is the sum of   the cutoff surface   B$_j$ and  of the string worldsheet $\mathbb  W$, i.e.
 $\partial \mathbb  S_j =$B$_j \cup\mathbb   W$\,. 
 The induced metrics are denoted by  hats. 
   The $K_j$  are  traces  of the extrinsic curvatures  on  each  slice
  computed with the inward-pointing normal vector.
  Finally, in addition to the standard  Gibbons-Hawking-York boundary terms,  
  one must  add the Hayward  term  \cite{Hayward:1993my, Takayanagi:2019tvn} 
 at   corners of   $\partial \mathbb  S_j$  denoted by C.\,\footnote{These play no role  here, but they
 can be important in the case of string junctions.}
 There is at least one such  corner at  the cutoff surface,    B$_1\cap$B$_2$, 
where  
 $\theta -\pi$ is the sum of  the angles $\theta_j$  defined in figure \ref{fig:3}.

  Let us break the   
  action into an interior and a  conformal boundary  term, $I_{\rm gr} = I_{\rm int} + I_{\rm B}$, 
  with  the former  including  contributions from the worldsheet W.  
Using  the field equations $R_j = - {6/\ell_j^2}$ and $K_1\vert_{\rm W}+K_2\vert_{\rm W}  = -2\lam$, 
and  the volume elements
that follow from  eqs.\,\eqref{charts}    and (\ref{00}, \ref{11}), 
$$   \sqrt{g_j}\,d^3x   =\ell_j r_i dr_j dx_j dt\,   \quad    {\rm and} \quad 
  \,\sqrt{\hat g_w}\,d^2s =  \sqrt{fg}\,  d\sigma dt\ , 
$$
we can write the interior on-shell action as follows\,: 
  \begin{equation}\label{B3}
 I_{\rm int} \, =  \,    \frac{2}{\ell_1}  \int_{{\Omega}_1} r_1\,   dr_1  dx_1 dt   
  + \frac{2}{\ell_2}  \int_{{\Omega}_2} r_2\,    dr_2  dx_2 dt   -
\lambda \int_{\mathbb{W}} \sqrt{fg}\, d \sigma dt \ .  
 \end{equation}
We have been  careful to distinguish 
the spacetime slice S$_j$ from the   coordinate chart $\Omega_j$, because we will 
now use   Stoke's theorem  treating $\Omega_j$ as part of flat Euclidean space, 
\bea\label{Stokes}
   \sum_{j=1,2}\,  {2\over \ell_j}  \int_{{\Omega}_j} r_j   dr_j  dx_j dt \  = \  \sum_{j=1,2}\, 
 {1\over \ell_j} \oint_{{\partial \Omega}_j} r_j^2 (\hat r_j\cdot d\hat n_j)  dt\ , 
\eea
  with $d\hat n_j dt$ the   
 surface element on the boundary  $\partial \Omega_j$. 
 Crucially,   the boundary of $\Omega_j$  
 may  include a horizon which is a  regular 
 interior submanifold  of the Euclidean spacetime and is not therefore part
 of $\partial \mathbb  S_j$\,. In particular, there is no Gibbons-Hawking-York  
 contribution there. 
 
     The boundary integral in eq.\,\eqref{Stokes} receives  contributions from
    the three  pieces of  ${\partial \Omega}_{1,2}$\,:  the cutoff surface  
    B$_1\cup$B$_2$, the horizon if there is one,   and the worldsheet $\mathbb{W}$.  
    Conveniently,  this last term precisely   cancels the 
    third  term in  \eqref{B3} by virtue of 
  the Israel-Lanczos equation \eqref{Israel}. Thus, after all the dust has settled,  the  action
  can be written as  the sum of terms evaluated either at   the black-hole horizon or  at the cutoff.
 After integrating over  periodic time the interior part of the action, eq.\,\eqref{B3}\,,  reads
   \bea\label{B5bis}
 I_{\rm int} \, =   \ {1\over \ell_1 T }  \, \Bigl[ r_1^2 \,\Delta x_1   \Bigr]_{\rm Hor}^{{\rm B}_1} +\,  
 {1\over \ell_2  T}  \, \Bigl[ r_2^2 \,\Delta x_2  \Bigr]_{\rm Hor}^{{\rm B}_2}  
 \eea
  where we  employ  
   the shorthand notation 
   $[X]_a^b = X\vert_b  - X\vert_a$\,, and 
       $X\vert_a$ for $X$ evaluated at $a$\,.
       If  the slice $\mathbb  S_j$ does not contain a horizon the corresponding contribution is
  absent.

  
      We now   turn to the conformal-boundary contributions from the lower line in the action \eqref{B1}.
  For a fixed-$r_j$  surface, 
    the   inward-pointing unit normal  expressed as a 1-form   is 
   {\bf n}$_j =  - dr_j / \sqrt{r_j^2 - M_j \ell_j^2}$\,. One finds  after a little algebra
   (we here drop the index $j$  for simplicity)
\bea
   K_{xx} = K_{tt} = - {r\over \ell}  \sqrt{r^2 - M\ell^2} \, \Longrightarrow\, 
   \sqrt{\hat g}\, K =  \, - {1 \over \ell} (2r^2 - M\ell^2)\ . 
\eea
  Combining the Gibbons-Hawking-York terms and the counterterms   gives
  \bea
  I_{\rm B} =  {1\over \ell_1 T} ( r_1 \sqrt{r_1^2 - M_1\ell_1^2} - 2r_1^2 +  M_1\ell_1^2)\, 
  \Delta x_1\,\Bigl\vert_{{\rm B}_1}\,+\,(1 \rightarrow 2)\ . 
  \eea
  Expanding  for large  cutoff radius, $r_j\vert_{{\rm B}_j} \to \infty$,   and dropping 
  the terms that vanish in the limit 
  we obtain
   \bea\label{B8}
  I_{\rm B} =  {1\over \ell_1 T} (  - r_1^2 +  {1\over 2} M_1\ell_1^2)\, 
  \Delta x_1\,\Bigl\vert_{{\rm B}_1}\,+\,(1 \rightarrow 2)\ . 
  \eea
 Upon adding up    \eqref{B5bis} and \eqref{B8} the leading divergent term cancels, giving
 the following result for the renormalized on-shell action\,: 
\bea\label{finalI}
 I_{\rm gr} \ = \  {M_1\ell_1 \over 2 T}  \bigl( 
L_1   -  2  \Delta x_1 \bigl\vert_{{\rm Hor}} \bigr) \,+\,{M_2\ell_2 \over 2 T}  \bigl( 
L_2   -  2  \Delta x_2 \bigl\vert_{{\rm Hor}} \bigr) \,\  . 
\eea
 We used here the fact that  $\Delta x_j\vert_{{\rm B}_j} = L_j$,  and that $r_j^2 = M_j\ell_j^2$
 at the horizon when  one exists. We also used implicitly  the fact that for smooth strings the 
 Hayward term receives  no contribution  from  the interior and is removed by the counterterm
 at the boundary. 
   
 
As a check of this on-shell action  let us compute the  entropy. 
 Using our   formula for the internal energy
$ \langle E\rangle  = {1\over 2}  (M_1 \ell_1 L_1 + M_2 \ell_2 L_2)$, see section \ref{sec:2}, 
and  $ I_{\rm gr} =    \langle E\rangle/T  - S$  
we find  
\bea
 S =  {1\over  T}&&  \hskip -10mm \bigl(M_1 \ell_1  \Delta x_1 \bigl\vert_{{\rm Hor}} + 
 M_2 \ell_2  \Delta x_2 \bigl\vert_{{\rm Hor}}\bigr)  \nonumber 
\\
&=& \hskip -2mm  4\pi^2 T \bigl(  \ell_1  \Delta x_1 \bigl\vert_{{\rm Hor}} +  
 \ell_2  \Delta x_2 \bigl\vert_{{\rm Hor}}\bigr)
\, =\, {A({\rm horizon})\over 4G}\ . 
\eea
In the lower line we   used  the fact that  $M_j = (2\pi T)^2$ and $r_j^{\rm H} = 2\pi T \ell_j$ for
slices  with horizon,
plus our choice of  units $8\pi G=1$.  
The  calculation thus reproduces correctly the  Bekenstein-Hawking entropy.


 \section{Opening arcs as  elliptic integrals}
 \label{app:2}

 In this appendix we   express the opening arcs, eqs.\,\eqref{Dir},  in terms of complete  elliptic integrals 
  of  the first, second  and third kind, 
  \bea
  {\rm\bf K}(\nu) = \int_0^1 \frac{dy}{\sqrt{(1-y^2)(1-\nu y^2)}}  
  \eea
  \vskip -3mm
  \bea
  {\bf E}(\nu)= \int_0^1\frac{  \sqrt{1-\nu y^2}\,dy }{ \sqrt{1-y^2}} \ . 
  \eea
  \vskip -3mm
  \bea
  {\bf \Pi}(u,\nu)= \int_0^1\frac{dy}{(1- uy^2)\sqrt{(1-y^2)(1-\nu y^2)}} \ . 
  \eea
 Consider the boundary conditions\,\eqref{Dira}. The other conditions\,(\ref{Dirb},\,\ref{Dirc})
   differ only  by the  constant periods or horizon arcs, $P_j$ or $\Delta x_j\vert_{\rm hor}$. 
Inserting  the expression  \eqref{soln} for $x_1^\prime$   gives 
 \begin{equation}
\begin{aligned}
   L_1 = - \int_{\sigma_+}^\infty \frac{\ell_1 \,d\sigma }{(\sigma+M_1\ell_1^2)}\, 
    \frac{(\lambda^2+ \lambda_{0}^2)\,\sigma    +M_1-M_2}
    {\sqrt{ A \sigma  (\sigma-\sigma_+)(\sigma-\sigma_-)}} \ ,  
\end{aligned}
\end{equation} 
  and likewise  for  $L_2$.  The roots $\sigma_\pm$ are given 
  by eqs.\,(\ref{ABC},\,\ref{spm}). 
   We assume that we are not in the  
   case  {\small [H2, H2]} where
   $M_1=M_2>0$,   nor in the fringe  case   $\sigma_+ = - M_j \ell_j^2$
   when  the string goes through an AdS center. These cases will be treated separately.

     Separating  the integral in two parts,  and trading  the integration variable $\sigma$ for  $y$,  with
     $y^2 := \sigma_+/\sigma$, we obtain 
$$
    L_1 =  - \frac{2\ell_1}{\sqrt{A\,\sigma_+}}\bigg[ {M_1 - M_2\over M_1\ell_1^2} 
   \int_0^1 \frac{dy}{\sqrt{(1-y^2)(1-  \nu
     y^2  )}}
$$\vskip -4mm
\bea
    +\Bigl(  (\lambda^2+ \lambda_{0}^2)  -   {M_1 - M_2\over M_1\ell_1^2}  \Bigr) \,\int_0^1 \frac{y^2 dy}{
    (1-  u_1y^2) \sqrt{(1-y^2)(1-  
     \nu y^2  )}} \, \bigg]\ \  \
\eea
   where 
$\nu =   \sigma_-/\sigma _+ $ and $u_1 = - M_1\ell_1^2/\sigma_+$\,.  Identifying the elliptic integrals 
finally gives
\bea\label{B5} 
 L_1 =  - \frac{2\ell_1}{\sqrt{A\,\sigma_+}}\bigg[ \frac{M_1-M_2}{M_1\ell_1^{\,2}}\, 
 \Bigl( {\rm\bf K}(\nu) - {\bf \Pi}(u_1, \nu)\Bigr)  + (\lambda^2+ \lambda_{0}^2)\, {\bf \Pi}(u_1, \nu)\bigg]\ ,   \ \  \
\eea    
 and a corresponding expression for $L_2$ 
\bea
 L_2 =  - \frac{2\ell_2}{\sqrt{A\,\sigma_+}}\bigg[ \frac{M_2-M_1}{M_2\ell_2^{\,2}}\, 
 \Bigl( {\rm\bf K}(\nu) - {\bf \Pi}(u_2, \nu)\Bigr)  +
  (\lambda^2 - \lambda_{0}^2)\, {\bf \Pi}(u_2, \nu)\bigg]\  \  \
\eea             
   with 
  $u_2 = - M_2\ell_2^2/\sigma_+$\,.   The prefactors in  \eqref{B5}  diverge when $M_1\to 0$ 
  but the singilarity  is removed  by  expanding   ${\bf \Pi}(u_1, \nu)$ around $u_1=0$.   In this limit
 \begin{subequations}
   \bea
   L_1(M_1=0)  = - \frac{ 2\ell_1}{\sqrt{A\,\sigma_+}}
   \bigg[ \frac{M_2}{\sigma_-}({\rm\bf E}(\nu)-{\rm\bf K}(\nu))
   + (\lambda^2 +\lambda_0^2) {\rm\bf K}(\nu) \bigg]
           \eea  \vskip -2mm        
{\rm  and similarly}\vskip -6mm
   \bea
   L_2(M_2=0)  = - \frac{ 2\ell_2}{\sqrt{A\,\sigma_+}}
   \bigg[ \frac{M_1}{\sigma_-}({\rm\bf E}(\nu)-{\rm\bf K}(\nu))
   + (\lambda^2 - \lambda_0^2) {\rm\bf K}(\nu) \bigg]
           \eea  
 \end{subequations}                 
with  ${\rm\bf E}(\nu)$   the complete elliptic integral of the second kind. 
 
 \smallskip      
  The  $M_1=M_2>0$  geometries  correspond  to 
  the high-temperature phase  where    $M_j  = (2\pi T)^2$,   $\sigma_+=0$
  and  $\sigma_- =  - (4\pi T \lambda)^2/A$. 
    The integrals 
  \eqref{Dirc}  simplify to elementary functions in this case: 
   $$
  L_1 - \Delta_1^{\rm Hor}\  =\    - {\ell_1 (\lambda^2+ \lambda_{0}^2) \over \sqrt{A \, \vert \sigma_- \vert\,} }
  \underbrace{\int_0^\infty {ds \over  (s + a)\,\sqrt{s+1}\,}}_{\textstyle\begin{array}{c}
  = {2\over \sqrt{1-a}}{\rm arctanh}({  \sqrt{1-a}})  \end{array}
  }
$$ 
with $a = A\ell_1^2/4\lambda^2$. Using the expression \eqref{ABC} for $A$, and going through the same
steps   for $j=2$,  
gives  after a little  algebra
\begin{subequations}\label{arctanh}
    \bea
  L_1 - \Delta_1^{\rm Hor}\  =\   - {1\over  \pi T}\, {\rm tanh}^{-1} \left( {\ell_1  (\lambda^2 + \lambda^2_0)
  \over 2\lambda}\right)\ , 
  \eea\vskip -3mm
    \bea
  L_2 - \Delta_2^{\rm Hor}\  =\   - {1\over  \pi T}\, {\rm tanh}^{-1}  \left( {\ell_2 (\lambda^2 -  \lambda^2_0)
  \over 2\lambda}\right)\ . 
  \eea  
\end{subequations}  
  Interestingly,  since $\Delta_2^{\rm Hor}$  must be  positive,  $T L_2$
  is bounded from below in the range  $\lambda < \lambda_0$ as discussed in section \ref{sec:62}.  
  
   In the high-temperature  phase the on-shell action,  eq.\,\eqref{finalI},   reads
 \bea
 I_{\rm gr}^{\rm (high-T)} =  4 \pi^2 T\Bigl[ -{1\over 2} 
  (\ell_1 L_1 + \ell_2 L_2) +  \ell_1 (L_1-  \Delta_1^{\rm Hor})+  \ell_2 (L_2-  \Delta_2^{\rm Hor})\Bigr]\,. \  \ 
 \eea
 Using  the expressions \eqref{arctanh} and rearranging the arc-tangent functions gives
 \bea
  I_{\rm gr}^{\rm (high-T)} :=   {E\over T}  - S  =  - 2\pi^2 T(\ell_1 L_1 + \ell_2 L_2)  - \log\, g_I
 \eea
 where  the interface entropy  $S=\log\, g_I$ is given by eq.\,\eqref{vanR}. 
 

 \section{Sweeping is continuous}
 \label{app:3}
  
  In this appendix we show that sweeping  transitions are continuous.  
 
 We focus for 
  definiteness on the  sweeping of the $j=2$ AdS  center  at zero temperature
(all other cases work out   the same).  
 The transition  takes place when   
 $\mu$  crosses   the critical value $\mu_2^*$  given by eq.\,\eqref{mucr}. 
Setting    $\mu = \mu_2^*(1 - \delta)$ in expression \eqref{mub} gives
$$
f_2(\mu) = \,     \frac{\ell_2}{ \sqrt{A}} 
\int_{s_+}^\infty ds\, 
\frac{(\lam^2-\lam_0^2)(s-\mu \ell_2^2)\,+ \delta }
{(s - \mu \ell_2^{ 2})\,\sqrt{A s ( s- s_+)(s-s_-)} }
$$
\bea\label{J}
= \, \frac{2\ell_2(\lam^2-\lam_0^2)}{\sqrt{As_+}}\,{\bf K} \bigl(\frac{s_-}{s_+}\bigr)\,+\,\frac{\ell_2 \,\delta}{\sqrt{A}}\,
\underbrace{\int_{s_+}^\infty
 \frac{ds }{(s-\mu \ell_2^2)\sqrt{s (s- s_+)(s-s_-)}}}_{J}\ . \ \ 
\eea
The first term is continuous at $\delta=0$, but the second requires some care because the integral
$J$  diverges. This is because for  small $\delta$ 
\bea
 s_+  -   \mu \ell_2^2 \,=\,   \frac{\delta^2}{4 \lam^2 \mu_2^*  }+\cal{O}(\delta^3)\ , 
\eea
as one finds by explicit computation of  the expression 
 \eqref{s++cold}. 
If we set $\delta=0$,  $J$  diverges near the lower integration limit. To bring the singular behavior to $0$
we perform the change of variable $u^2 =s -s_+$, so that 
\bea
 J = \int_0^\infty \frac{2 du}{(u^2 +  \delta^2/4\lam^2 \mu_2^* )\sqrt{ (u^2+s_+^*)(u^2 +s_+^*-s_-^*)}}\ , 
\eea
where we   kept only the leading order in $\delta$,  and $s_\pm^*$ are the roots at $\mu=\mu_2^*$. 
    Since $s_+^*$ and $s_+^* -s_-^*$ are positive and finite, the small-$\delta$ behavior of the integral  is
(after rescaling appropriately $u$)
\bea
J  =  {4 \lambda \vert \mu_2^* \vert \over \vert\delta\vert \sqrt{s_+^*(s_+^*-s_-^*)}}
\, \underbrace{\int_0^\infty {du\over u^2+1}}_{\pi/2} 
\ + \ {\rm finite}\ .  
\eea
Inserting in expression \,\eqref{J} and doing  some  tedious algebra
   leads finally to a discontinuity  of  the function   $f_2(\mu)$ equal to   sign$ (\delta)\, \pi/\sqrt{\mu_2^*}$\,.  
This is precisely what is required  for $L_2$, 
eq.\,\eqref{61}, to be  continuous when  the red ($j=2$) slice goes from type {\small E1} at negative $\delta$ to
type {\small E2} at positive $\delta$.


 \section{Bubbles   exist}
 \label{app:4}

  We   show here that the bubble phenomenon of section \ref{sec:Faraday} is indeed realized in a 
region of the parameter space of the holographic model.

This is the region of non-degenerate gravitational vacua ($\ell_2$ strictly bigger than $\ell_1$) 
and a sufficiently light domain wall. 
Specifically,  we  will show that for  $\lambda$ close to its minimal value,  $\lambda_{\rm min}$,  
the arc $L_1(\mu=0)$ is negative, so the wall self-intersects and $\mu_0$ is necessarily finite. 
\smallskip

Let  $\lambda =  \lambda_{\rm min}(1+  \delta)$  with $\delta \ll 1$. Setting $\mu=0$ and expanding 
eqs.\,\eqref{s++cold} for small  $\delta$ gives
 $$
A = {8 \lambda_{\rm min}^2\, \delta  \over  \ell_1\ell_2} +\cal{O}(\delta^2)  \ , \quad 
 s_+ =  {\ell_2\over 4 \lambda_{\rm min}}  +\cal{O}(\delta) \ , 
\quad 
 s_-  = - { \ell_1 \over  \ 2 \lambda_{\rm min} \delta } +\cal{O}(1)\ . 
$$
Plugging   into eq.\,(\ref{B5})  with $ M_2 = \mu M_1\approx 0$ we find :
\begin{equation}
    \sqrt{\vert M_1\vert}\,  
L_1 =  -\frac{2 }{\ell_1 \sqrt{A s_+ }}\left[ {\bf K}\left({s_-\over s_+}\right) +  (1-\frac{2\ell_1}{\ell_2})
{\bf \Pi}\left({\ell_1^2\over s_+},\, {s_-\over s_+}\right)\right]
\end{equation}
where we have only kept leading orders in $\delta$.  
Now we need the asymptotic form  of the elliptic integrals when their argument diverges
\begin{equation}
        {\bf K} \left[-\frac{a}{\delta}\right] \approx   {\bf \Pi}\left[u ,-\frac{a}{\delta}\right] \approx 
 -\frac{\ln(\delta)\sqrt{\delta}}{2\sqrt{a}}+\cal{O}(\sqrt{\delta}) 
\end{equation}
 for $\delta\to 0_+$ with $u, a$ fixed.  Using $a = 2\ell_1/\ell_2$ finally gives
\bea
  \sqrt{\vert M_1\vert}\, L_1\, \approx\,  \bigl(\frac{\ell_2}{\ell_1}-1\bigr)^{1/2} \ln(\delta)\, +\,
{\rm subleading}\ . 
 \eea
For  $\delta \ll 1 $ this is negative, proving our claim. Note that we took the green slice to be of type
{\small E2}, as follows from our analysis of the sweeping transitions for light domain walls --
 see  section \ref{sec:71}.


\end{document}